\documentclass[trackchanges]{aastex701}

\usepackage{multirow}
\usepackage{amsmath}

\usepackage[normalem]{ulem}
\usepackage{xcolor}

\begin{document}

\title{Helical Magnetic Field in the Acceleration--Collimation Zone of the M87 Jet}

\author[0000-0001-6558-9053]{Jongho Park}
\affiliation{School of Space Research, Kyung Hee University, 1732, Deogyeong-daero, Giheung-gu, Yongin-si, Gyeonggi-do 17104, Republic of Korea}
\affiliation{Institute of Astronomy and Astrophysics, Academia Sinica, P.O. Box 23-141, Taipei 10617, Taiwan, R. O. C.}
\email[show]{jparkastro@khu.ac.kr}

\author[0000-0002-8314-1946]{Kazuya Takahashi}
\affiliation{Research Center for the Early Universe, Graduate School of Science, University of Tokyo, Bunkyo, Tokyo 113-0033, Japan}
\affiliation{Frontier Research Institute for Interdisciplinary Sciences, Tohoku University, Sendai 980-8578, Japan}
\email[show]{k.takahashi.azure@gmail.com}

\author[0000-0002-7114-6010]{Kenji Toma}
\affiliation{Frontier Research Institute for Interdisciplinary Sciences, Tohoku University, Sendai 980-8578, Japan}
\affiliation{Astronomical Institute, Graduate School of Science, Tohoku University, Sendai 980-8578, Japan}
\email{toma@astr.tohoku.ac.jp}

\author[0000-0001-6906-772X]{Kazuhiro Hada}
\affiliation{Graduate School of Science, Nagoya City University, Yamanohata 1, Mizuho-cho, Mizuho-ku, Nagoya, 467-8501, Aichi, Japan}
\affiliation{Mizusawa VLBI Observatory, National Astronomical Observatory of Japan, 2-12 Hoshigaoka-cho, Mizusawa, Oshu, 023-0861, Iwate, Japan}
\email{hada@nsc.nagoya-cu.ac.jp}

\author[0000-0001-6081-2420]{Masanori Nakamura}
\affiliation{Department of General Science and Education, National Institute of Technology, Hachinohe College, Hachinohe City, Japan}
\affiliation{Institute of Astronomy and Astrophysics, Academia Sinica, P.O. Box 23-141, Taipei 10617, Taiwan, R. O. C.}
\email{nakamrms-g@hachinohe.kosen-ac.jp}

\author[0000-0001-9270-8812]{Hung-Yi Pu}
\affiliation{Department of Physics, National Taiwan Normal University, No. 88, Section 4, Tingzhou Road, Taipei 116, Taiwan, R. O. C.}
\affiliation{Centre of Astronomy and Gravitation, National Taiwan Normal University, No. 88, Section 4, Tingzhou Road, Taipei 116, Taiwan, R. O. C.}
\affiliation{Institute of Astronomy and Astrophysics, Academia Sinica, P.O. Box 23-141, Taipei 10617, Taiwan, R. O. C.}
\email{hypu@gapps.ntnu.edu.tw}

\author[0000-0001-6988-8763]{Keiichi Asada}
\affiliation{Institute of Astronomy and Astrophysics, Academia Sinica, P.O. Box 23-141, Taipei 10617, Taiwan, R. O. C.}
\email{asada@asiaa.sinica.edu.tw}

\author[0000-0002-3412-4306]{Paul T. P. Ho}
\affiliation{Institute of Astronomy and Astrophysics, Academia Sinica, P.O. Box 23-141, Taipei 10617, Taiwan, R. O. C.}
\email{pho@asiaa.sinica.edu.tw}

\author[0000-0002-2709-7338]{Motoki Kino}
\affiliation{Kogakuin University of Technology \& Engineering, Academic Support Center, 2665-1 Nakano-machi, Hachioji, Tokyo 192-0015, Japan}
\affiliation{National Astronomical Observatory of Japan, Osawa 2-21-1, Mitaka, Tokyo 181-8588, Japan}
\email{motoki.kino@gmail.com}

\author[0000-0001-8527-0496]{Tomohisa Kawashima}
\affiliation{National Institute of Technology, Ichinoseki College, Takanashi, Hagisho, Ichinoseki, Iwate, 021-8511}
\affiliation{Institute for Cosmic Ray Research, The University of Tokyo, 5-1-5 Kashiwanoha, Kashiwa, Chiba 277-8582, Japan}
\email{kawashima@g.ichinoseki.ac.jp}

\author[0000-0001-9799-765X]{Minchul Kam}
\affiliation{Institute of Astronomy and Astrophysics, Academia Sinica, P.O. Box 23-141, Taipei 10617, Taiwan, R. O. C.}
\email{mkam@asiaa.sinica.edu.tw}

\author[0009-0007-8554-4507]{Kunwoo Yi}
\affiliation{School of Space Research, Kyung Hee University, 1732, Deogyeong-daero, Giheung-gu, Yongin-si, Gyeonggi-do 17104, Republic of Korea}
\email{kunwoo@khu.ac.kr}

\author[0000-0001-6083-7521]{Ilje Cho}
\affiliation{Korea Astronomy and Space Science Institute, Daedeok-daero 776, Yuseong-gu, Daejeon 34055, Republic of Korea}
\affiliation{Yonsei University, Department of Astronomy, Seoul, Republic of Korea}
\affiliation{Instituto de Astrof\'{i}sica de Andaluc\'{i}a—CSIC, Glorieta de la Astronom\'{i}a s/n, E-18008 Granada, Spain}
\email{icho@kasi.re.kr}

\begin{abstract}

Relativistic jets from supermassive black holes are expected to be magnetically launched and guided, with magnetic energy systematically converted to bulk kinetic energy throughout an extended acceleration--collimation zone (ACZ). A key prediction of magnetohydrodynamic (MHD) models is a transition from poloidally dominated fields near the engine to toroidally dominated fields downstream, yet direct tests within the ACZ are hampered by weak polarization and strong Faraday rotation. We report quasi-simultaneous, high-sensitivity, multifrequency very long baseline interferometric polarimetry of M87 spanning 1.4--24.4\,GHz. We present high-fidelity, Faraday rotation-corrected maps of intrinsic linear polarization that continuously resolve the ACZ in the de-projected distance range of $\sim9.0 \times 10^3$ to $\sim3.6 \times 10^5$ gravitational radii from the black hole. The maps reveal pronounced north--south asymmetries in fractional linear polarization and electric vector position angle (EVPA), peaking in the inner ACZ at a projected distance of $\sim$20\,mas along the jet and remaining prominent out to $\sim$100\,mas. These signatures are best reproduced by models with a large-scale, ordered helical field that retains a substantial poloidal component--contrary to the rapid toroidal dominance expected under steady, ideal MHD. This tension implies ongoing magnetic dissipation that limits toroidal buildup over the ACZ. The handedness of the helix provides an independent constraint on the black hole's spin direction, supporting a spin vector oriented away from the observer, consistent with the orientation inferred from horizon-scale imaging. Farther downstream, the asymmetries diminish, and the EVPA and fractional polarization distributions become more symmetric; we tentatively interpret this as evolution toward a more poloidally dominated configuration, while noting current sensitivity and dynamic-range limits.

\end{abstract}


\section{Introduction} 
\label{sec:introduction}

A fraction of active galactic nuclei (AGN) launch relativistic jets, which are powered by accretion onto a supermassive black hole (SMBH) and accumulation of large-scale magnetic flux \citep[e.g.,][]{Blandford2019}. Horizon-scale imaging by the Event Horizon Telescope (EHT) detected a ring-like structure surrounding a dark shadow at the center of the giant elliptical galaxy M87 \citep{EHT2019a,EHT2019b,EHT2019c,EHT2019d,EHT2019e,EHT2019f,EHT2024,EHT2025}, interpreted as gravitationally lensed synchrotron emission from plasma surrounding the black hole. EHT polarimetric observations of M87 showed azimuthally distributed linear polarization around the ring \citep{EHT2021a, EHT2025b} and relatively low levels of circular polarization \citep{EHT2023}. These results are consistent with a magnetically arrested disk (MAD) state in M87 \citep{BR1974, Igumenshchev2003, Narayan2003, Tchekhovskoy2011}, in which the magnetic flux near the horizon saturates and significantly affects the dynamics of the flow around the black hole.

Jets are not fully formed at launch; they develop over an extended region where acceleration and collimation proceed gradually to $\sim10^{4}$--$10^{6}\,R_{\rm g}$ \citep{VK2004, Komissarov2007, Lyubarsky2009}, where $R_{\rm g} \equiv GM_{\rm BH}/c^2$ is the gravitational radius, with $G$, $M_{\rm BH}$, and $c$ the gravitational constant, black hole mass, and speed of light, respectively. This regime is referred to as the acceleration–collimation zone (ACZ; \citealp{Marscher2008}). Jets are continuously collimated in the ACZ, implying that the jet's opening angle decreases with increasing distance from the black hole, exhibiting a parabolic or quasi-parabolic shape. This collimation is believed to occur due to the interplay between the jet's internal pressure and the pressure of the external medium surrounding the jet \citep{Komissarov2009, Lyubarsky2009, Nakamura2018, Park2019a, Nokhrina2019, Kovalev2020}. In this region, jets are gradually accelerated to relativistic speeds by converting the Poynting flux into the jet's kinetic energy flux, partly through the magnetic nozzle effect \citep{BL1994, VK2004, Komissarov2007, Komissarov2009, Lyubarsky2009, Vlahakis2015}.

There is strong observational support for an extended ACZ in multiple systems. In M87, a well-defined parabolic jet geometry, co-spatial with gradual acceleration on comparable spatial scales, has been reported in a series of Very Long Baseline Interferometry (VLBI) studies (e.g., \citealp{AN2012, NA2013, Hada2013, Mertens2016, Walker2018, Nakamura2018, Park2019b}). The jet geometry transitions into a conical shape at a deprojected distance of $\sim 8 \times 10^5~R_g$ ($\sim$246~pc) from the black hole, where it also starts to decelerate \citep{Park2019b}. This transition point marks the end of the ACZ (e.g., \citealp{AN2012, NA2013}). Similar co-spatial jet collimation and acceleration have been observed in the giant radio galaxies Cygnus A \citep{Boccardi2016a, Boccardi2016b} and NGC 315 \citep{Park2021b, Park2024, Boccardi2021, Ricci2022, Ricci2024}, the narrow-line Seyfert 1 galaxy 1H 0323+342 \citep{Doi2018, Hada2018}, and the flat-spectrum radio quasar 1928+738 \citep{Yi2024}. These results strongly support the view that magnetic fields play a key role in the formation of highly collimated relativistic jets in AGN.

Theoretically, in steady, axisymmetric, ideal magnetohydrodynamics (MHD), rotation of the black hole and/or accretion disk imposes isorotation along magnetic flux surfaces \citep{Camenzind1986}. As the flow approaches and crosses the Alfv\'en surface, it can no longer corotate with the field pattern; the field lines are wound into a predominantly toroidal configuration, marking a transition from poloidal to toroidal dominance\footnote{In steady, Poynting-dominated MHD jets, the Alfv\'en surface is nearly coincident with the (outer) light surface; we therefore use the two interchangeably here \citep[e.g.,][]{VK2003, Komissarov2007, PT2020}.} \citep[e.g.,][]{Spruit1996, CS2002, Komissarov2009, Gelles2025}. The toroidal component supplies hoop stress that contributes to jet collimation under external pressure confinement \citep[e.g.,][]{Komissarov2007, Lyubarsky2009}, while the poloidal gradient of the toroidal magnetic pressure drives gradual acceleration of the plasma \citep[e.g.,][]{VK2003, VK2004, Komissarov2007}. As the flow proceeds, the magnetization decreases and the bulk Lorentz factor increases, and approaching equipartition can promote a transition toward conical expansion \citep[e.g.,][]{Nokhrina2019, Nokhrina2020}. Changes in external pressure and possible recollimation shocks can further facilitate this transition \citep[e.g.,][]{AN2012, Nakamura2018, Kovalev2020}. Thus, the poloidal-to-toroidal magnetic field transition develops progressively throughout the ACZ as rotation winds the field lines, with the toroidal component typically becoming dominant beyond the Alfv\'en surface.

However, the magnetic field geometry in ACZs of AGN jets has been poorly constrained by observations; testing the expected field-line transition in the ACZ has therefore been challenging \citep[e.g.,][]{PA2022}. The primary reason is that the ACZs of most AGN are unpolarized or very weakly polarized \citep[e.g.,][]{Hodge2018, Lister2018}. Even if some parts of the ACZs are significantly polarized, the linearly polarized emission experiences substantial Faraday rotation \citep[e.g.,][]{ZT2002, Park2019a}. The observed Faraday rotation is often consistent with Faraday rotation by an external screen \citep[e.g.,][]{ZT2004, Hovatta2012, Park2019b}. Accurate determination of the Faraday rotation measure (RM) is therefore required to derotate the vectors and infer the intrinsic, RM-corrected linear polarization, which is believed to be directly associated with the jet's magnetic field.

With this backdrop, M87 is uniquely suited for quantitative constraints on the magnetic structure of the ACZ \citep{Hada2024} owing to (i) its proximity (at a distance of $\approx16.8$\,Mpc; \citealt{Blakeslee2009, Bird2010, Cantiello2018}) and large black hole mass of $\approx6.5\times10^9\,M_\odot$ (\citealp{EHT2019f}; see also \citealp{Gebhardt2011}), which make it easier to resolve the fine-scale structure of the jet, and (ii) a pronounced limb-brightened jet traceable across multiple decades in deprojected distance, out to $\sim10^{6}\,R_{\rm g}$, throughout the ACZ \citep{AN2012,Mertens2016,Nakamura2018,Park2019b}. 

Early Very Large Array (VLA) polarization observations reported large and structured RMs over the jet and lobes out to kpc scales, implying a strongly magnetized environment (e.g., \citealp{Owen1990}). Hubble Space Telescope observations at optical/UV wavelengths subsequently revealed high fractional linear polarization reaching up to $m_L \sim50\%$ ($m_L \equiv \sqrt{Q^2 + U^2}/I$) and ordered EVPAs in knots and edges consistent with shock compression and shear alignment \citep{Perlman1999}, with HST-1 showing correlated flux–polarization variability tied to particle acceleration and field ordering \citep{Perlman2011}. More recently, broadband VLA (4--18\,GHz) polarimetric observations resolved the jet width and uncovered a double-helix morphology with edge-enhanced fractional polarization and oppositely signed Faraday-depth gradients across the flow, interpreted as evidence for a large-scale helical magnetic field persisting to $\sim$1\,kpc \citep{Pasetto2021}.

On parsec and subparsec scales, within the jet ACZ, the jet is largely unpolarized. Nevertheless, some regions of the jet have been reported to be significantly linearly polarized. Early Very Long Baseline Array (VLBA) observations at 5 and 8\,GHz detected linear polarization in the inner jet of M87 at a projected distance of $\approx20$\,mas \citep{junor2001}. Multifrequency VLBA observations at 8--15\,GHz revealed an extreme RM distribution that varies from $-4000\ {\rm rad\ m^{-2}}$ to $+9000\ {\rm rad\ m^{-2}}$ in a similar $\approx20$\,mas region, with localized sign reversals \citep{ZT2002}. \citet{Park2019a} extended this study by utilizing multifrequency VLBA archival data sets at 2--8\,GHz and discovered that $|{\rm RM}|$ systematically decreases with increasing distance from the black hole within the Bondi radius, interpreted as a signature of winds from hot accretion flows that confine the jet in the ACZ. At 43\,GHz, complex linear polarization structures with smoothly rotating EVPAs around the core have been reported with the VLBA \citep{Walker2018, Kravchenko2020}. However, a reanalysis of the same data sets with improved polarization calibration \citep{Park2021a} suggested that the core is actually very weakly polarized ($m_L\simeq0.2$--$0.6\%$) and exhibits a simple polarization structure with EVPAs near $70^\circ$--$90^\circ$ \citep{Park2021c}. At a higher frequency, High Sensitivity Array (HSA) observations at 86\,GHz detected a polarized feature with high fractional polarization ($\sim20\%$) at a projected distance of $\sim0.1$\,mas downstream \citep{Hada2016}. Polarimetric observations of M87’s core with the Atacama Large Millimeter/submillimeter Array (ALMA) show a few percent linear polarization with large Faraday rotation: in Band~6 (\(\sim\)230\,GHz; 2017--2018) the core exhibits $m_L\approx2$--$3\%$ and RM spanning $-4.11\times10^5$ to $+1.51\times10^5\ {\rm rad\ m^{-2}}$, including a sign change \citep{Goddi2021}, while in Band~7 (\(\sim\)345\,GHz; 2021) the core has $m_L\sim3\%$ and \(\mathrm{RM}=(1.1\pm0.6)\times10^{5}\ {\rm rad\,m^{-2}}\) \citep{Goddi2025}. These connected-array measurements have limited angular resolution, so the precise origin of the reported core polarization remains ambiguous.

Very recently, \citet{Nikonov2023} presented observations of M87 with a combined array (VLBA + one VLA antenna + the Effelsberg 100\,m telescope) at 8 and 15\,GHz, detecting significant linear polarization at projected distances of $\approx15$--$30$\,mas at both frequencies and deriving a detailed RM distribution together with an intrinsic, RM-corrected EVPA map for that region. They further reported transverse asymmetries in both RM and EVPA, which they interpreted as signatures of a helical magnetic field in the jet. However, because only two frequencies were used, the $n\pi$ ambiguity was resolved by choosing the integer that brings the mean RM into agreement with previous measurements \citep{ZT2002,Park2019a} and by adopting a uniform $n$ across the jet. Thus, the RM-corrected EVPA field implicitly assumes negligible temporal and spatial RM variability between epochs and across the flow—an assumption that may not always hold.

Another limitation in previous studies of M87 is that relativistic aberration is often not fully incorporated into the interpretation. For relativistically moving, optically thin synchrotron flows, relativistic aberration rotates the observed EVPA, so the observed EVPA is generally not perpendicular to the sky-projected magnetic field in the observer's frame \citep[e.g.,][]{BK1979,Pariev2003,Lyutikov2005}. Robust inference of the intrinsic field geometry therefore requires forward modeling that includes projection and relativistic aberration effects. Such modeling has been applied mainly to blazars \citep[e.g.,][]{Clausen-Brown2011, Murphy2013, Zamaninasab2013}, typically on parsec scales and often without transverse resolution of the ACZ, leaving the field geometry within ACZs still challenging to constrain.

In this study, we present new high sensitivity, multifrequency VLBI observations of M87 and derive detailed, Faraday rotation--corrected linear polarization maps that trace the jet over the ACZ. To our knowledge, this provides the first continuous, high-fidelity mapping of intrinsic linear polarization throughout a black hole jet’s ACZ. We then construct a stationary, axisymmetric, ideal MHD model with optically thin synchrotron emission from the jet edge, including projection and relativistic aberration effects; comparison with the data constrains the jet’s intrinsic magnetic field geometry in the ACZ.

This Letter is organized as follows. Section~\ref{sec:observations} describes the observations and data analysis; Section~\ref{sec:modeling} presents the modeling framework; Section~\ref{sec:discussion} discusses the physical implications. Using the fiducial distance and mass stated above and adopting a viewing angle $\theta=163^\circ$ (e.g., \citealt{Mertens2016,Walker2018,EHT2019e}), $1$\,mas corresponds to $0.081$\,pc ($\simeq262\,R_{\rm g}$) on the sky, implying a deprojected linear scale along the jet of $\simeq0.279$\,pc ($\simeq896\,R_{\rm g}$) per mas.

\section{Observations and Data Analysis} 
\label{sec:observations}

We observed M87 with the HSA, which consists of 10 VLBA stations, the phased VLA, and the Effelsberg 100\,m telescope. The observations were conducted in late March 2020 at frequencies between 1.4 and 24.4\,GHz. Standard data reduction was carried out using the NRAO Astronomical Image Processing System (AIPS; \citealp{Greisen2003}), following procedures from previous VLBI studies \citep[e.g.,][]{Park2021b}. We then employed an iterative imaging and self-calibration procedure with Difmap to obtain total intensity images and self-calibrated visibilities. For instrumental polarization calibration, we used GPCAL \citep{Park2021a} to estimate and remove antenna polarimetric leakage. To mitigate potential time-dependent polarization leakage, as evidenced by antenna gain solutions deviating substantially from unity in some scans (e.g., due to pointing offsets), we applied the time-dependent leakage calibration implemented in GPCAL \citep{Park2023}. We verified that our main conclusions remain unchanged by this correction. For a more comprehensive description of the observations, data reduction, and imaging, see Appendix~\ref{appendix:observation}.

We conducted two primary analyses to recover the intrinsic linear polarization structure in the ACZ of the M87 jet (see Appendix~\ref{appendix:analysis} for more details). First, the core shift effect \citep[e.g.,][]{BK1979, Lobanov1998, Hirotani2005, NP2024}—in which the apparent base of the jet shifts with frequency due to synchrotron self-absorption—was measured and corrected to align images obtained at different frequencies. The measured shift is consistent with previous astrometric observations \citep{Hada2011}, enabling accurate co-alignment of the multifrequency data sets for the subsequent analysis.

Second, we derived intrinsic EVPAs by modeling and correcting for Faraday rotation. To balance sensitivity, $(u,v)$ coverage, and bandwidth depolarization, we analyzed three frequency sets: (i) 2.24, 2.27, 2.30, 2.40, and 4.87\,GHz; (ii) 8.18, 8.30, 8.43, 8.56, and 12.37\,GHz; and (iii) 12.37, 15.37, 21.77, and 24.39\,GHz. These sets probe the polarization at projected distances of $\sim$140--400\,mas, 20--200\,mas, and 10--30\,mas, respectively, with set~(iii) providing the highest angular resolution and the most detailed intrinsic polarization structure. Consequently, the 10--30\,mas region provides the most robust data for the detailed comparison with our theoretical model (Section~\ref{sec:modeling}). 

We fitted a $\lambda^2$ law to EVPAs at pixels with linear polarization detected at all frequencies within each set. The fits generally follow the $\lambda^2$ relation, and the EVPA rotation across the frequency range of each set mostly exceeds $45^{\circ}$. We do not observe the strong wavelength-dependent depolarization expected from internal Faraday rotation. Taken together, the large EVPA rotations over much of the jet and the absence of significant depolarization indicate that, for the analyzed pixels, Faraday rotation arises predominantly in an external screen \citep{Burn1966, Sokoloff1998}. This external screen is thought to be magnetized winds launched from the hot accretion flow, which are also responsible for the jet collimation in the ACZ \citep{Park2019a, Yuan2022}. Consequently, the RM-corrected intrinsic EVPAs can be obtained by extrapolating the fitted $\lambda^2$ relations to $\lambda=0$.

\begin{figure}[t!]
\centering
\includegraphics[width=\linewidth]{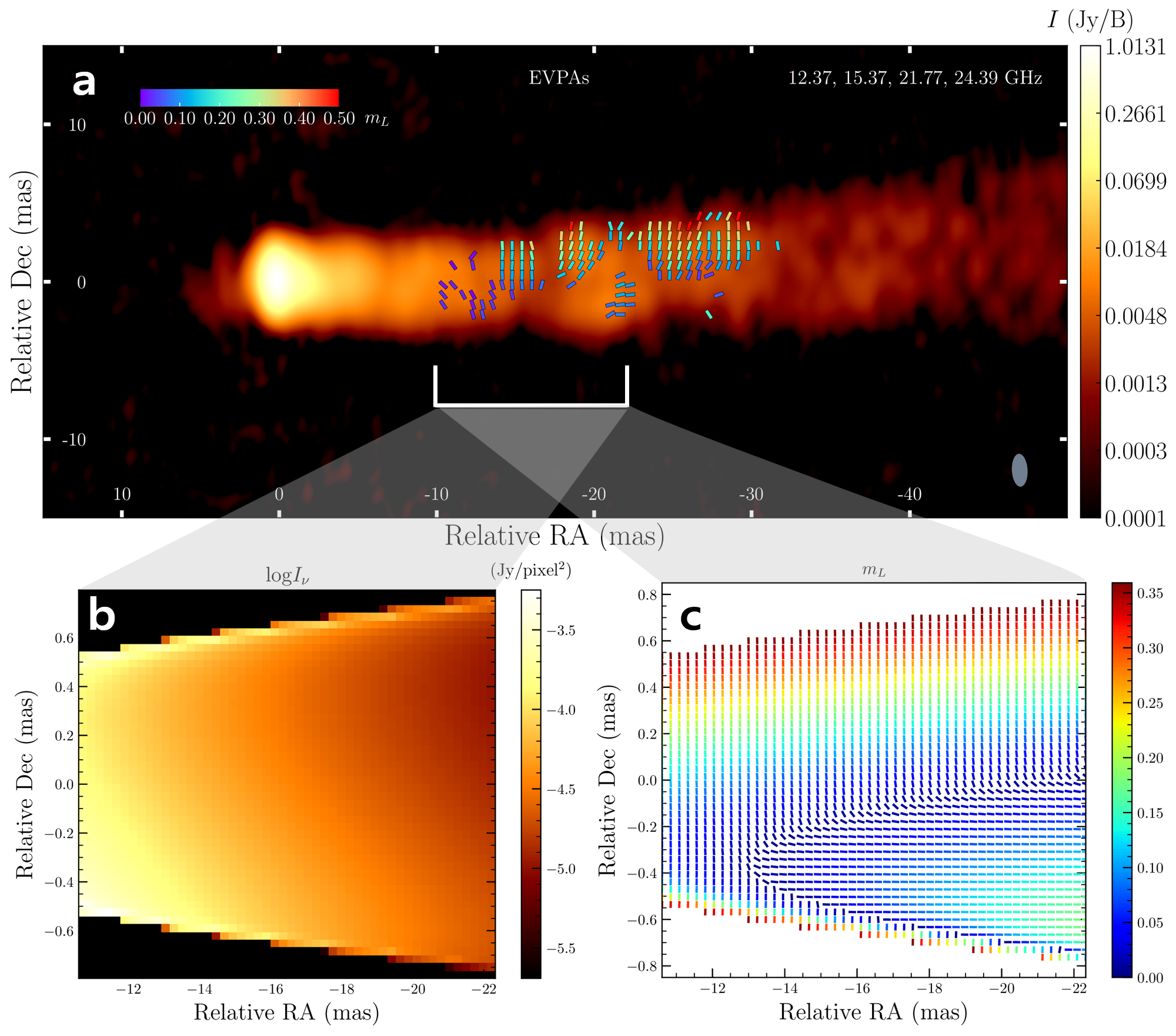}
\caption{Observed and model images of the M87 jet. Top (a): Background color scale: Total intensity emission of the M87 jet (logarithmic scale) at 12.4 GHz in units of Jy per beam. Colored ticks: Intrinsic, RM-corrected EVPAs derived using the 12.4 -- 24.4 GHz frequency combination, with tick color indicating the fractional polarization ($m_L$) derived at 12.4 GHz. The image is rotated clockwise by 18 degrees to compare with the model images. The synthesized beam, corresponding to the lowest frequency (12.4 GHz) used in this combination, is indicated in the bottom right corner. All images within the combination were convolved to this resolution before deriving the intrinsic EVPAs. The white bracket in panel (a) and the connecting regions indicate the spatial range covered by the model images in panels (b) and (c). Bottom left (b): Model image: Total intensity emission in logarithmic scale using the fiducial parameter set, incorporating projection and relativistic aberration effects. Bottom right (c): Corresponding model polarization image: Linear polarization vectors, with color indicating the fractional polarization. It is noted that the black hole spin of $|a| = 0.998$ is used for this model (see Appendix~\ref{appendix:model} for more details). These model results successfully reproduce key observed features: (i) nearly symmetric total intensity structure; (ii) higher fractional polarization in the northern than the southern jet part; and (iii) predominantly perpendicular EVPAs in the northern part, contrasted with mixed perpendicular/parallel EVPAs in the southern part.}
\label{fig:m87jet}
\end{figure}

Figure~\ref{fig:m87jet}a presents the intrinsic linear polarization map of the innermost 10–30 mas region derived from the highest-frequency set (iii); maps for farther downstream regions (20–200 mas and 140–400 mas) are shown in Appendix~\ref{appendix:observation}. These high-fidelity, RM-corrected maps throughout the ACZ reveal three robust characteristics: (1) the total-intensity emission is nearly symmetric between the northern and southern parts; (2) the northern part exhibits a higher and broader distribution of fractional polarization (peaking at $m_L \approx 16\%$ with a significant tail extending to $\sim 40\%$) compared to the southern part (which shows a narrower distribution peaking at $m_L \approx 8\%$; see Appendix~\ref{appendix:modelcomp} for a detailed analysis); and (3) the northern EVPAs are predominantly perpendicular to the jet axis, whereas the southern EVPAs are either perpendicular or parallel. These features are clearest in the 10–30 mas data (set iii), which we use for the comparison with our model in Section~\ref{sec:modeling}. Downstream, the same polarization characteristics remain evident to at least $\sim100$ mas; beyond this distance, the patterns trend toward greater symmetry—the southern EVPAs become predominantly perpendicular to the jet axis, and the fractional polarization becomes comparable across the two sides (Appendices~\ref{appendix:observation} and~\ref{appendix:analysis}).

\section{Modeling} 
\label{sec:modeling}

We have developed a model that qualitatively explains both the observed total intensity and linear polarization images of the M87 jet by employing a steady, axisymmetric, cold ideal MHD jet model \citep{TT2003, PT2020}. Our model incorporates projection and relativistic effects, which can cause the observed EVPAs to deviate from being perpendicular to the jet's magnetic field \citep{Lyutikov2005}. We take into account the observed edge-brightened jet structure in M87 by considering synchrotron radiation emitted by non-thermal electrons confined to a geometrically thin edge of the jet. This emitting region is defined not as a fixed fraction of the local jet width, but as the volume threaded by the outermost magnetic field lines that touch the black hole event horizon. We assume that the jet is optically thin and the non-thermal electrons have a power-law energy distribution and an isotropic\footnote{It is noted that a recent study has also explored the effects of anisotropic electron distributions in modeling the edge-brightening phenomenon of the M87 jet \citep{Tsunetoe2025}.} pitch angle distribution in the fluid rest frame, with their number density following the continuity equation \citep{BroderickLoeb2009, Takahashi2018}. See Appendix~\ref{appendix:model} for details. 

Our model has two key parameter combinations: $\Omega_F/|\Omega_{\rm BH}|$ and $L\Omega_F/(c^2\mathcal{E})$, where $\Omega_F$ and $\Omega_{\rm BH}$ denote the field-line angular velocity and the black hole's angular velocity, respectively, $L$ is the ratio of angular momentum flux to mass flux, and $\mathcal{E}$ is the ratio of energy flux to rest-mass energy flux.\footnote{$L$ and $\mathcal{E}$ are respectively the same as $\hat{L}$ and $\hat{E}$ in \citet{PT2020}.} We conducted a systematic survey of the parameter combinations and found that the model with parameters $\Omega_F/|\Omega_{\rm BH}| = -0.065$ and $L\Omega_F/(c^2\mathcal{E}) = -0.6$ reproduced the observed total intensity and linear polarization features. The negative signs indicate our sign convention that the black hole's spin vector points away from the observer. In this model, the rest-frame field pitch angle is $\sim45^\circ$ (i.e., $|B'_\phi/B'_p| \sim 1$, where $B'_\phi$ and $B'_p$ are the toroidal and poloidal magnetic field components in the rest frame, respectively) at deprojected distances of $z \gtrsim 10^4~r_g$ (see Appendix~\ref{appendix:model} for more details). Our final projected model images in Stokes $I$, $Q$, and $U$ were obtained by solving the radiative transfer equation along rays through the jet's emitting region. To phenomenologically account for depolarization effects caused by additional turbulent magnetic fields \citep{Laing1980, Laing1981}, we scale the fractional polarization by a factor of 1/2. While complex depolarization might affect the absolute polarization levels, it should not produce the systematic north–south asymmetry observed (see Section~\ref{sec:discussion} and Appendix~\ref{appendix:model} for further discussion). 

\begin{figure}[t!]
\centering
\includegraphics[width=0.7\columnwidth]{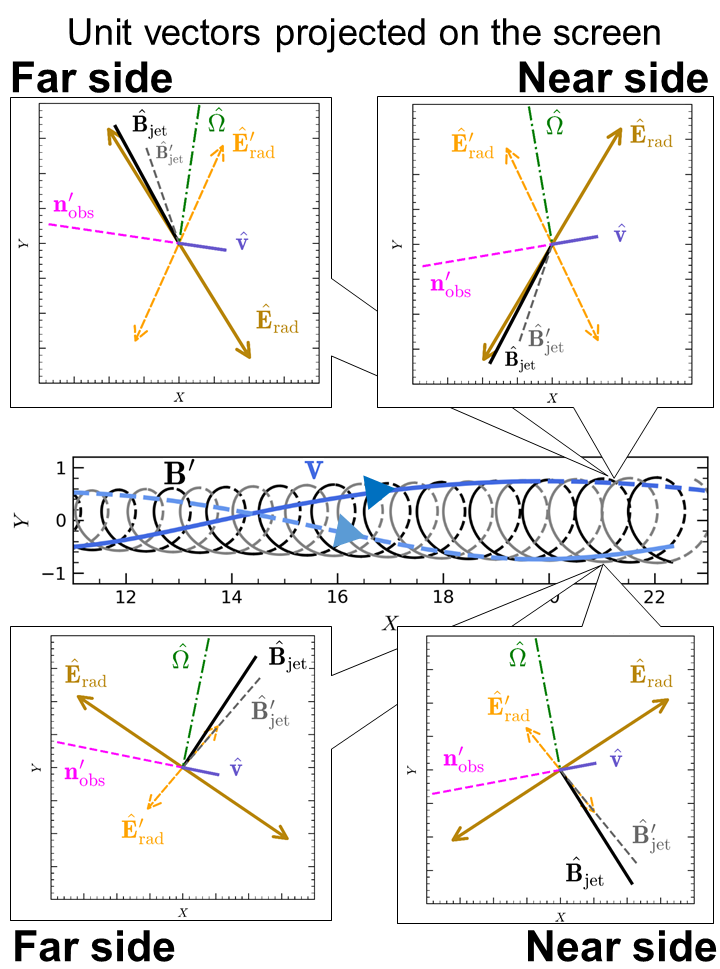}
\caption{Geometric overview of the fiducial model and polarization vectors. The helical magnetic fields and jet flows of our fiducial jet model are shown in the observer's view, where the line of sight is perpendicular to the paper. The black and gray lines show two representative magnetic field lines measured in the fluid rest frame. The blue and light blue lines show two representative stream lines. The solid parts lie on the side near the observer while the dashed parts lie on the opposite side far from the observer. As indicated by arrows, the jet flows from left to right. Within each pair (upper/lower), the left and right panels display two different positions with the same coordinates $(X,Y)$, where the $X$-axis is aligned with the jet axis projected on the sky plane and the units of the $X$ and $Y$ axes are mas, but at different depths with respect to the observer as designated in the figure. Specifically, $(X/\mathrm{mas},Y/\mathrm{mas},z/R_g) = (20.9, 0.40,1.82\times10^4), (20.9, 0.40,1.93\times10^4), (20.9, -0.40,1.82\times10^4), (20.9, -0.40,1.93\times10^4)$, where $z$ is the physical height from the equatorial plane, for upper left, upper right, lower left, and lower right panels, respectively. Colored lines represent: purple, velocity field $\hat{\boldsymbol{v}}$; yellow, electric field of radiation $\hat{\boldsymbol{E}}_\mathrm{rad}$; black, magnetic field of the jet $\hat{\boldsymbol{B}}_\mathrm{jet}$, where we assumed $B'_p > 0$ and $B'_\phi > 0$, which is one possible choice consistent with the constraint $B'_p B'_\phi > 0$; pink, line of sight measured in the fluid rest frame $\boldsymbol{n}_\mathrm{obs}^\prime$; and green, rotational axis of the Lorentz transformation $\hat{\boldsymbol{\Omega}}\propto \boldsymbol{n}_\mathrm{obs}^\prime\times \hat{\boldsymbol{v}}$. The rotational angle is the same as the angle between $\boldsymbol{n}_\mathrm{obs}$ and $\boldsymbol{n}_\mathrm{obs}^\prime$, where quantities with prime denote those measured in the fluid rest frame. The solid and dashed lines are assigned to quantities measured in the laboratory and fluid rest frames, respectively. The velocity $\hat{\boldsymbol{v}}$ is determined by the jet structure of our model and determines the direction of $\boldsymbol{n}_\mathrm{obs}^\prime$ through the relativistic aberration effect, both of which fix the rotational axis $\hat{\boldsymbol{\Omega}}$ and angle of the Lorentz transformation of other vectors. The magnetic field is determined by our jet model and determines the plane of oscillating electric fields of radiation so that $\hat{\boldsymbol{E}}_\mathrm{rad}^\prime \propto \boldsymbol{n}_\mathrm{obs}^\prime \times \hat{\boldsymbol{B}}_\mathrm{jet}^\prime$ in the fluid rest frame. $\hat{\boldsymbol{E}}_\mathrm{rad}^\prime$ is Lorentz-transformed into $\hat{\boldsymbol{E}}_\mathrm{rad}$ by the rotation determined above. The observed radiation (total intensity and polarization) is obtained as a superposition of local electric field vectors $\hat{\boldsymbol{E}}_\mathrm{rad}$ along a ray.}
\label{fig.theo.overview.fid}
\end{figure}

Figures~\ref{fig:m87jet}b,c present the model images of total intensity and linear polarization emission from the jet (see Appendix~\ref{appendix:modelcomp} for more detailed data--model comparisons). The model effectively replicates the three observed features explained above (1)--(3). Furthermore, the principles of our model, incorporating the helical field structure and relativistic effects, qualitatively account for the observed polarization characteristics extending out to distances of $\approx100$\,mas (see Appendix~\ref{appendix:observation} and~\ref{appendix:analysis}).

The primary reason why a helical magnetic field with a rest-frame pitch angle of $\sim45^\circ$ threading the relativistic jet can reproduce the observed asymmetric linear polarization emission is as follows. The M87 jet is believed to be viewed close to the pole \citep{Mertens2016, Walker2018}. From the observer's perspective, the emission that constitutes each of the observed north and south parts of the jet is a line-of-sight integration of emission originating from both the near side (regions of the jet at greater line-of-sight distances from the core) and the far side (regions at smaller line-of-sight distances from the core) within its three-dimensional structure (\citealp{Walker2018}; see Appendix~\ref{appendix:model} for more details). The magnetic field possesses the helicity $B'_p B'_{\phi} > 0$, which directly results from $\Omega_F < 0$. The helical magnetic field produces systematic variations in the rest-frame electric field directions of the synchrotron radiation, which are subsequently Lorentz-transformed into the observer's frame. In the northern jet part, the observed electric field vectors are predominantly vertical on both the near and far sides. In contrast, in the southern part, they are nearly orthogonal to each other, each pointing diagonally with a slight horizontal or vertical tilt, depending on the bulk Lorentz factors. This configuration results in EVPAs that are perpendicular to the jet axis with high fractional polarization in the northern part, due to a marginal cancellation effect of the polarization vectors. Meanwhile, in the southern part, EVPAs are either predominantly perpendicular or parallel to the jet axis with low fractional polarization, due to a stronger cancellation effect (see Figure~\ref{fig.theo.overview.fid} for more details). It is noteworthy that a similar model was previously introduced to explain observed EVPAs oriented either parallel or perpendicular to the jets of unresolved sources \citep[e.g.,][]{Lyutikov2005, Clausen-Brown2011}. However, this is the first instance in which the presence of a helical magnetic field is inferred from modeling well-resolved total intensity and linear polarization structures in the ACZs of AGN jets.

\section{Discussion and Conclusions} 
\label{sec:discussion}

\begin{figure}[t!]
\centering
\includegraphics[width=\linewidth]{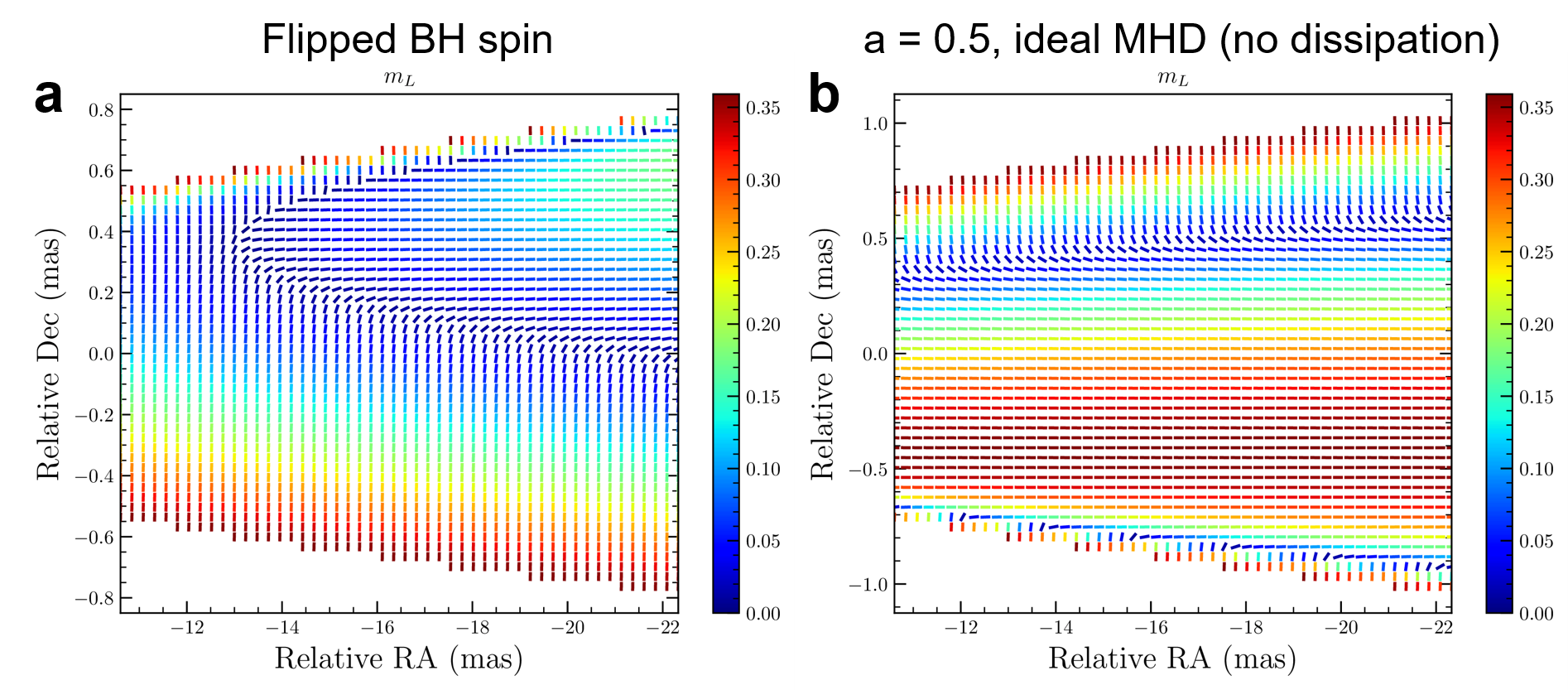}
\caption{Alternative model polarization images of the M87 jet. Left (a): Linear polarization image from the model assuming the same physical parameters as those in Figure~\ref{fig:m87jet}c but with an opposite black hole spin vector direction (toward the observer). This results in the image being flipped along the declination axis. Right (b): Linear polarization image from the model assuming the same physical parameters as presented in Figure~\ref{fig:m87jet}c but with an assumed $a=0.5$ and $\Omega_{F} = \Omega_{\rm H}/2$ (i.e., assuming no dissipation between the horizon and the ACZ). Thus, this represents the expected image under ideal MHD conditions throughout the jet, leading to a toroidally dominated magnetic field structure at this scale. While these models can reproduce symmetric total intensity emission similar to that shown in Figure~\ref{fig:m87jet}b, they do not replicate the observed linear polarization properties, particularly the asymmetry in the fractional polarization evident in Figure~\ref{fig:m87jet}a.
\label{fig:model_alternatives}}
\end{figure}

Our findings have significant physical implications for the supermassive black hole at the center of M87 and the physics of its relativistic jet. A key result is that the black hole's spin vector must point away from the observer. If the spin were directed toward the observer, the predicted handedness of the helical magnetic field would be reversed, yielding the opposite north–south linear-polarization asymmetries (see Figure~\ref{fig:model_alternatives}a for the expected model polarization image), in direct contradiction to our observations. This determination of the spin orientation, based entirely on the downstream jet properties measured in this work, provides compelling independent verification of conclusions drawn from EHT observations. Specifically, it is fully consistent with interpretations \citep{EHT2019e, EHT2021b, EHT2025} in which the brighter southern part of the ring-like structure in the EHT images is attributed to relativistic beaming by plasma orbiting a black hole with a spin vector pointing away from us. Furthermore, our modeling indicates that the jet flow rotates clockwise when viewed from the pole, i.e., from the top of the approaching jet. This is also consistent with previous kinematic studies \citep{Mertens2016, Walker2018}.

Furthermore, based on our fiducial model for the 10--30 mas region, our results indicate that the jet is threaded by a large-scale, ordered helical magnetic field at a projected jet distance greater than $\sim10$\,mas, corresponding to a deprojected jet distance of $\approx10^4\,R_g$. According to ideal MHD theory for the jet evolution from the event horizon to the far zone without taking account of any energy dissipation, the jet's magnetic field is expected to transition from a poloidally dominated structure to a toroidally dominated structure at the jet radius near the outer light surfaces (\citealp{TT2013, PT2020, Gelles2025}; also known as light cylinders). The light cylinder radius is given by $R_L = c/\Omega_{F}$, where $\Omega_F \sim \Omega_{\rm BH} / 2$ \citep{BZ1977, Tchekhovskoy2010}. Since $\Omega_{\rm BH} = ac/2r_H$, where $r_H = (1 + \sqrt{1-a^2}) R_g$ is the radius of the event horizon, it follows that $R_L \approx 4r_H/a$. In this derivation, $a$ is the dimensionless black hole spin. The M87 black hole is inferred to be spinning at mild or fast speeds with $|a|\gtrsim0.5$ \citep{EHT2019e, EHT2021b}. In this scenario, the transition is expected to occur at a jet radius of $R_L\lesssim0.06$\,mas, corresponding to a projected jet distance of $\lesssim0.03$\,mas \citep{AN2012, Nakamura2018, Lu2023}. This derivation aligns well with results from more realistic general relativistic magnetohydrodynamic (GRMHD) modeling and simulations \citep{PT2020, Cruz-Osorio2022, Gelles2025}. Indeed, polarization observations of M87 with the EHT, probing the scale inside the light cylinder, indicate the presence of a significant poloidal magnetic field component near the black hole \citep{EHT2021a, EHT2021b, EHT2023}, supporting theoretical predictions.

The ideal MHD models for the whole jet with no dissipation predict that the magnetic field will be continuously wound up, becoming a toroidally dominated structure in the far zone. The ratio of the amplitudes of the toroidal ($B_{\phi}$) to poloidal ($B_p$) field components in the lab frame is given by $|B_{\phi}/B_p| \approx R/R_L$ \citep{McKinney2006, TT2013}, where $R$ is the radius of the jet. The observed jet radius at the $\sim10$\,mas scale is $\sim2.2$\,mas \citep{AN2012, Nakamura2018}, which translates into $|B_{\phi}/B_p| > 37$, resulting in a toroidally dominated field structure. However, our fiducial model of the 10--30 mas region predicts a lab-frame field strength ratio of $|B_{\phi}/B_p| \sim 5.6$, which is significantly smaller than the lower limit predicted at this scale under ideal MHD evolution from the black hole vicinity. This contrast in magnetic field configurations is visually demonstrated by comparing the predicted linear polarization map of our fiducial model (Figure~\ref{fig:m87jet}c) with that expected under ideal MHD conditions for the whole jet, adopting $\Omega_F = \Omega_{\rm BH}/2$ and $a=0.5$ (Figure~\ref{fig:model_alternatives}b), the latter of which indeed exhibits the toroidally dominated signature discussed above. This stark discrepancy challenges the current (GR)MHD-based paradigm for relativistic jet formation by demonstrating that significant dissipation of the toroidal magnetic field component \citep{Nakamura2007, Chatterjee2019, Rieger2019, Sironi2021}—a process unaccounted for in standard ideal models—must occur within the jet's ACZ to explain the observed structure (see more discussion in Appendix~\ref{appendix:model}). This implies that while the jet may be launched with a high $|\Omega_F/\Omega_H|$ near the black hole, consistent with standard BZ-type models, significant dissipation of the toroidal magnetic field must occur upstream of the region we observe (i.e., within $\sim 10^4~r_g$) to produce the mildly toroidal field structure required by our data. A direct implication of this is that the standard scaling relationship for the toroidal magnetic field ($B_\phi \propto R^{-1}$) often assumed in VLBI studies \citep[e.g.,][]{Baczko2016, Ro2023}, which assumes ideal MHD, may not hold in the M87 ACZ.

One might question the reliability of this conclusion, as it partly relies on the observed $m_L$ asymmetry. Two alternative mechanisms, not included in our model, could theoretically produce such an asymmetry: (i) stronger turbulent magnetic fields in the southern part of the jet, or (ii) stronger internal Faraday rotation in the southern part. However, the former is unlikely, as it requires turbulent fields to be distributed preferentially on one side of the jet over a long distance range (10--100\,mas), and we are not aware of any physical mechanism that would produce such a large-scale, systematic effect. The latter (stronger internal Faraday rotation) is inconsistent with our data. As shown in our detailed EVPA vs. $\lambda^2$ fit plots (Appendix~\ref{appendix:analysis}), the detected regions in both the northern and southern parts of the jet exhibit clean $\lambda^2$ fits and show no strong frequency-dependent depolarization. This behavior is the opposite of what would be expected from strong internal Faraday rotation, which would predict both strong frequency-dependent depolarization and a breakdown of the $\lambda^2$ linearity \citep[e.g.,][]{Burn1966, Sokoloff1998}. This strengthens our conclusion that the observed $m_L$ asymmetry is an intrinsic feature of the ordered field geometry.

The observed linear polarization patterns become notably more symmetric between the northern and southern jet segments at larger distances from the core ($\gtrsim100$\,mas; see Appendices~\ref{appendix:observation} and~\ref{appendix:analysis}). While precisely modeling this extended region would require incorporating continuous dissipation processes, which is beyond the scope of our current framework, the model considerations above offer predictive insights into this behavior. Specifically, if the dissipation mechanisms that limit extreme toroidal winding in the inner region continue to operate downstream, the magnetic field structure would be expected to evolve toward a relatively stronger poloidal component. Such an evolution naturally accounts for the observed increase in symmetry in both the fractional polarization and the EVPA distributions farther down the jet, with EVPAs becoming predominantly perpendicular to the jet axis (see Appendix~\ref{appendix:poloidal_model}). This qualitative consistency supports a continuing role for upstream non-ideal processes in shaping the M87 jet over extended scales.

Our findings of a large-scale helical magnetic field with a significant poloidal component in the M87 jet ACZ offer a new perspective on jet formation. Helical magnetic fields have also been proposed in the jets of blazars, often based on observations of transverse Faraday rotation gradients \citep{Asada2002, Gabuzda2004, Hovatta2012, Gomez2016, Gabuzda2018, Livingston2025} or point-symmetric patterns in intrinsic EVPAs around the cores of blazars \citep{Zamaninasab2013, Gomez2016}. Studies focusing on transverse Faraday rotation gradients often note that a good $\lambda^2$ fit to observed EVPAs--a characteristic frequently present in such analyses--typically indicates Faraday rotation occurring in a screen external to the jet plasma \citep{Burn1966, Sokoloff1998}. Therefore, the connection between magnetic fields probed via such external Faraday screens and the intrinsic magnetic field structuring the jet may be inconclusive. Similarly, interpretations based on point-symmetric EVPA patterns have faced limitations; for instance, some have utilized EVPAs observed at a single frequency for key modeling comparisons \citep{Zamaninasab2013}, while others have not directly compared observed EVPAs with a comprehensive model incorporating projection and relativistic aberration effects \citep{Gomez2016}. Furthermore, in many of these cases, the regions probed are most likely beyond the primary jet ACZs as typically defined in the literature for systems such as M87 \citep{Marscher2008, Jorstad2013}. Consequently, the detailed, RM-corrected intrinsic linear polarization maps in the ACZ of the M87 jet presented in our study, coupled with direct modeling of the observed polarization features, provide stronger evidence of the magnetic field geometry within this crucial region. This approach enables a direct comparison with jet formation paradigms, highlighting the challenge our results pose to models predicting predominantly toroidal fields.

\begin{figure}[t!]
\centering
\includegraphics[width=0.7\linewidth]{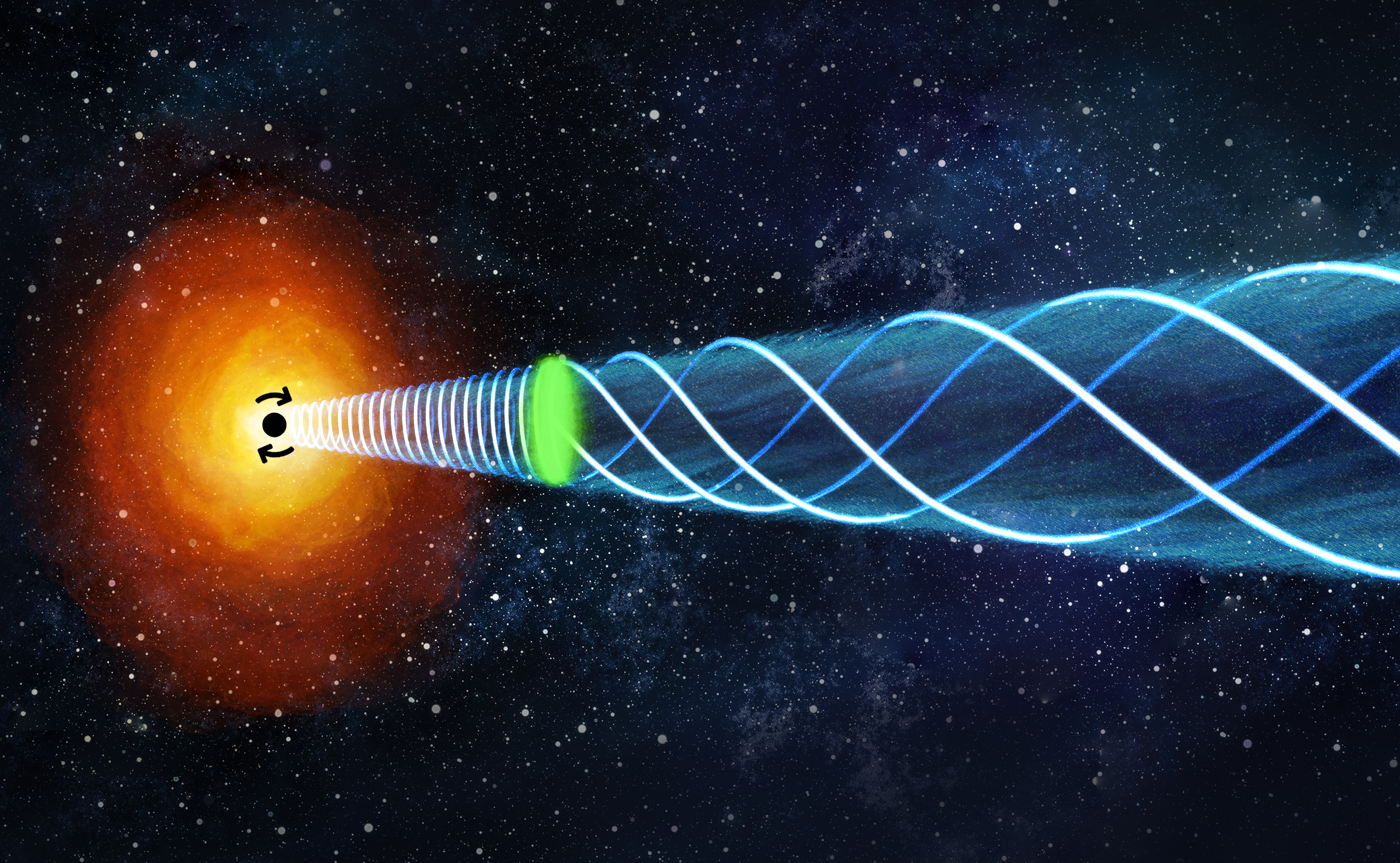}
\caption{Schematic diagram summarizing our findings and inferred physics. The black hole's spin vector points away from the observer, consistent with interpretations of EHT observations \citep{EHT2019e, EHT2025}. Within the jet ACZ (the 'downstream region' depicted), the jet is threaded by a large-scale helical magnetic field (rest-frame pitch angle $45^\circ$). The discrepancy between this structure and the prediction under the ideal MHD condition implies significant dissipation of the toroidal magnetic field component, conceptually illustrated by the light green zone representing the inferred region where significant toroidal magnetic field dissipation is required to occur 'upstream' (closer to the black hole than the region dominated by the observed helical field). Note that while a single conceptual zone is depicted for simplicity, the actual underlying dissipation is likely an extended process, potentially distributed over considerable distances or occurring in multiple regions, and is not explicitly modeled here. The jet exhibits slow clockwise rotation (viewed from top) superimposed on its dominant poloidal motion.}
\label{fig:schematic}
\end{figure}

Given the presumed universality of magnetic jet launching and acceleration processes across different black hole mass scales, these findings likely have significant implications for understanding jet physics in other relativistic outflow systems, such as gamma-ray bursts, tidal disruption events, and X-ray binaries, potentially indicating a widespread need to incorporate non-ideal MHD effects such as magnetic dissipation. We present a schematic diagram that summarizes our findings and the inferred physics in Figure~\ref{fig:schematic}.

\begin{acknowledgments}

The authors appreciate the referee’s constructive comments, which have improved the Letter. This work was supported by the National Research Foundation of Korea (NRF) grant funded by the Korea government (MSIT; RS-2024-00449206; RS-2025-02214038). This research has been supported by the POSCO Science Fellowship of POSCO TJ Park Foundation. This research was supported by Global-Learning \& Academic research institution for Master's \textperiodcentered~PhD students, and Postdocs(G-LAMP) Program of the National Research Foundation of Korea(NRF) grant funded by the Ministry of Education(RS-2025-25442355). This work was supported by the BK21 FOUR program through the National Research Foundation of Korea (NRF) under the Ministry of Education (Kyung Hee University, Human Education Team for the Next Generation of Space Exploration). This work was supported by MEXT/JSPS KAKENHI (grants 25H00660, 22H00157, 21H04488). This work was supported the Mitsubishi Foundation (grant 202310034). This work was supported by a University Research Support Grant from the National Astronomical Observatory of Japan (NAOJ). M.N. is supported by JSPS KAKENHI Grant Number 24K07100. The VLBA is an instrument of the National Radio Astronomy Observatory. The National Radio Astronomy Observatory is a facility of the National Science Foundation operated by Associated Universities, Inc. This work is partly based on observations with the 100-m telescope of the MPIfR (Max-Planck-Institut f\"{u}r Radioastronomie) at Effelsberg. This work made use of the Swinburne University of Technology software correlator, developed as part of the Australian Major National Research Facilities Programme and operated under licence \citep{Deller2011}. 

\end{acknowledgments}

\begin{contribution}

J.P. initiated the project and coordinated the research. J.P., K.H., M.N., and M.K. worked on the VLBI scheduling and coordination of the observations. J.P. worked on calibration and analysis of the data. K. Takahashi, K. Toma, M.N., and J.P. worked on the development of the model and its application to the observed data. K. Takahashi wrote the code that implements the model and conducts necessary calculations to produce model images. J.P., K. Takahashi, and K. Toma wrote the original manuscript. All authors contributed to the discussion of the results presented and commented on the manuscript.

\end{contribution}

\facilities{VLBA, VLA, Effelsberg:100m}

\software{AIPS \citep{Greisen2003}, DIFMAP \citep{Shepherd1997},
          DiFX \citep{Deller2011}, eht-imaging \citep{Chael2016, Chael2018, Chael2023} GPCAL \citep{Park2021a,Park2023}}

\appendix

\begin{deluxetable*}{llll}
\digitalasset
\tablewidth{0pt}
\tablecaption{Information about the HSA observations of M87 \label{tab:observations}}
\tablehead{
\colhead{Project Code} & \colhead{Observation Date} & \colhead{Starting Frequencies (GHz)} & \colhead{Station}
}
\startdata
\multirow{3}{*}{BP241D} & \multirow{3}{*}{23 Mar 2020} & 1.39, 1.42, 1.46, 1.49, 1.55$^*$, 1.58$^*$, 1.65, 1.71 & -EB, -FD, -MK$^{**}$ \\
        & & 4.63, 4.69, 4.76, 4.82, 4.88, 4.95, 5.01, 5.08 & -MK$^{**}$ \\
        & & 15.13, 15.19, 15.26, 15.32, 15.38, 15.45, 15.51, 15.58 & -MK$^{**}$ \\
        \hline 
        \multirow{3}{*}{BP241E} & \multirow{3}{*}{24 Mar 2020} & 2.16$^{*}$, 2.19$^{*}$, 2.22, 2.25, 2.28, 2.32$^{*}$, 2.35$^{*}$, 2.38 & -EB, -HN, -MK$^{**}$ \\
        & & 8.13, 8.19, 8.26, 8.32, 8.38, 8.45, 8.51, 8.58 & -EB, -HN, -MK$^{**}$ \\
        & & 12.13, 12.19, 12.26, 12.32, 12.38, 12.45, 12.51, 12.54 & -EB, -HN, -MK$^{**}$ \\
        \hline 
        \multirow{3}{*}{BP241E1} & \multirow{3}{*}{31 Mar 2020} & 2.16$^{*}$, 2.19$^{*}$, 2.22, 2.25, 2.28, 2.32$^{*}$, 2.35$^{*}$, 2.38 & -EB, -MK$^{**}$ \\
        & & 8.13, 8.19, 8.26, 8.32, 8.38, 8.45, 8.51, 8.58 & -MK$^{**}$ \\
        & & 12.13, 12.19, 12.26, 12.32, 12.38, 12.45, 12.51, 12.54 & -MK$^{**}$ \\
        \hline 
        \multirow{2}{*}{BP241F} & \multirow{2}{*}{30 Mar 2020} & 21.53, 21.59, 21.66, 21.72, 21.78, 21.85, 21.91, 21.94 & -MK$^{**}$ \\
        & & 24.26, 24.29, 24.32, 24.36, 24.39, 24.42, 24.45, 24.48 & -MK$^{**}$ \\
        \hline 
\enddata
\tablecomments{$^*$ indicates frequency bands significantly affected by radio-frequency interference (RFI) or RFI filters. $^{**}$ The VLBA Mauna Kea station experienced issues recording some data frames, resulting in only $\sim$10--20\% of the scheduled on-source visibilities being recorded. The BP241E1 observations were rescheduled from the BP241E observations because the Effelsberg 100\,m telescope, a key station due to its superior sensitivity, was unable to participate in the BP241E observations.}
\end{deluxetable*}

\section{Detailed Description of Observations and Data Reduction}
\label{appendix:observation}

We observed M87 with the HSA, which consists of 10 VLBA stations, the phased VLA, and the Effelsberg 100\,m telescope. The observations took place in late March 2020 at frequencies between 1.4 and 24.4\,GHz. Detailed information about the observations is summarized in Table~\ref{tab:observations}.

\begin{figure}[t!]
\centering
\includegraphics[width=\linewidth]{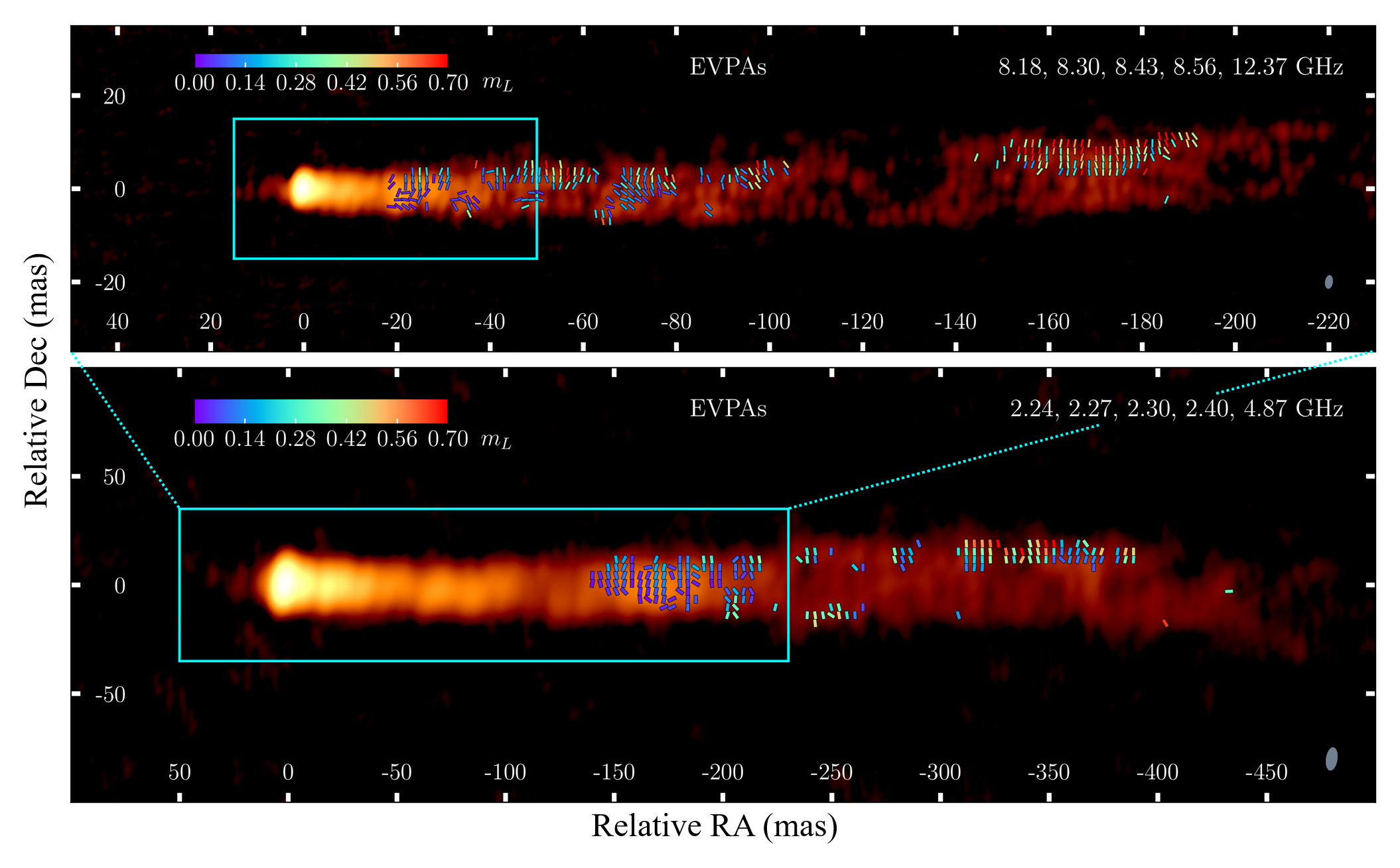}
\caption{Intrinsic polarization maps for lower frequency combinations. Similar to Figure~\ref{fig:m87jet}a, these panels show the total intensity emission (background color scale, logarithmic) and intrinsic, RM-corrected EVPAs (colored ticks; color indicates fractional polarization derived at the lowest observing frequency for each frequency combination) for the M87 jet, derived using lower frequency combinations. Top: Results from the 8.2--12.4\,GHz combination, which probe the jet structure at larger distances. The image has been rotated clockwise by $21^\circ$ relative to the observed images to align the jet axis with the $x$-axis. The cyan box indicates the region shown in Figure~\ref{fig:m87jet}a. Bottom: Results from the 2.2--4.9\,GHz combination. The image has been rotated clockwise by $23^\circ$ relative to the observed images. Note the different spatial scale used in each panel compared to Figure~\ref{fig:m87jet}a. The synthesized beam for each combination (corresponding to the lowest frequency in that set) is shown in the bottom right corner of each panel.}
\label{fig:m87jet_large}
\end{figure}

We applied standard data reduction procedures using the NRAO Astronomical Image Processing System (AIPS; \citealp{Greisen2003}), following previous VLBI studies \citep[e.g.,][]{Park2021b}. The calibrated data were averaged in time over 10\,s in Difmap \citep{Shepherd1997}. An iterative imaging and self-calibration procedure was employed to obtain total intensity images and self-calibrated visibilities. Additional amplitude and phase self-calibration was performed with a 10\,s solution interval using the task \texttt{CALIB} in AIPS to correct potential gain ratio offsets between the two polarizations that are not accounted for by Difmap imaging. After this additional self-calibration, the iterative imaging and self-calibration procedure with Difmap was performed again until the improvement in the $\chi^2$ value of the CLEAN model became negligible. The resulting total intensity images (Figure~\ref{fig:m87jet} and Figure~\ref{fig:m87jet_large}) reveal complex, knotty structures along the jet that are not solely attributable to the convolution by the synthesized beam. These features may reflect underlying physical processes within the jet, or may partially arise from residual artifacts inherent to the nonlinear CLEAN deconvolution process \citep{Lu2023,Park2024,Kim2025}.

To estimate and remove antenna polarimetric leakage from the dataset, we used the instrumental polarization calibration pipeline GPCAL \citep{Park2021a}. Weakly polarized sources such as OQ~208 and M87 were used for initial leakage estimation using the similarity approximation, followed by 10 iterations of instrumental polarization self-calibration using calibrators with higher signal-to-noise ratios, such as OJ~287 and J1239+0730.

We conducted EVPA calibration, i.e., correction of the gain phase ratio between the two polarizations at the reference antenna, using the EVPA calibrator 3C~286 (for frequencies below 5\,GHz) and VLA-only observations obtained as a by-product of our HSA observations. For each frequency, we used at least two calibrators for EVPA calibration, achieving an accuracy of $\sim2^\circ$ in most cases and $\sim3^\circ$ in some. Given that these estimates were derived from a limited number of data points, we conservatively assumed a $1\sigma$ uncertainty of $3^\circ$ for the EVPA calibration at each frequency.

Some scans exhibited large antenna gains in self-calibration, suggesting significant effects from factors such as pointing offsets, solar-induced antenna deformation, and related issues. Because such effects can introduce non-negligible time-dependent polarization leakage, we applied the time-dependent leakage calibration implemented in GPCAL \citep{Park2023}. Our main conclusions are unchanged with or without this correction.

For the 2, 8, and 12\,GHz bands, we used images averaged between the BP241E (observed on 24 Mar 2020) and BP241E1 (observed on 31 Mar 2020) observations for each Stokes parameter in the subsequent analysis (Table~\ref{tab:observations}). Since we found no noticeable differences between the images from the two epochs, we used the averaged images to enhance the signal-to-noise ratio.

\begin{figure}[t!]
\centering
\includegraphics[width=0.32\linewidth]{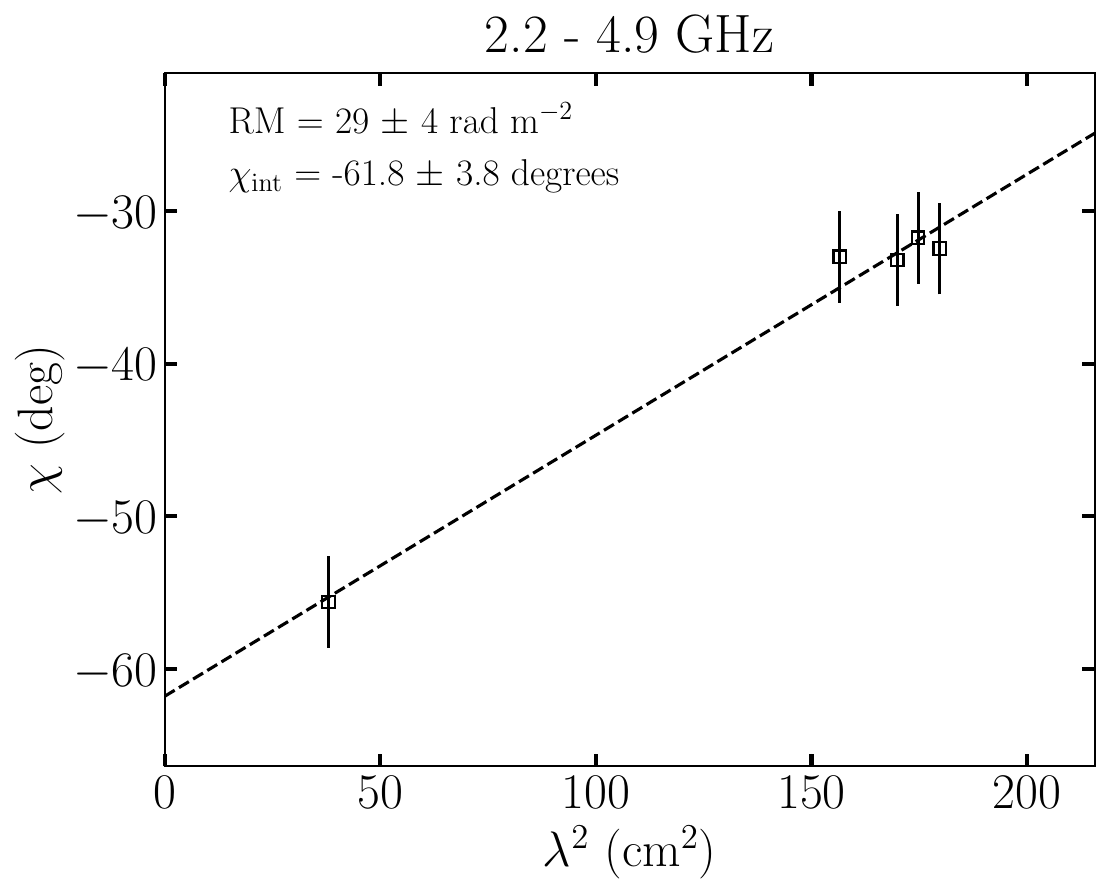}
\includegraphics[width=0.32\linewidth]{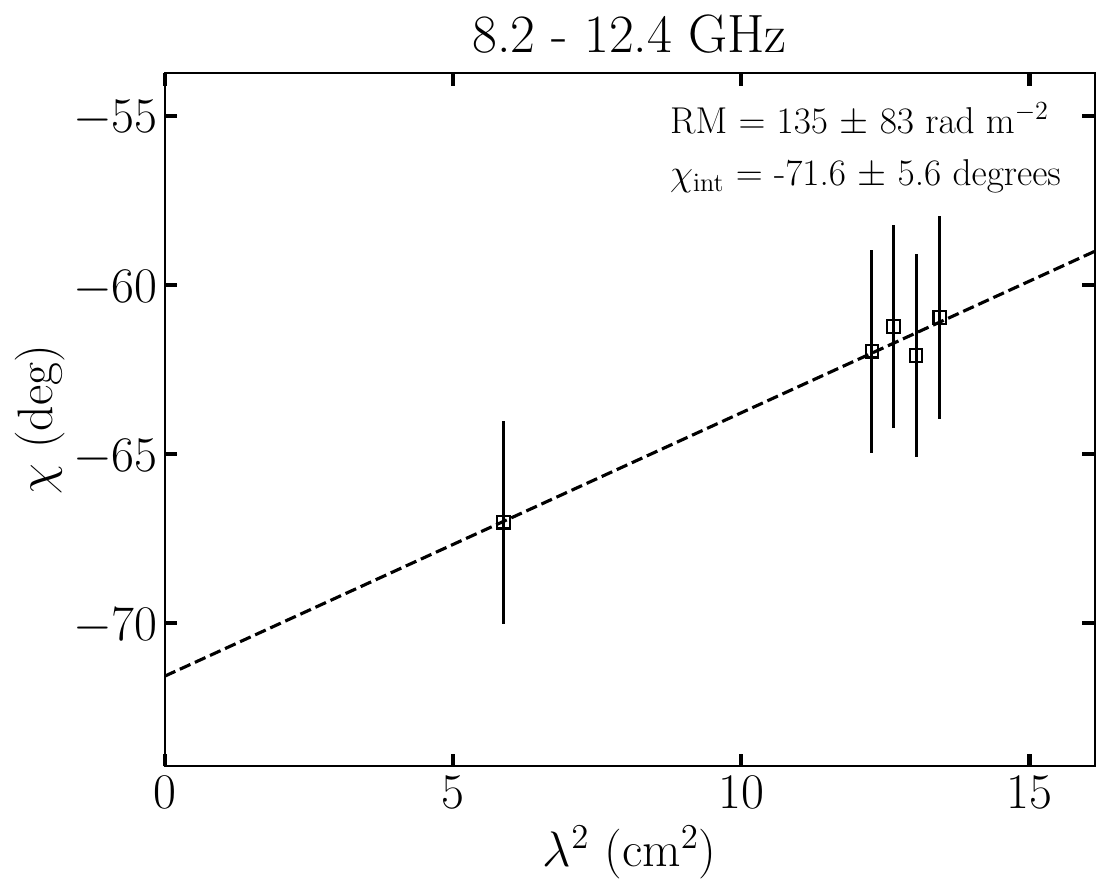}
\includegraphics[width=0.32\linewidth]{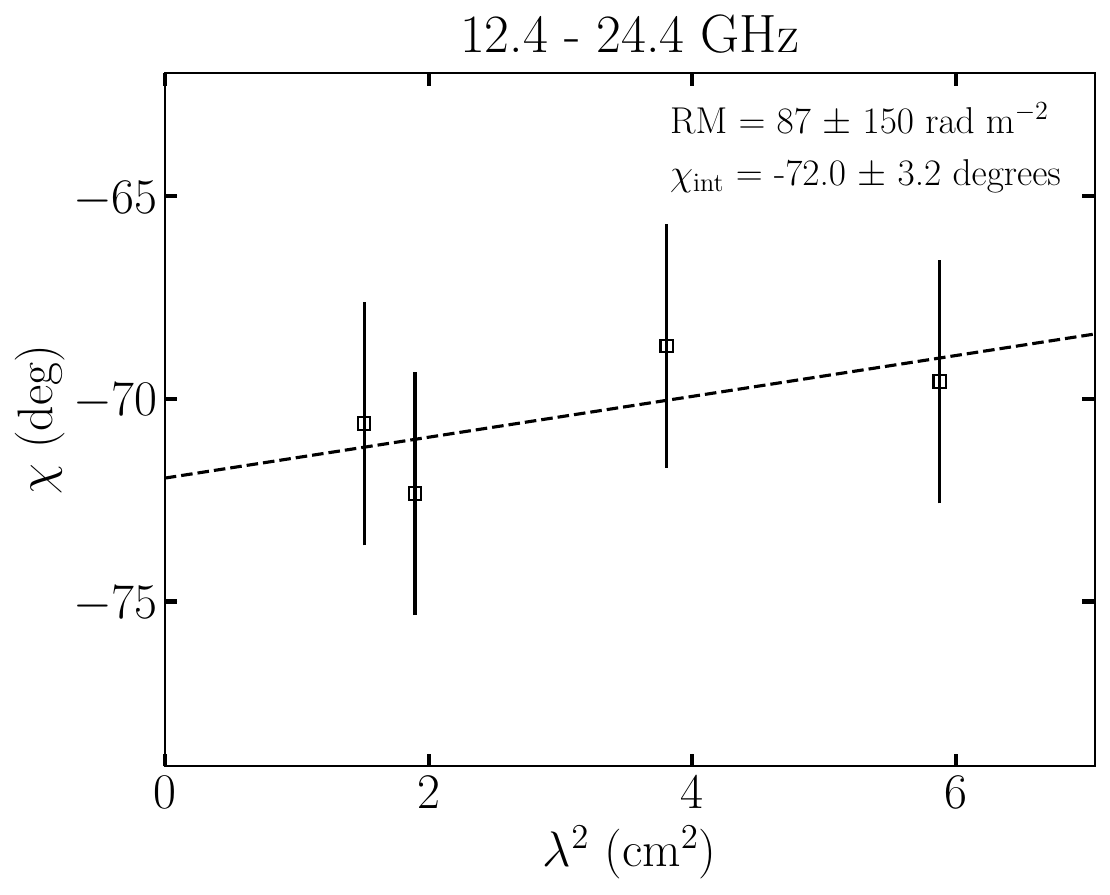}
\caption{EVPAs as functions of $\lambda^{2}$ at the core of the calibrator OJ 287, which demonstrate our robust calibration. These data were calibrated, imaged, and analyzed using the same methods as for our target source, M87. The dashed lines represent the best-fit $\lambda^2$ models to the observed EVPAs. The best-fit RMs and intrinsic EVPAs are indicated in the upper right corner of each panel.}
\label{fig:oj287_evpa}
\end{figure}

In Figure~\ref{fig:oj287_evpa}, we present the EVPAs at the center of the core of the calibrator OJ~287, following the same analysis performed for the target source M87, except for the core-shift correction. The EVPAs are well fit by $\lambda^2$ relations with relatively small values of Faraday rotation measure in the core, consistent with previous observations at centimeter wavelengths \citep{Hovatta2012, Park2018}. Similar results hold for the other calibrators in our experiment. This confirms that our data calibration and imaging procedures are robust.

\section{Data Analysis}
\label{appendix:analysis}

\subsection{Core shift effect}

In VLBI observations of AGN jets, the apparent base of the jet, known as the radio ``core'', often appears at different positions along the jet axis when observed at different frequencies. This phenomenon, termed the ``core shift'' \citep{Lobanov1998, Hirotani2005}, arises primarily from synchrotron self-absorption within the jet plasma \citep{BK1979}. Lower frequency radio emission cannot escape from the denser regions closer to the jet base due to this absorption, meaning the $\tau=1$ surface (the apparent core) is located farther downstream at lower frequencies. Conversely, higher frequency emission originates from deeper within the jet, closer to the central engine. This frequency-dependent opacity causes the apparent core position to shift systematically with observing frequency, typically moving closer to the black hole at higher frequencies \citep{Pushkarev2012}.

\begin{figure}[t!]
\centering
\includegraphics[width=0.48\linewidth]{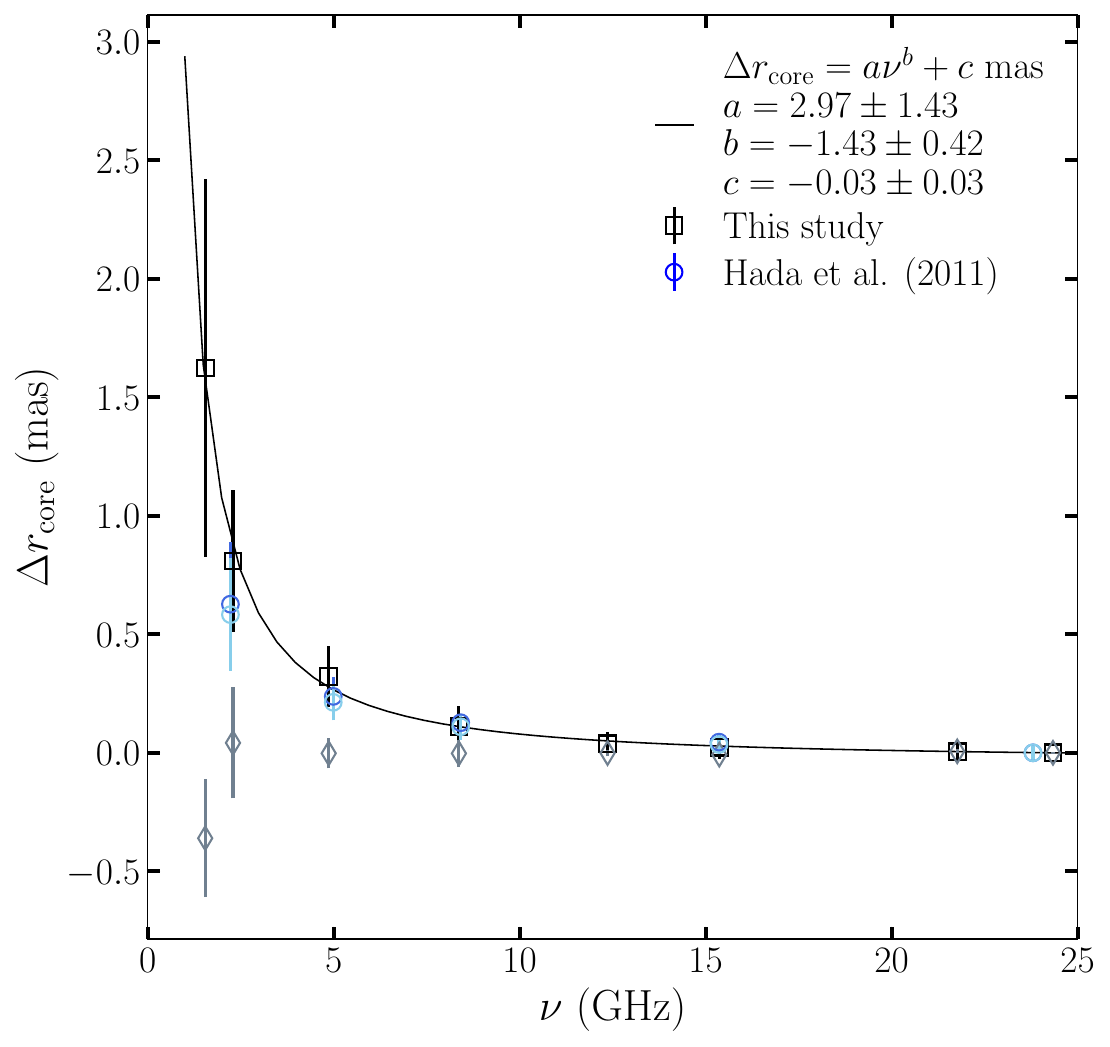}
\includegraphics[trim={0 -0.8cm 0 0},clip,width=0.48\linewidth]{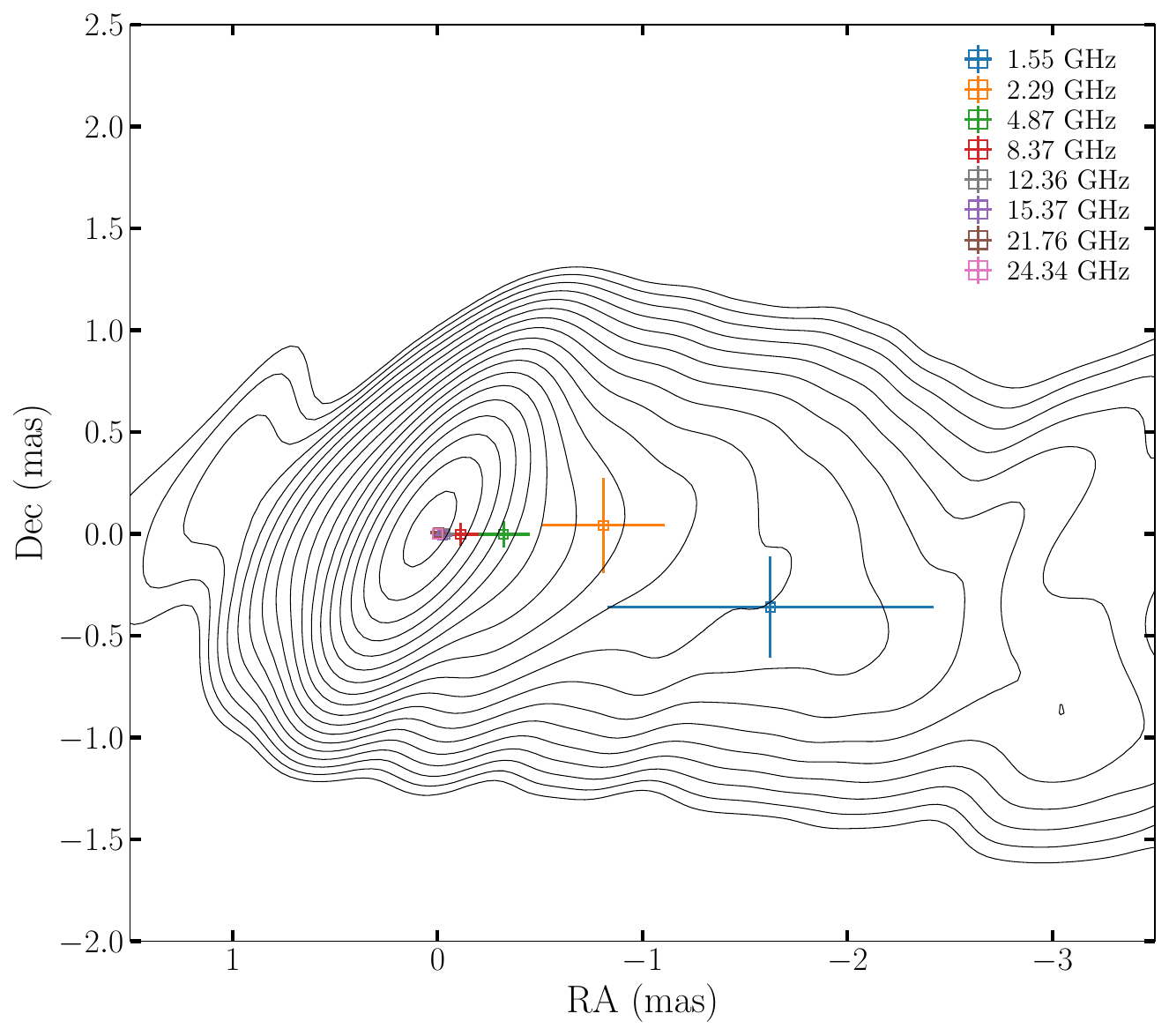}
\caption{M87 core-shift measurement. Left: core shift as a function of frequency derived from the cross-correlation method of the optically thin jet emission \citep{CG2008} along the jet axis (black square data points) and perpendicular to the jet axis (gray diamond data points). The black solid line represents the best-fit power-law function fitted to the black data points, with the best-fit parameters noted in the upper right corner. The blue data points show the core-shift measurements from \citet{Hada2011}, which are plotted by aligning their 22~GHz core position to zero for a direct comparison with our results. Note that the two data points at each frequency represent the measurements from two observations conducted on 8 and 18 Apr 2010. The good agreement is evident. Right: the best-fit core positions at various observing frequencies are shown in different colors overlaid on the Stokes~$I$ contours rotated clockwise by $23^\circ$ at 24.34\,GHz.}
\label{fig:coreshift}
\end{figure}

To quantify this effect for M87, we measured the core shift using the cross-correlation method on optically thin jet emission \citep{CG2008}, following our previous core-shift analysis of the NGC~315 jet \citep{Park2021b}. The derived core shifts are primarily aligned with the jet direction (Figure~\ref{fig:coreshift}), as expected from synchrotron self-absorption predominantly upstream in the jet \citep{BK1979}. Our results are consistent with previous measurements from astrometric observations \citep{Hada2011}. To properly align images obtained at different frequencies for subsequent intrinsic-EVPA analysis, we corrected for the core shift.

Recently, \citet{NP2024} have shown that the core shift of AGN jets is expected to follow the relationship $\Delta r_{\rm core} \propto \nu^{-1/k_r}$ with distinct behavior in $k_r$ for three different cases based on the relationship between the jet viewing angle ($\theta$) and the beaming cone ($\Gamma^{-1}$). The core-shift effect in M87 occurs at projected distances of $\lesssim1.5$\,mas at the frequencies of interest in our study (Figure~\ref{fig:coreshift}). In this region, jet kinematic results suggest that the jet is still sub-relativistic ($\Gamma \sim 1$; \citealt{Mertens2016, Park2019b}). This kinematic measurement is robust, as it is based not only on monitoring observations but also on the brightness ratio between the approaching and receding jet (assuming the difference is due to relativistic aberration; \citealt{Park2019b}). This mitigates the possibility that the measured slow speed is merely a pattern speed, suggesting it represents the bulk plasma velocity. Therefore, the M87 jet, which is believed to have a jet viewing angle of $\theta \approx 17^\circ$ \citep{Mertens2016, Walker2018}, corresponds to the $\theta \ll \Gamma^{-1}$ case, as $\Gamma^{-1} \approx 57.3^\circ$.

In this case, \citet{NP2024} predicts $k_r = k(3.5+\alpha)/(2.5+\alpha)$, where $k$ is the power-law index in the jet collimation profile $R \propto z^k$ (where $R$ is the jet radius at a distance $z$), and $\alpha$ is the spectral index in the optically thin regime (defined as $S_\nu \propto \nu^{-\alpha}$). \citet{Lu2023}, based on observations of M87 with the Global Millimeter VLBI Array (GMVA) in conjunction with the Atacama Large Millimeter/submillimeter Array (ALMA) and the Greenland Telescope (GLT; \citealt{Chen2023}) at 86\,GHz, have derived $k = 0.58$ in the innermost jet region where the core-shift effect occurs. Similarly, \citet{Ro2023} derived a spectral index corresponding to $\alpha \approx 0.7$ in our notation for the optically thin jet between 22 and 43\,GHz. Using these observationally constrained parameters, the model predicts $k_r \approx 0.76$. Our core-shift measurements indicate $k_r = 0.70 \pm 0.21$, which is consistent with the model prediction.

\subsection{Deriving the intrinsic EVPAs of the M87 jet}

Deriving the intrinsic orientation of the electric vectors in the M87 jet required correcting for significant Faraday rotation caused by an external Faraday screen. To accurately estimate the Faraday rotation and the intrinsic EVPA at a given jet location, detections of linear polarization at widely separated frequencies are required. However, if the observing frequencies are too different, those datasets probe completely different regions in the $(u,v)$ plane, making cross-frequency image comparisons challenging.

We selected three sets of observing frequencies to derive the intrinsic EVPAs: (i) 2.24, 2.27, 2.30, 2.40, and 4.87\,GHz; (ii) 8.18, 8.30, 8.43, 8.56, and 12.37\,GHz; and (iii) 12.37, 15.37, 21.77, and 24.39\,GHz. These frequencies are sensitive to the linear polarization of the jet at projected distance ranges of approximately 140--400\,mas, 20--200\,mas, and 10--30\,mas, respectively. The highest frequency set (iii), probing the innermost 10--30\,mas region, yields the highest angular resolution and the most detailed intrinsic polarization structure presented in Figure~\ref{fig:m87jet}a. Consequently, this region provides the most robust data for detailed comparison with our theoretical model presented in Section~\ref{sec:modeling}.

The intrinsic polarization maps derived from the lower frequency sets (i and ii), which probe the jet structure further downstream, are presented in Figure~\ref{fig:m87jet_large}. While qualitatively informative, the larger spatial scales and challenges in robustly deriving intrinsic polarization over the entire structure in these downstream regions make detailed modeling complex and beyond the scope of the primary quantitative analysis of this work.

\begin{figure}[t!]
\centering
\includegraphics[width=\linewidth]{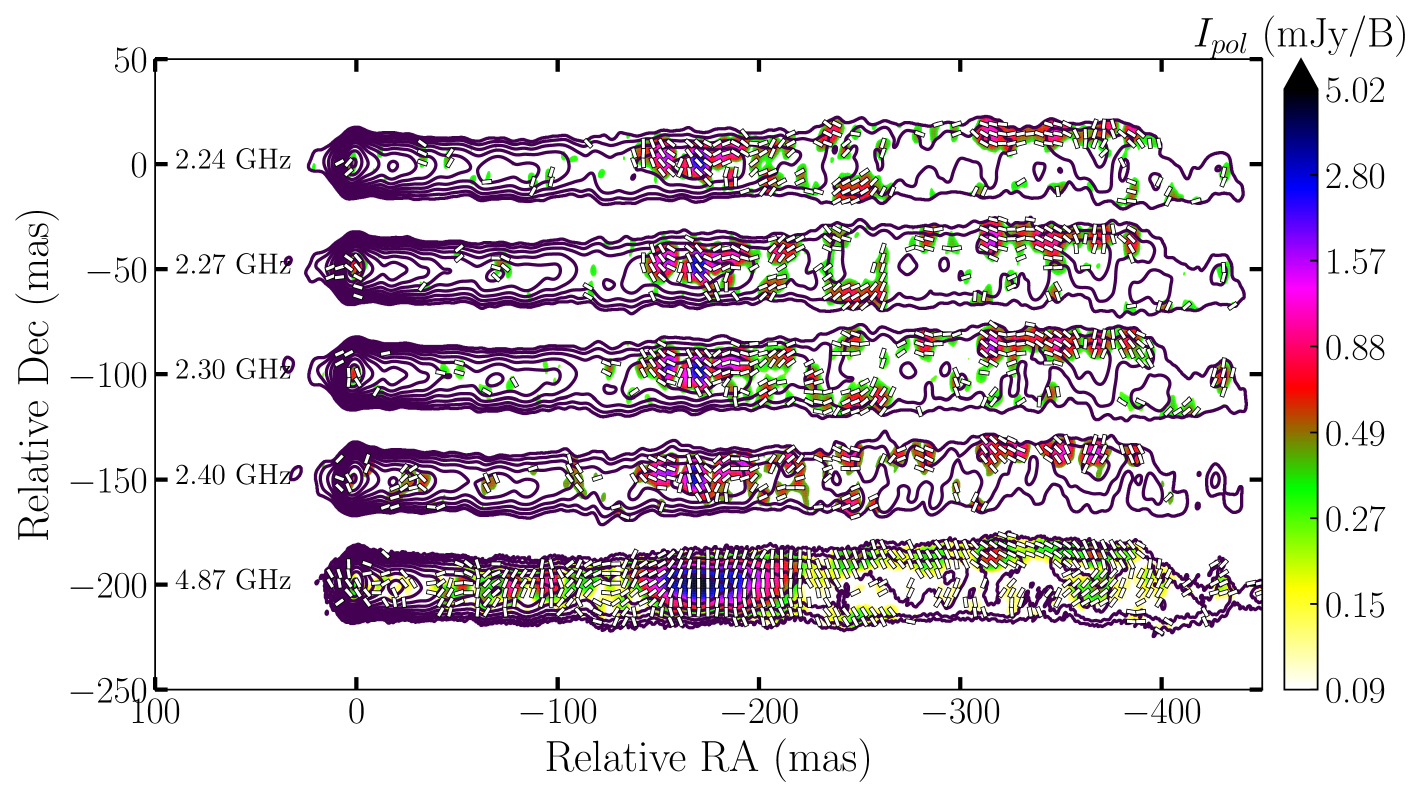}
\caption{Observed polarization maps (2.2--4.9\,GHz). Linear polarization images of the M87 jet observed at 2.24, 2.27, 2.30, 2.40, and 4.87\,GHz from top to bottom, from which the intrinsic EVPAs of the jet in the distance range of 150--400\,mas were derived (Figure~\ref{fig:m87jet_large}, top). Contours represent the total intensity distributions of the jet, starting at $7\sigma_{\rm rms}$ and increasing in steps of 2. The colors (linear polarization intensity) and white ticks (observed EVPAs) are shown only for pixels where $I_{\rm pol} > 3\sigma_{\rm rms}$. The images have been rotated clockwise by $23^\circ$ relative to the observed images to align the jet axis with the $x$-axis.}
\label{fig:m87pol_map_low}
\end{figure}

\begin{figure}[t!]
\centering
\includegraphics[width=\linewidth]{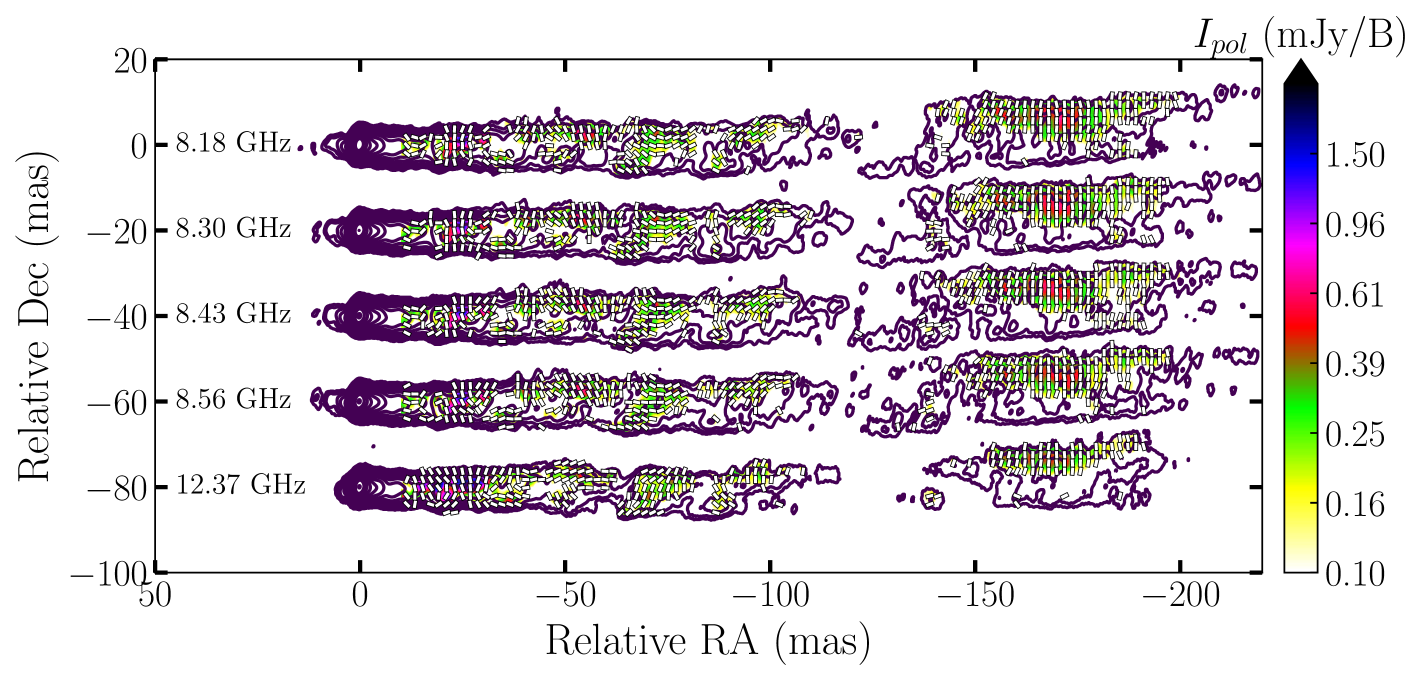}
\caption{Observed polarization maps (8.2--12.4\,GHz). Continuation of figures for the frequencies 8.18, 8.30, 8.43, 8.56, and 12.37\,GHz, from which the intrinsic EVPAs of the jet in the distance range of 15--200\,mas were derived (Figure~\ref{fig:m87jet_large}, bottom). The images have been rotated clockwise by $21^\circ$ relative to the observed images to align the jet axis with the $x$-axis.}
\label{fig:m87pol_map_mid}
\end{figure}

\begin{figure}[t!]
\centering
\includegraphics[width=\linewidth]{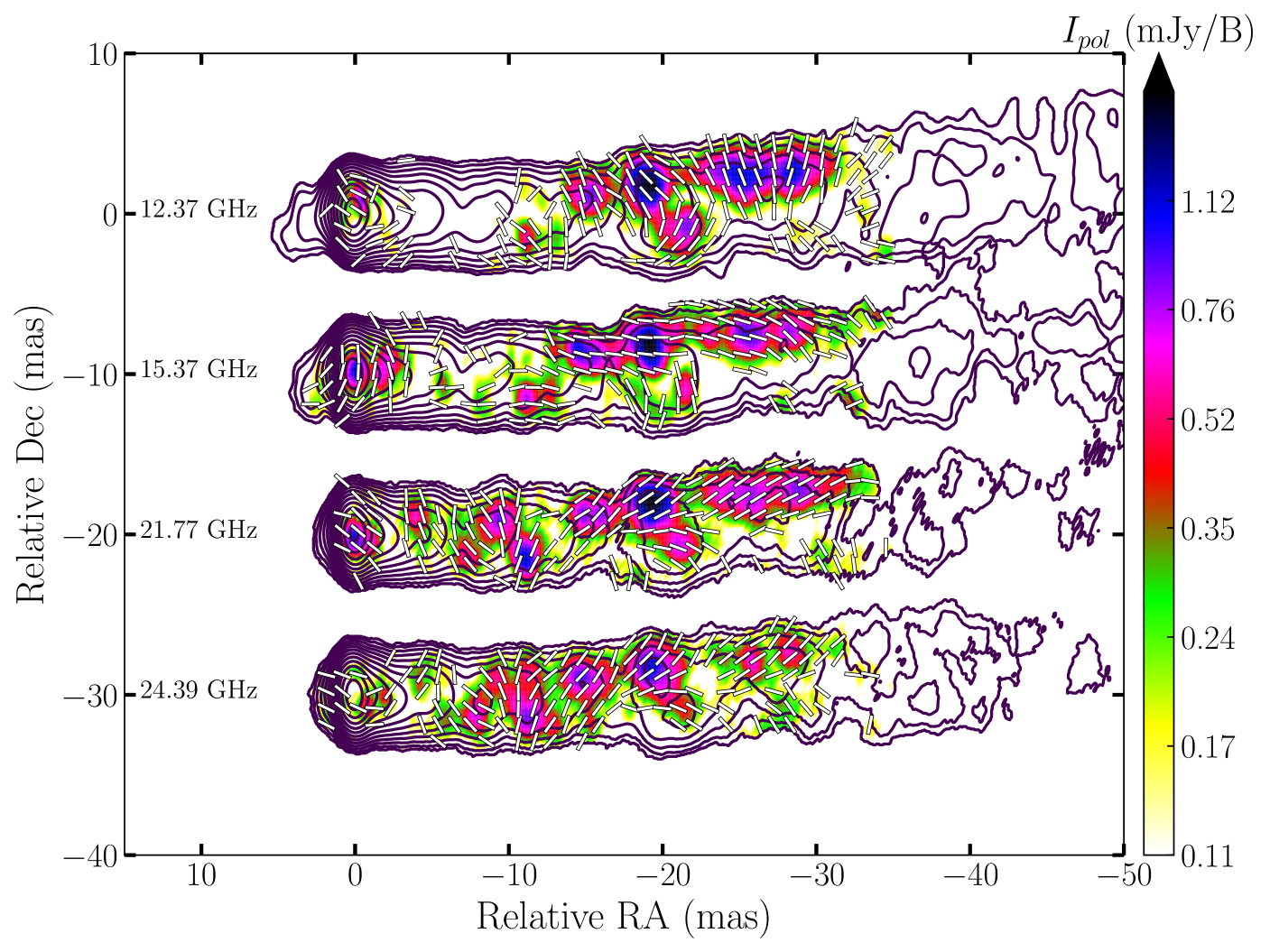}
\caption{Observed polarization maps (12.4--24.4\,GHz). Continuation of figures for the frequencies 12.37, 15.37, 21.77, and 24.39\,GHz, from which the intrinsic EVPAs of the jet in the distance range of 10--30\,mas were derived (Figure~\ref{fig:m87jet}a). The images have been rotated clockwise by $18^\circ$ relative to the observed images to align the jet axis with the $x$-axis.}
\label{fig:m87pol_map_high}
\end{figure}

To optimize sensitivity while managing Faraday depolarization across the observing bands, different approaches were used for image generation. The images for the 4.87, 12.37, 15.37, 21.77, and 24.39\,GHz bands were derived by combining all baseband channels (often referred to as IFs). In contrast, images for the lower frequency bands were derived either from each baseband channel separately (2.24, 2.27, 2.30, 2.40\,GHz) or from pairs of neighboring channels (8.18, 8.30, 8.43, 8.56\,GHz). This strategy maximizes sensitivity by using wider bandwidths where possible (e.g., at higher frequencies or larger distances where Faraday rotation within the band is minimal), while avoiding significant bandwidth depolarization where Faraday rotation between sub-bands is substantial (typically at lower frequencies).

In this paper, we present results primarily from the three selected frequency sets detailed above. While analyses using other combinations of frequencies were performed and confirmed to yield qualitatively similar outcomes, they are not included here as they largely overlap with the presented results and provide minimal additional insight.

\begin{figure}[t!]
\centering
\includegraphics[trim={0.15cm 0.15cm 0.15cm 0.15cm},clip,width=\linewidth]{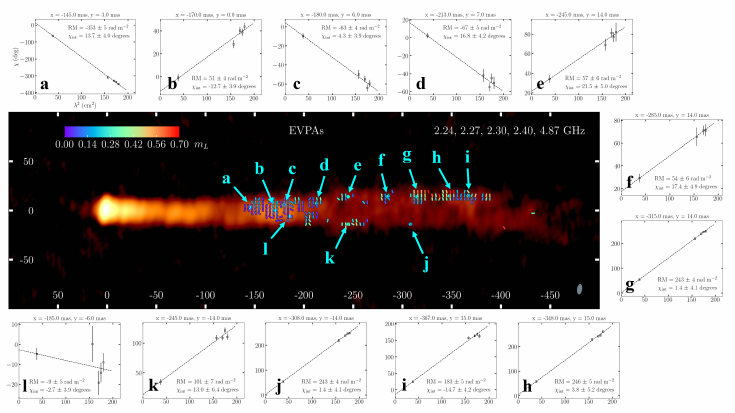}
\caption{EVPA vs.\ $\lambda^2$ fits (2.2--4.9\,GHz). Intrinsic EVPAs of the jet derived from the low-frequency bands (2.24--4.87\,GHz), as shown in Figure~\ref{fig:m87jet_large} (top), but with additional subplots displaying the observed EVPAs as functions of $\lambda^2$ at various locations on the jet, indicated by cyan arrows. Each subplot notes the values of the best-fit RM and the intrinsic, RM-corrected EVPA.}
\label{fig:m87pol_prop_low}
\end{figure}

\begin{figure}[t!]
\centering
\includegraphics[trim={0.15cm 0.15cm 0.15cm 0.15cm},clip,width=\linewidth]{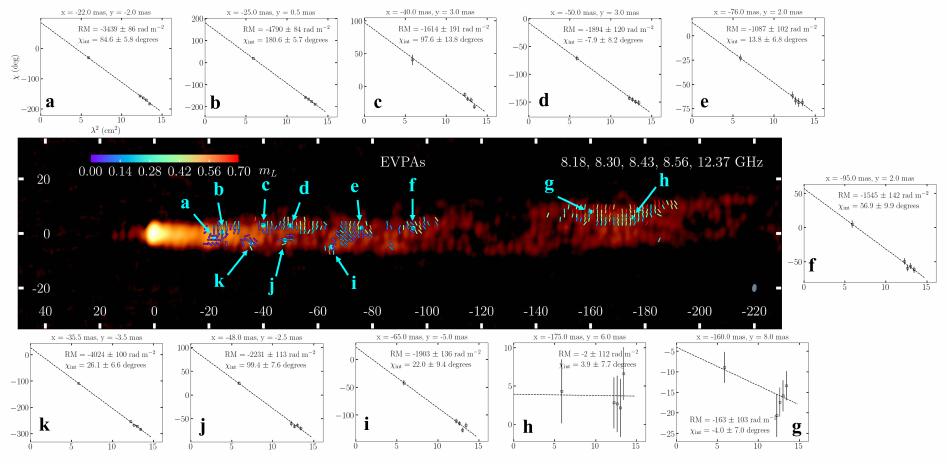}
\caption{EVPA vs.\ $\lambda^2$ fits (8.2--12.4\,GHz). Continuation of figures for the mid-frequency bands (8.18--12.37\,GHz).}
\label{fig:m87pol_prop_mid}
\end{figure}

\begin{figure}[t!]
\centering
\includegraphics[trim={0.15cm 0.15cm 0.15cm 0.15cm},clip,width=\linewidth]{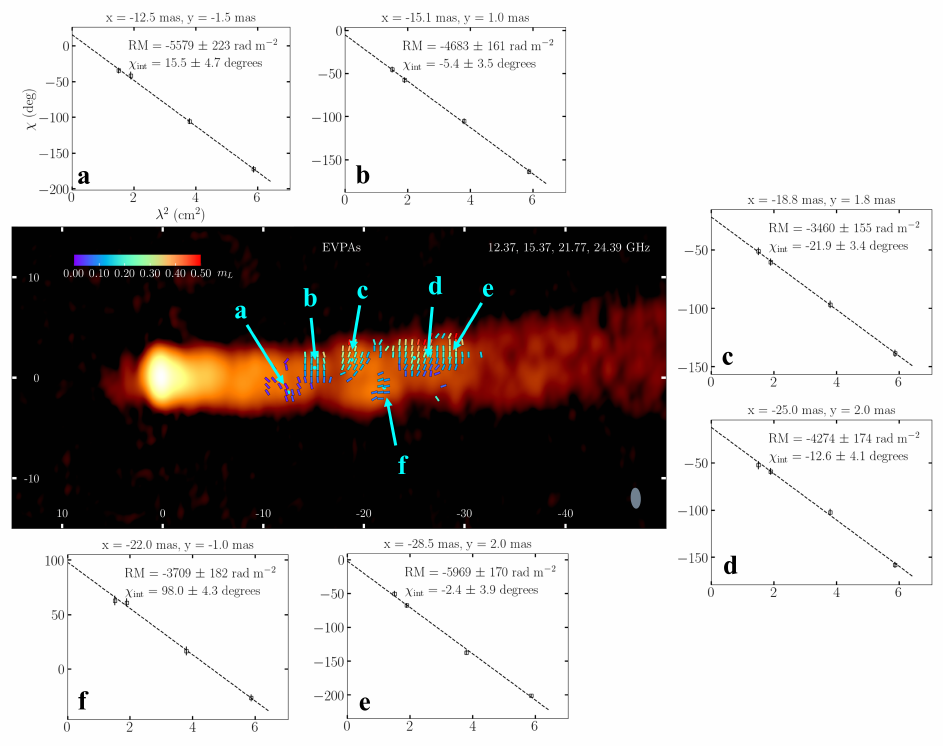}
\caption{EVPA vs.\ $\lambda^2$ fits (12.4--24.4\,GHz). Continuation of figures for the high-frequency bands (12.37--24.39\,GHz).}
\label{fig:m87pol_prop_high}
\end{figure}

We convolved the CLEAN models for Stokes $I$, $Q$, and $U$ in each frequency set with the synthesized beam of the lowest frequency in the set. The convolved linear polarization images for each frequency combination are presented in Figures~\ref{fig:m87pol_map_low},~\ref{fig:m87pol_map_mid},~\ref{fig:m87pol_map_high}. We selected pixels where the Ricean de-biased linear polarization intensity ($I_{\rm pol}$) is greater than three times the rms noise level derived from the off-source region in the CLEAN images at each frequency. For the error analysis, we used the off-source rms noise level instead of the method suggested by a previous study \citep{Hovatta2012}, because we found that the errors in the linear polarization images caused by imperfect antenna leakage calibration are not preferentially distributed on source; instead, they appear off source. The $1\sigma$ uncertainty of the EVPA, $\sigma_{\chi}$, was computed following \citet{Hovatta2012} by adding the statistical and systematic uncertainties in quadrature: $\sigma_{\chi} = \sqrt{\sigma_{\rm stat}^2 + \sigma_{\rm cal}^2}$. The statistical uncertainty $\sigma_{\rm stat}$ (in radians) is given by $\sigma_{\rm stat} = \sigma_{\rm rms} / (2 I_{\rm pol})$, where $I_{\rm pol}$ is the linearly polarized intensity and $\sigma_{\rm rms} = (\sigma_Q + \sigma_U) / 2$ is the average off-source rms noise, where $\sigma_Q$ and $\sigma_U$ are the off-source rms noise levels in the Stokes $Q$ and $U$ images, respectively. The systematic calibration uncertainty is $\sigma_{\rm cal} = 3^\circ$ (in degrees; Appendix~\ref{appendix:observation}). We fitted a $\lambda^2$ law to the observed EVPAs derived at the selected pixels. To resolve the $n\pi$ ambiguity, we rotated the EVPAs at each frequency by integer multiples of $\pi$ (up to $7\pi$) and selected the combination that minimized $\chi^2$. We found that the EVPAs are generally well described by the $\lambda^2$ law (see Figures~\ref{fig:m87pol_prop_low},~\ref{fig:m87pol_prop_mid},~\ref{fig:m87pol_prop_high} for example fits), but show large $\chi^2$ values at the edges of the polarized regions where the linear polarization is weak at one or more frequency bands (usually at lower frequency bands). This behavior can be interpreted as detecting only parts of a complex linear polarization structure in those frequency bands, while the pixels in the edge regions passed the intensity cutoff criteria due to the convolution of the CLEAN beam. Therefore, we removed the pixels with $\chi^2$ values exceeding the 95\% confidence limit of the $\chi^2$ distribution for each frequency set, following the approach used in a previous systematic study of Faraday rotation in various AGN jets \citep{Hovatta2012}.

\begin{figure}[t!]
\centering
\includegraphics[width=0.5\linewidth]{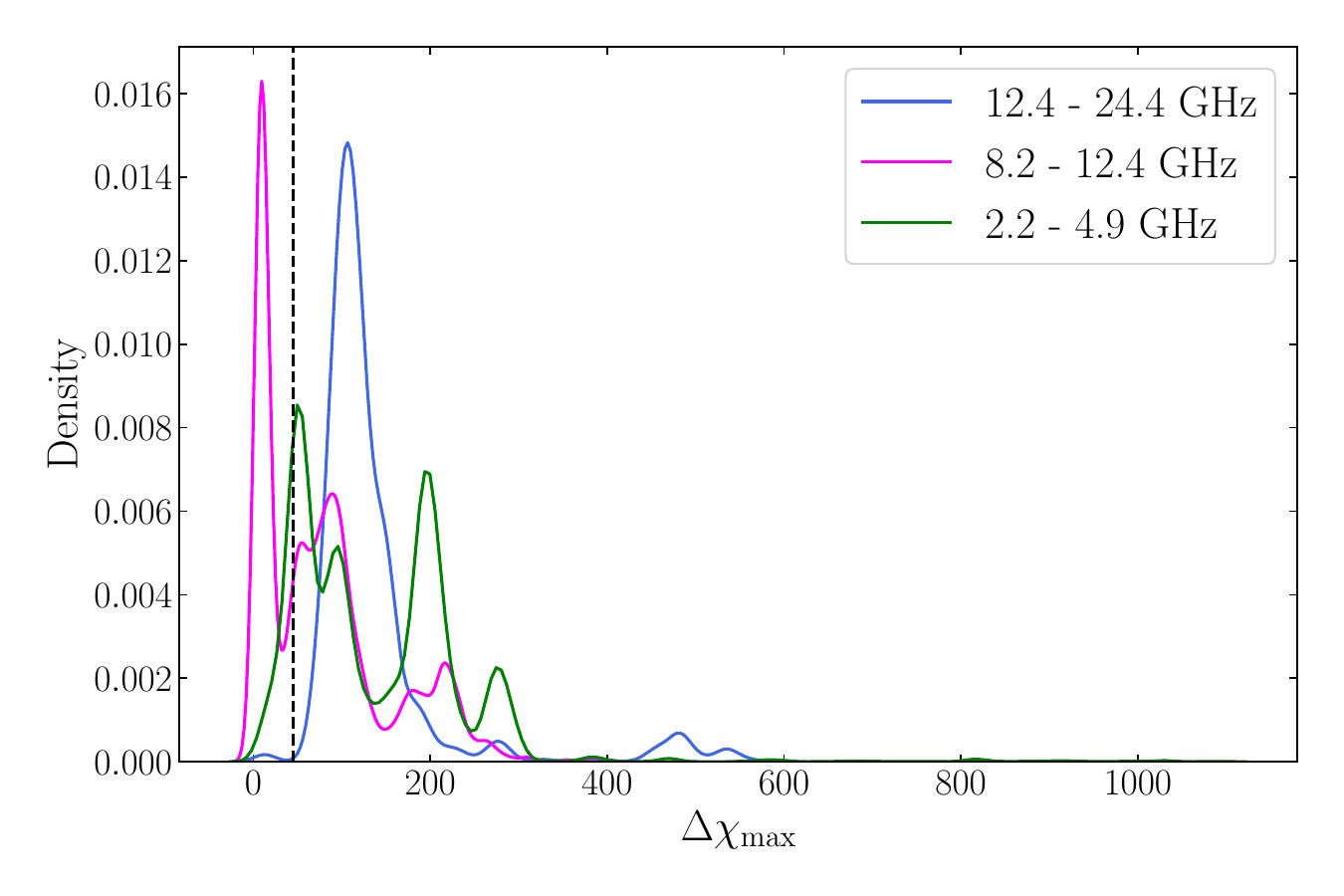}
\caption{Distribution of maximum EVPA rotations. The density distributions of the maximum EVPA rotations within the bandwidths observed in the 12.4--24.4\,GHz, 8.2--12.4\,GHz, and 2.2--4.9\,GHz bands are depicted in blue, magenta, and green, respectively. The vertical black dashed line marks a rotation of $45^\circ$. Notably, most of the observed EVPA rotations within the respective bandwidths exceed $45^{\circ}$, suggesting that the observed Faraday rotations are caused by an external magnetized medium. It is important to note that the peak smaller than $45^\circ$ in the 8.2--12.4\,GHz band is due to the region where the observed RMs are close to zero.}
\label{fig:maxrot}
\end{figure}

\begin{figure}[t!]
\centering
\includegraphics[width=\linewidth]{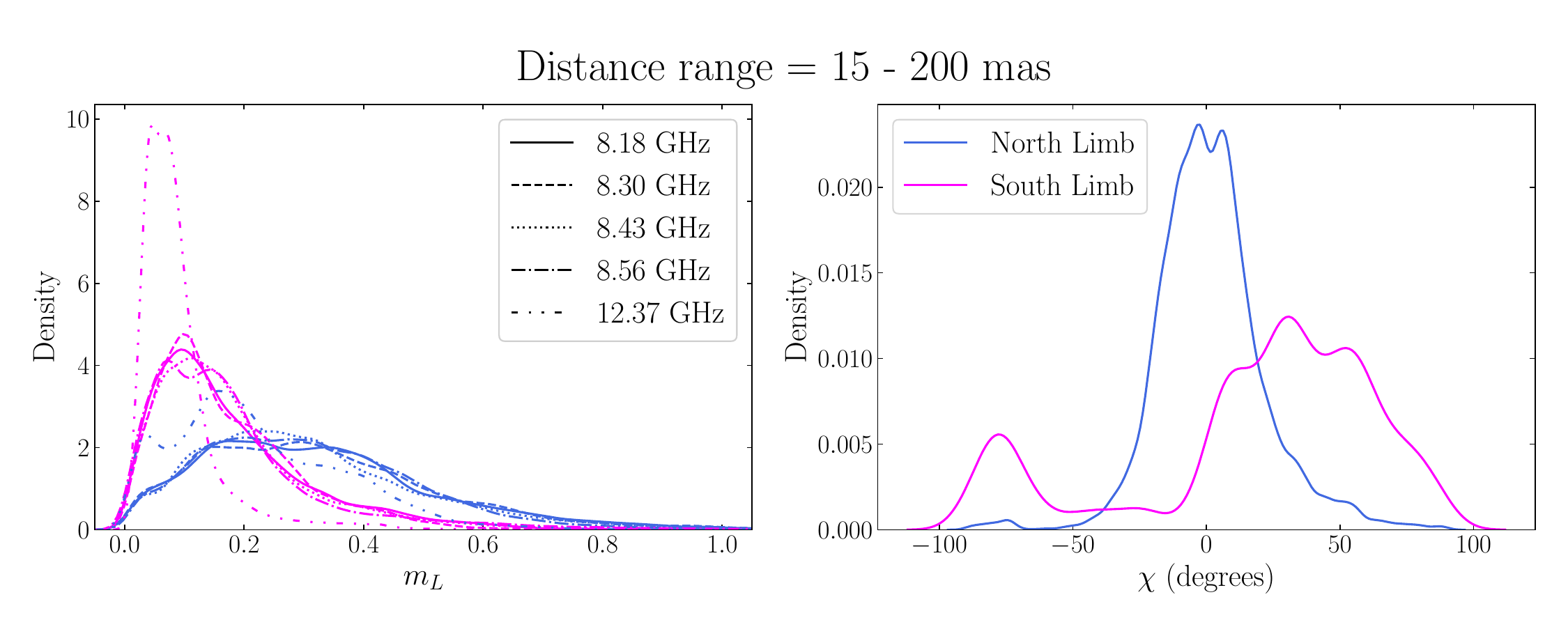}
\includegraphics[width=\linewidth]{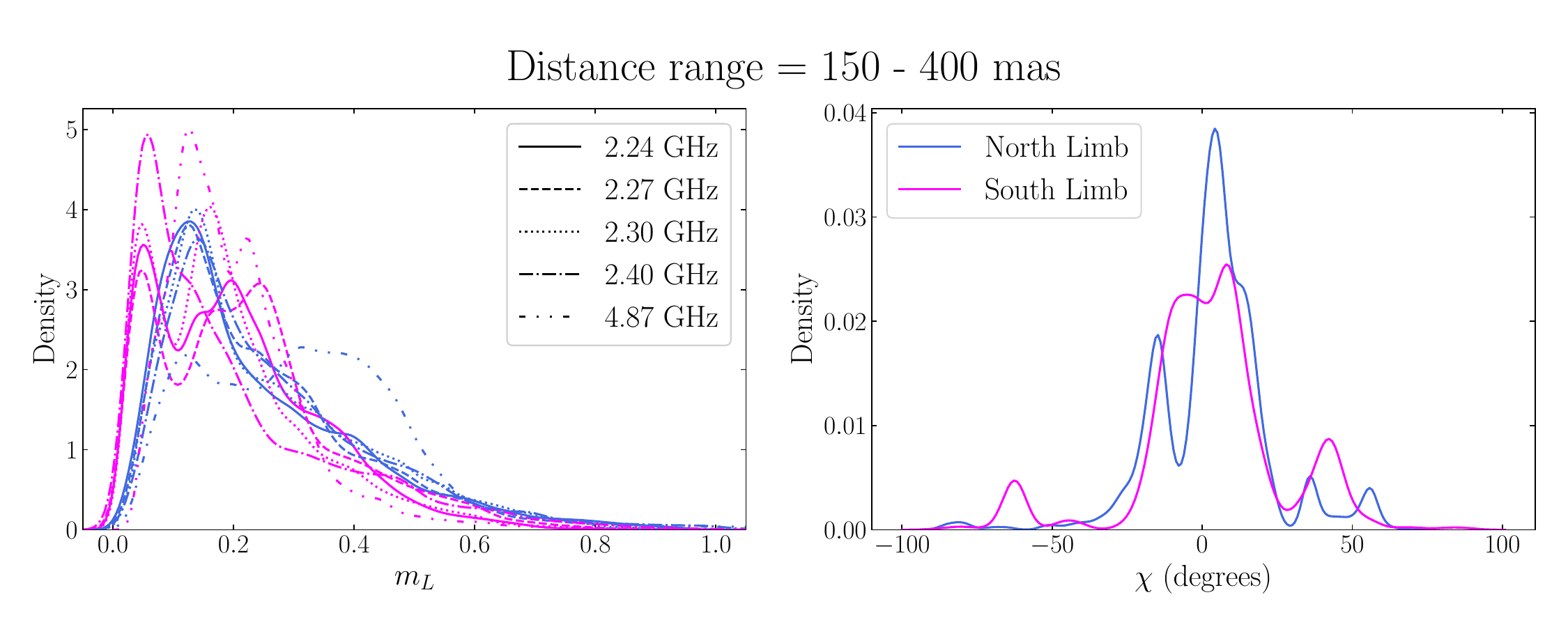}
\caption{Polarization property distributions for lower frequencies. This figure is similar to Figure~\ref{fig:modelcomp}, but it presents data for the 8.2--12.4\,GHz bands (upper) and the 2.3--4.9\,GHz bands (lower). Unlike Figure~\ref{fig:modelcomp}, the density distributions of the fractional polarization for different observing frequencies are distinguished by various line styles.}
\label{fig:frac_evpa_histo_larger}
\end{figure}

Figure~\ref{fig:maxrot} shows the distribution of maximum EVPA rotations for pixels meeting both the intensity and $\chi^2$ cutoff criteria. Most detected EVPA rotations across the observed bandwidth exceed $45^{\circ}$. However, in the 8.2--12.4\,GHz frequency range, the distribution peaks at an angle less than $45^\circ$. This peak corresponds to the jet region between 150 and 200\,mas from the core. This region is characterized by small RMs of $\lesssim100\ {\rm rad\ m^{-2}}$ \citep{Park2019a}; correspondingly, the EVPA rotations here are also minimal within this frequency range. Furthermore, internal Faraday rotation can cause significant depolarization due to the partial cancellation of linear polarization vectors experiencing varying amounts of rotation along the line of sight. This effect, which is typically stronger at longer wavelengths, is not observed in our data (Figure~\ref{fig:frac_evpa_histo_larger}). Both the large magnitudes of the observed EVPA rotations and the lack of significant depolarization suggest that the observed Faraday rotation occurs mainly in an external Faraday screen \citep{Burn1966, Sokoloff1998}, at least for the analyzed pixels. This finding is consistent with our previous study \citep{Park2019a} but is based on a significantly expanded area of detected Faraday rotation. Moreover, because the rotation occurs in an external screen, we can determine the RM-corrected, intrinsic jet EVPAs by extrapolating the fitted $\lambda^2$ laws to $\lambda=0$.

In Figures~\ref{fig:m87pol_prop_low},~\ref{fig:m87pol_prop_mid},~\ref{fig:m87pol_prop_high}, we present the linear polarization images and the EVPAs as functions of $\lambda^2$ and their corresponding best-fit $\lambda^2$ fits at several locations on the jet.

\section{Comparison between the Data and Model} 
\label{appendix:modelcomp}

Figure~\ref{fig:m87jet} presents a qualitative comparison in which the observed images are shown after CLEAN–beam convolution, whereas the model images are left unconvolved so that intrinsic model morphology is not overly blurred in the main figure. A natural concern is that observational filtering (beam convolution and thermal noise) could modify the apparent north–south asymmetry. To address this, here we perform a forward, like-for-like test for the innermost jet (10--30\,mas) at 12.4--24.4\,GHz: we convolve the fiducial MHD model images with the CLEAN restoring beam, match the pixel scale, inject thermal noise, and then compare them with the data as follows.

The model images were convolved with the synthesized beam of the 12.4\,GHz data, which was used to convolve all the CLEAN images. We added random noise to each pixel in the model image, drawn from Gaussian distributions with zero means and standard deviations equal to the observed off-source rms noise levels for the Stokes $I$, $Q$, and $U$ model images, respectively. Additionally, we added random noise to the model EVPA in each pixel, drawn from Gaussian distributions with zero means and standard deviations equal to the mean 1$\sigma$ uncertainty of the intrinsic EVPAs estimated from the data, which is approximately $4.8^{\circ}$.

\begin{figure}[t!]
\centering
\includegraphics[width=0.49\linewidth]{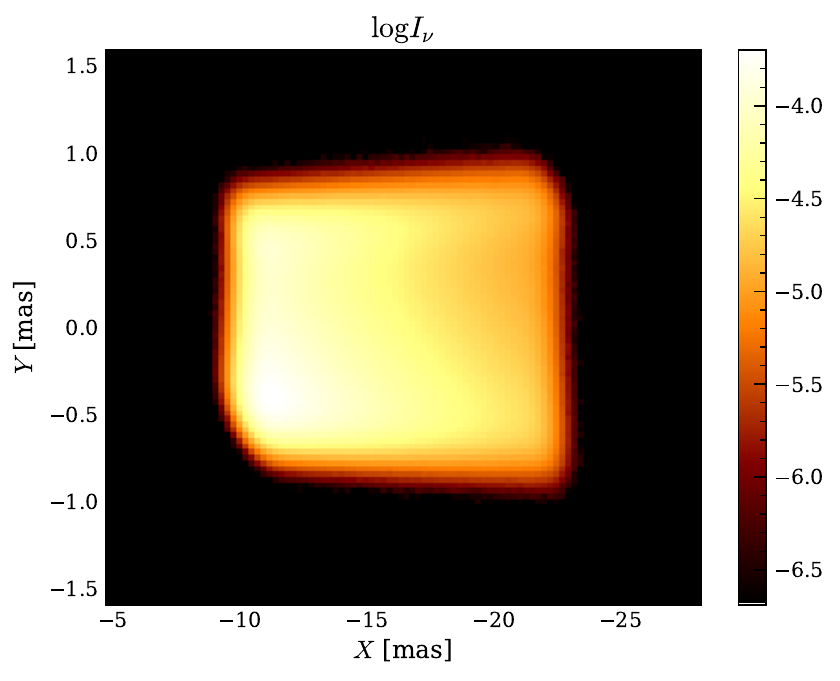}
\includegraphics[width=0.49\linewidth]{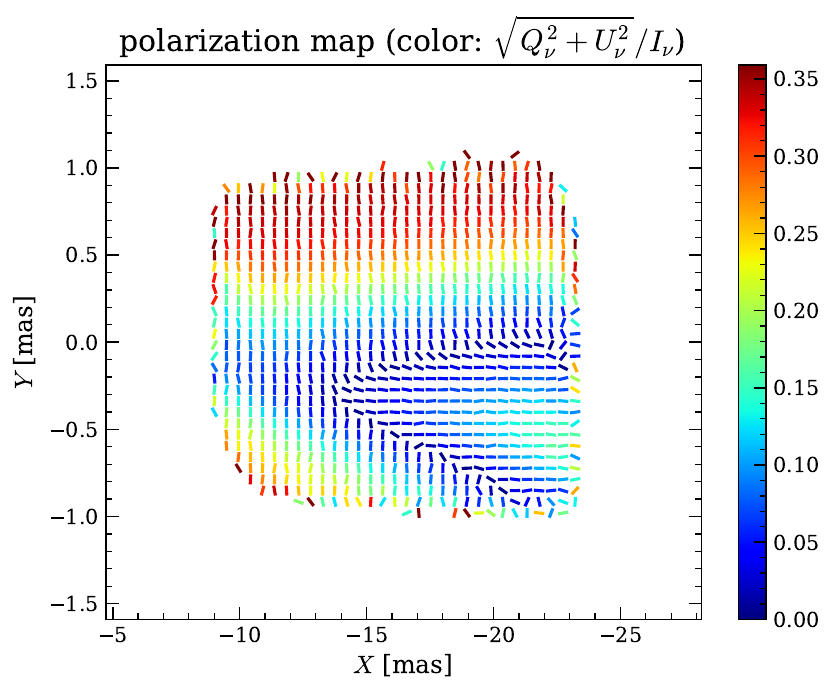}
\caption{Processed fiducial model images for comparison. Same as Figure~\ref{fig:m87jet}b,c but after (i) scaling the total flux to match the observed CLEAN flux within the same jet distance range, (ii) convolving the images with the observed synthesized CLEAN beam, (iii) adding Gaussian random noise to each Stokes image, with the noise level set by the observed data, and (iv) introducing additional noise to the intrinsic EVPAs, with a level derived from the observed intrinsic EVPA images.}
\label{fig:model_images_blurred}
\end{figure}

\begin{figure}[t!]
\centering
\includegraphics[width=\linewidth]{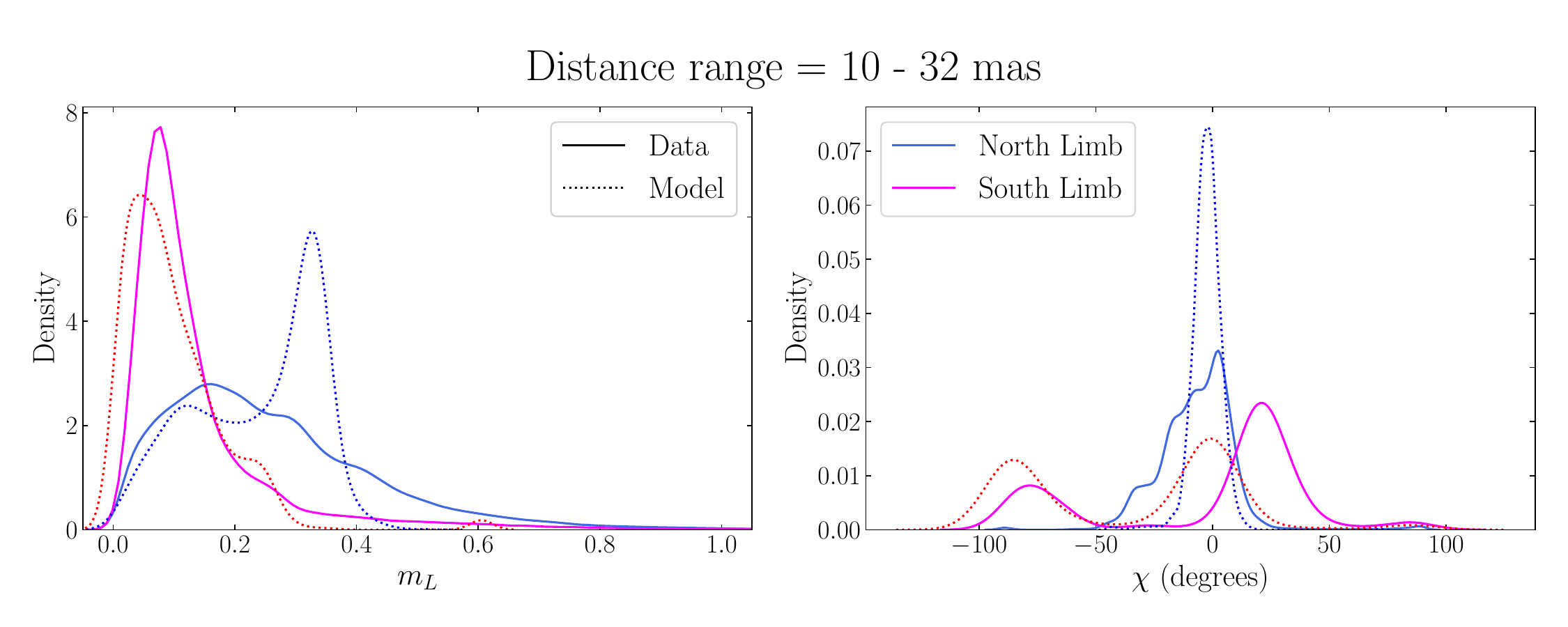}
\caption{Data--model comparison for inner jet polarization properties (12.4--24.4\,GHz). The distributions of the fractional polarization (left) and intrinsic, RM-corrected EVPAs (right) in the north (blue) and south jet parts (magenta), obtained from the frequency combination of 12.4--24.4\,GHz, are sensitive to the emission in the distance range of 10--32\,mas. The data are represented by solid curves, while the model is depicted with dotted curves.}
\label{fig:modelcomp}
\end{figure}

\begin{figure}[t!]
\centering
\includegraphics[width=0.49\linewidth]{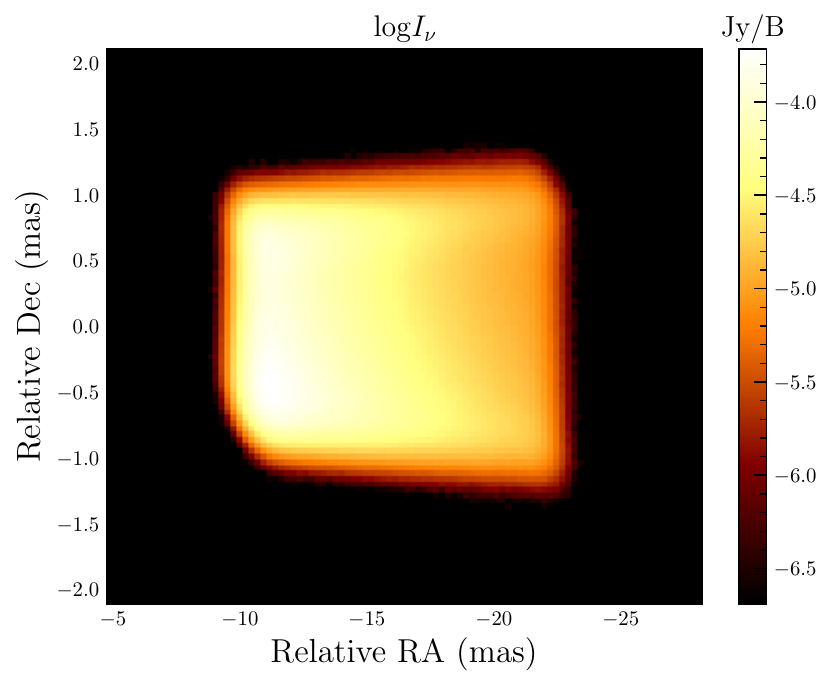}
\includegraphics[width=0.49\linewidth]{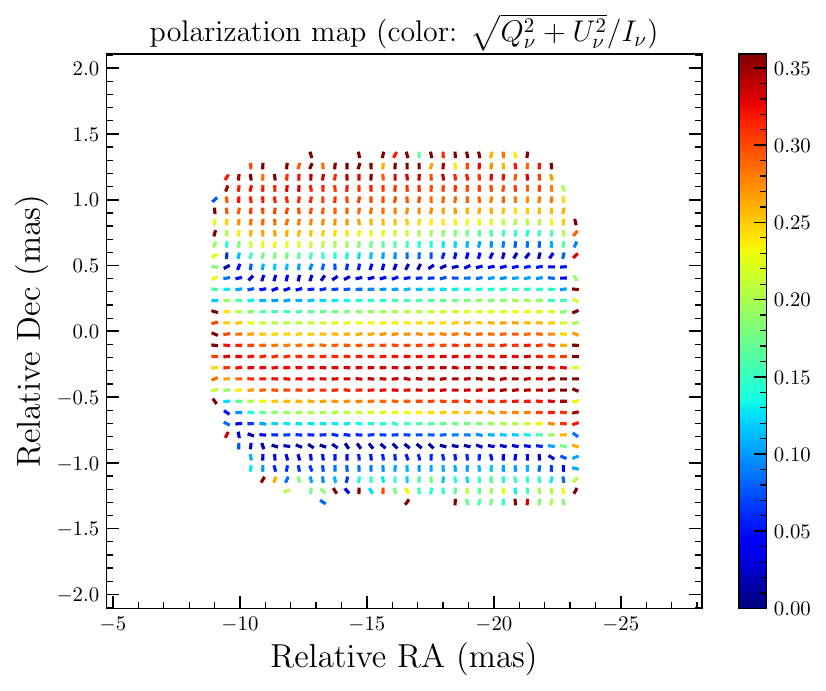}
\caption{Same as Figure~\ref{fig:model_images_blurred}, but for the "no dissipation" model with $a=0.5$ (as presented in Figure~\ref{fig:model_alternatives}b). The jet appears wider than the fiducial model (Figure~\ref{fig:model_images_blurred}) because the event horizon radius, $r_H = (1 + \sqrt{1 - a^2})R_g$, is larger for a smaller spin $a$, resulting in a wider location where the outermost magnetic field line touches the horizon.}
\label{fig:no_dissipation_model_images_blurred}
\end{figure}

To define a consistent reference axis for comparing polarization properties along the jet, we computed the intensity-weighted average of each transversely sliced intensity profile and defined this averaged point as the local jet axis. We derived the distributions of the fractional linear polarization and the intrinsic EVPAs in the north and south jet parts, defined as the regions above and below the jet axis, respectively. For the model image, the jet axis was precisely defined as the $x$-axis of the image, and we computed the distributions of the north (positive $y$-values) and south (negative $y$-values) jet parts. The processed model images of our fiducial model are presented in Figure~\ref{fig:model_images_blurred}. The comparison of the distributions of linear polarization properties between the north and south jet parts from the data and this processed fiducial model is shown in Figure~\ref{fig:modelcomp}. For a direct comparison, the same processing was applied to the alternative "no dissipation" model (Figure~\ref{fig:model_alternatives}b), and the results are presented in Figure~\ref{fig:no_dissipation_model_images_blurred}.

\section{Model}
\label{appendix:model}

\begin{figure}[t!]
\centering
\includegraphics[width=0.5\columnwidth]{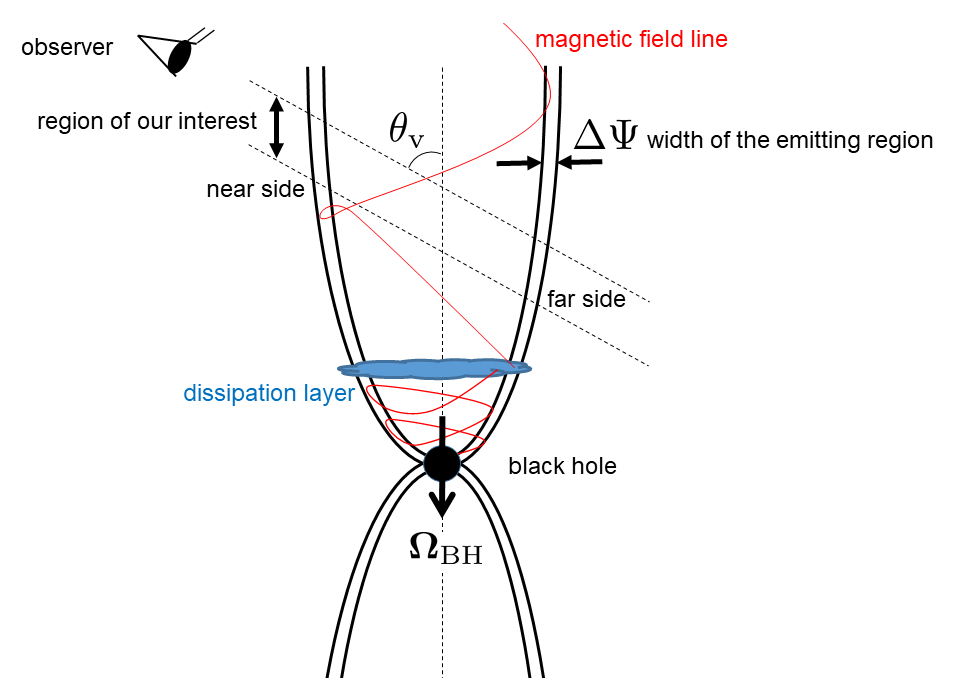}
\caption{A schematic representation of our jet model. Our model incorporates the net effects of an inferred upstream magnetic dissipation process, conceptually depicted here as a ``dissipation layer'' (see also Figure~3), occurring upstream of our main region of interest ($\gtrsim10$ mas).  This upstream dissipation is required by our model to transform the magnetic field, facilitating the observed helical structure downstream. The depicted layer is a simplified illustration for our numerical treatment, which focuses on the result of this dissipation (e.g., a modified $\Omega_F$ and magnetic field geometry) rather than its intrinsic microphysics. The actual dissipation is likely an extended and potentially complex physical process, possibly occurring over a wide range of distances or in multiple layers; resolving its detailed nature will be key for future advancements in jet theory. The black hole spin is assumed to be oriented away from the observer, with the M87 jet viewed at an angle of $163^\circ$ \citep{Mertens2016, Walker2018}. The observed edge-brightened jet structure results from the integration of emissions from the near side (upper left in the figure) and the far side (lower right part).}
% \hyp{[it would be super helpful to indicate the "near side" and "far side" in the figure.}\kt{Revised as suggested.}
\label{fig.theo.schematic}
\end{figure}

Figure~\ref{fig.theo.schematic} presents a schematic picture of our jet model. The observed polarization and intensity patterns have a complicated structure and hence require a non-trivial jet structure that is different from either toroidally or poloidally dominant structures. In our jet model, under the assumption of steady, axisymmetric, cold ideal MHD flows, a jet structure is ultimately given by several parameters that include $\Omega_F$ as shown below, where $\Omega_F$ is the field-line angular velocity and is conserved along a magnetic field line. The model parameters are determined at the boundary where a magnetic field line originates. Usually, boundary conditions are given around the central black hole, where magnetic field lines penetrate the black hole and the Blandford--Znajek (BZ; \citealp{BZ1977}) process launches the jet. The BZ mechanism predicts $\Omega_F \sim \Omega_{\rm BH}/2$ \citep{Tchekhovskoy2010}, where $\Omega_{\rm BH}$ is the angular velocity of the black hole. However, a large $|\Omega_{\rm BH}|$ quickly winds the field lines, producing toroidal dominance on $\sim$10\,mas scales that fails to reproduce the observed polarization and intensity patterns. Hence, we assume that there is a dissipation layer upstream of the observed region at a distance of $\approx10^4\,R_g$ from the black hole, where a toroidally dominant magnetic field is dissipated into a helical magnetic field while maintaining the shape of the jet, which effectively reduces the magnitude of $\Omega_F$ and could change other parameter values as well. Following this picture, we set the upstream boundary at the dissipation layer and study the jet structure by freely changing the magnitude of $\Omega_F$ and other parameters irrespective of the boundary conditions near the central black hole.

A jet structure is constructed so that the accelerated flow passes through a fast critical point by following the prescription of a previous study for a purely paraboloidal jet \citep{PT2020}. We assume that a parameter in the prescription, $\xi$, the ratio of the electric field strength to the toroidal magnetic field strength, is constant along a magnetic field line, while adopting $\xi = \sqrt{1 - \Gamma_\infty^{-2}}$ ($\Gamma_\infty$ is the terminal Lorentz factor), which is a unique choice when taking into account the asymptotic region consistently. Since we primarily focus on a geometrically thin edge of a jet, we assume that the model parameters that are conserved along a magnetic field line are constant across the field line. 

We adopt a purely paraboloidal jet shape ($z\propto R^2$, where $z$ is the distance along the jet direction), which enables us to bypass solving the trans-field equation (the Grad--Shafranov equation) and achieve a trans-sonic flow using the method described in a previous study \citep{TT2003}. Note that the actual jet shape is semi-parabolic ($z\propto R^{1.6}$) at this scale \citep{AN2012, Hada2013, Nakamura2018}, resulting in a slightly smaller jet width in the model image than the observed jet image (Figure~\ref{fig:m87jet}). We assume that the black hole's spin vector is directed away from us, determining the winding direction of the frame-dragged magnetic fields and, consequently, the sign of $\Omega_F$. As explained in Section~\ref{sec:modeling}, we focus on two key parameter combinations: $\Omega_F/\Omega_{\rm BH}$ and $L\Omega_F/(c^2\mathcal{E})$.

For a given jet structure, the model EVPAs are obtained by calculating Stokes parameters in the same manner as in previous studies \citep{Takahashi2018,Lyutikov2005}. The electric field vectors are obtained via the Lorentz transformation (i.e., rotation depending on the Lorentz factor of a jet and observer direction) of those in the fluid rest frame. In the fluid rest frame, these vectors are determined by the jet's magnetic field structure and the observer direction.

To reflect the observed edge-brightened structure of the M87 jet \citep{Hada2016, Walker2018}, we model synchrotron emission from non-thermal electrons confined to a geometrically thin sheath at the jet boundary (see Figure~\ref{fig.theo.schematic}). We assume that the jet is optically thin, with a non-thermal electron energy distribution that follows a power law and an isotropic pitch-angle distribution in the fluid rest frame, and a number density that evolves according to the continuity equation without any additional particle loading or injection downstream \citep{BroderickLoeb2009, Takahashi2018}. Plasma effects, such as internal Faraday rotation, are neglected. This is justified because the model is compared to intrinsic EVPAs after correcting for external Faraday rotations, and the data show no indication of significant internal Faraday rotation (see Section~\ref{sec:observations} and Appendix~\ref{appendix:analysis} for more details). 

In our model, the outermost magnetic field line in the emitting region threads the event horizon on the equatorial plane (i.e., the plane orthogonal to the black hole spin axis), while the innermost magnetic field line in the emitting region penetrates the black hole and effectively touches the horizon at $0.9 r_H$ on the equatorial plane.

We adopt the viewing angle $\theta = 163^\circ$ and the Kerr parameter $a=0.998$. The terminal Lorentz factor of the jet is set to $\mathcal{E}=\Gamma_\infty=10$. For the non-thermal electron energy distribution, we adopt a power-law with index $p=2.4$, the minimum Lorentz factor $\gamma_\mathrm{m}' = 100$, and the maximum Lorentz factor $\gamma_\mathrm{M}' = \infty$.

A systematic survey of combinations of the key model parameters shows that the model with parameters $\Omega_F/|\Omega_{\rm BH}| = -0.065$ and $L\Omega_F/(c^2\mathcal{E}) = -0.6$ reproduces the observed total intensity and linear polarization features, as presented in Figure~\ref{fig:m87jet}b,c; other combinations of these parameters tested in our survey did not match the observations as well. The negative sign of $\Omega_F$ indicates that the black hole's spin vector points away from the observer. We note that a recent study \citep{Kino2022} suggested a small value of $\Omega_F$ in the M87 jet based on modeling of the jet kinematic result \citep{Park2019b} under the ideal MHD assumption without including dissipation processes. Projected Stokes $I$, $Q$, and $U$ model images are obtained by solving the radiative transfer equation along rays through the emitting region. As a simple phenomenological treatment of depolarization from additional turbulent fields \citep{Laing1980, Laing1981}, we scale the computed fractional polarization by a factor of 1/2. 

We acknowledge that the true depolarization process is likely more complex and may vary spatially within the jet, potentially explaining why some pixels in the observed linear polarization maps (Figures~\ref{fig:m87jet}a and~\ref{fig:m87jet_large}) lack detectable polarization. However, it is physically unlikely that these depolarization effects would be significantly stronger preferentially in the southern part of the jet compared to the northern part at a given projected distance. Consequently, although such depolarization can shift the absolute polarization levels, it cannot plausibly produce the persistent north--south asymmetry over a long distance range (10-100 mas) we observe. Our main conclusions, derived from this systematic difference, should therefore be robust to unmodeled depolarization effects.

\begin{figure}[t!]
 \centering
 \includegraphics[width=0.49\columnwidth]{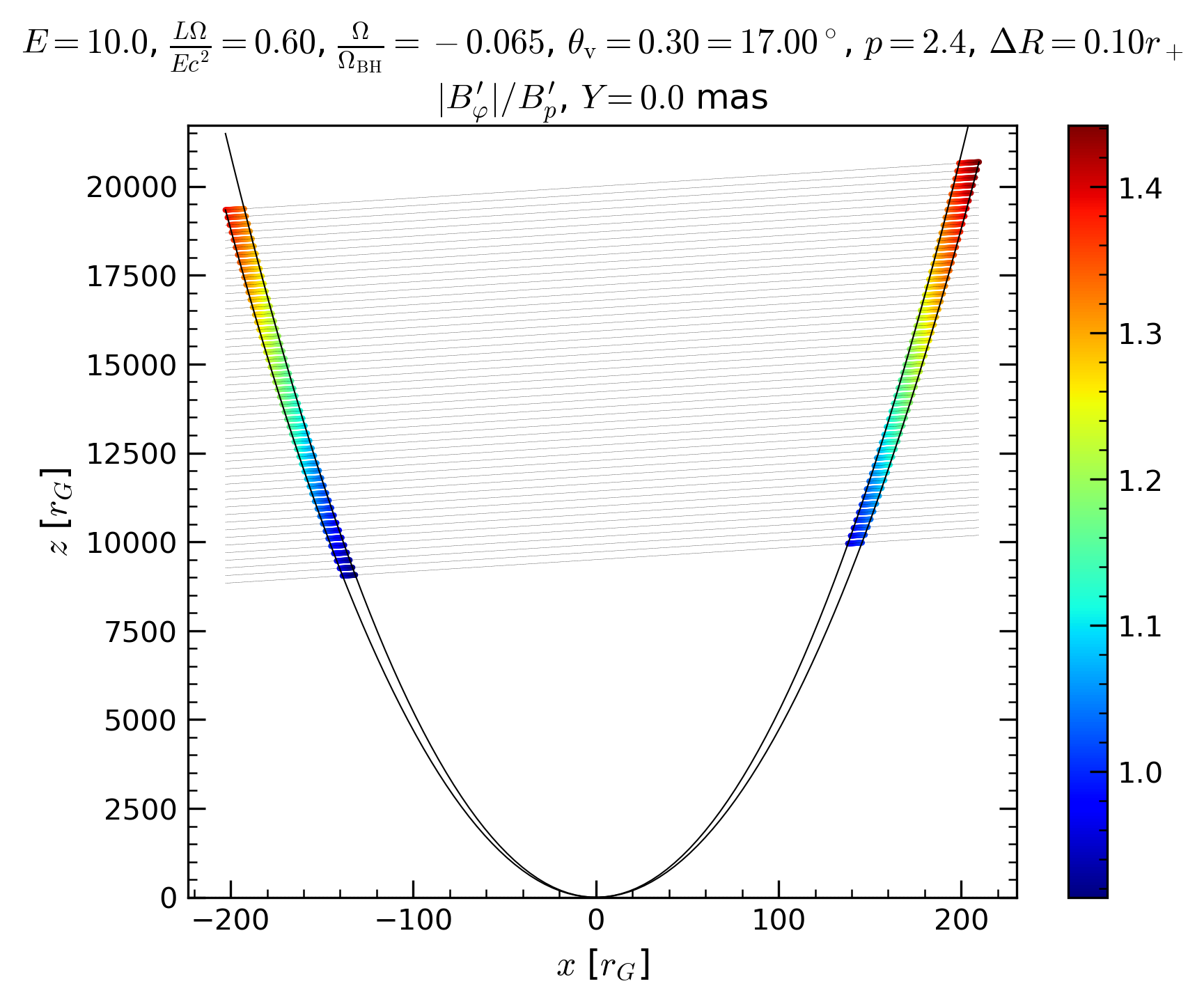}
 \includegraphics[width=0.49\columnwidth]{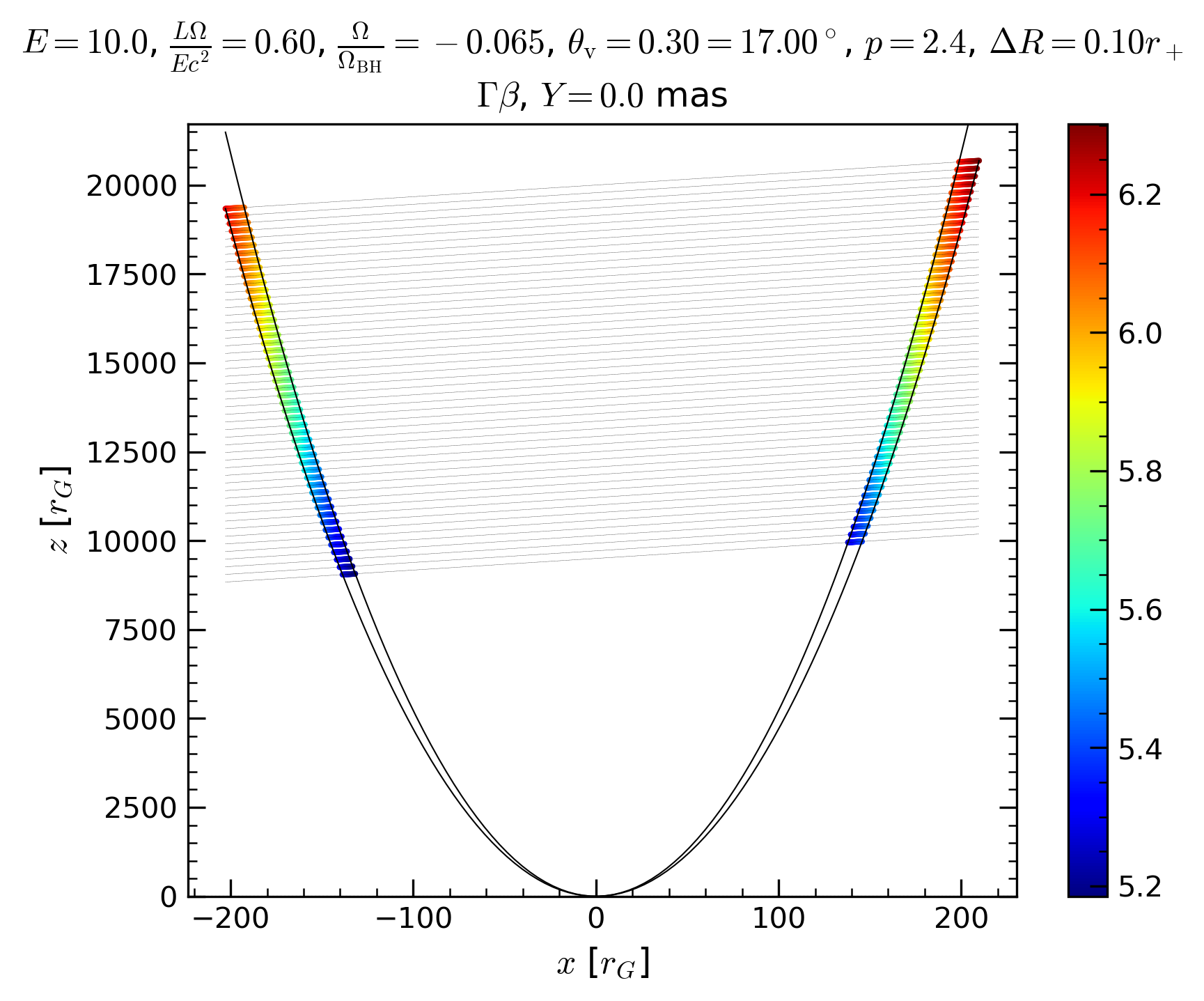}
 \caption{Physical properties of the fiducial jet model at the cross-section of $Y=0$. See Figure~\ref{fig.theo.overview.fid} for coordinate $Y$. The outer and inner parabolas show the outer and inner boundaries of the emitting region where non-thermal particles exist. The gray thin lines show rays arriving at the observer. Left: Ratio of the toroidal magnetic field strength to the poloidal magnetic field strength. The magnetic field strength is measured in the fluid rest frame. Right: Flow velocity ($\Gamma \beta$) in the fiducial model, showing the 4-velocity of flow $\Gamma \beta$, where $\Gamma$ is the Lorentz factor and $\beta$ is the speed normalized by the speed of light.}
 \label{fig.theo.pitch.speed.fid}
\end{figure}

Reproducing the observed polarization pattern requires $|B'_\phi/B'_p| \sim 1$ due to the subtle cancellation effect between the near and far sides of the jet (see Section~\ref{sec:modeling}). Although the jet shape in our model is narrower than the observed jet image, this does not affect this conclusion. Even if the jet were wider, the same condition $|B'_\phi/B'_p| \sim 1$ would still produce near--far limb polarization cancellation (with nearly constant $\Gamma$), remaining consistent with our observational data. Considering that the bulk Lorentz factor $\Gamma \sim$ several, then the ratio of the lab-frame magnetic field strengths is $|B_\phi/B_p| \sim \Gamma \sim$ several. This ratio is much smaller than the prediction in the ideal MHD jet model $|B_\phi/B_p| > 37$ (see Section~\ref{sec:modeling}). The spatial distributions of $|B'_\phi/B'_p|$ and $\Gamma\beta$ in the fiducial model are shown in Figure~\ref{fig.theo.pitch.speed.fid}. We note that alternative combinations of \(\Omega_F/|\Omega_{\rm H}|\) and \(L\Omega_F/(c^2\mathcal{E})\), distinct from our fiducial choice, also reproduce \(|B'_\phi/B'_p|\sim 1\)—a condition necessary to reproduce the observed polarization asymmetry. However, for these parameter sets, either \(\Gamma\beta\) is too high (inconsistent with the observed proper motions) or the toroidal (azimuthal) velocity is too high, producing overly asymmetric total intensity profiles.

Our semi-analytical jet model, based on standard assumptions, offers distinct advantages over approaches relying on numerical simulations. Current (GR)MHD simulations face challenges in reaching large distances (e.g., $10^5\,R_g$) from the black hole vicinity while maintaining the spatial resolution required to resolve small-scale kinetic and dissipation processes, particularly at the jet edge. Furthermore, such simulations often require ad hoc prescriptions for the synchrotron-emitting particle distribution and typically rely on hyperparameters (e.g., $R_\mathrm{low}$ and $R_\mathrm{high}$, as used in modeling EHT results; \citealp{EHT2019e, EHT2025}). In contrast, our model, while incorporating several simplifying assumptions, is physically well motivated and allows us to compute the linear polarization emission in the far downstream regions, yielding results consistent with observational constraints.

While our model defines an effective upstream boundary for the resulting helical field configuration (see Figure~\ref{fig.theo.schematic}), we anticipate that the actual underlying dissipation processes are likely more complex. For instance, dissipation might be distributed over substantial distances along the jet or occur in several distinct layers, potentially involving mechanisms such as current-driven kink instability in the jets \citep{Nakamura2007, Mizuno2012, Bodo2021}, magnetic reconnection \citep{Hardcastle2016, Sironi2021, Fromm2022}, particle acceleration through velocity shear \citep{Rieger2019}, and mass loading onto the jets from the surrounding medium \citep{Chatterjee2019}. The dissipated energy could be converted to non-thermal electrons and positrons with hard energy spectra and magnetic turbulence, and the non-thermal energy could be efficiently released from the jet by synchrotron emission, mainly in the X-ray band \citep{Kuze2024}. In this case, our assumption of a cold jet for the observed region is reasonable, and our model adequately captures characteristics of the present observation. Examining the detailed spatiotemporal distributions of dissipation and the combination of dissipation, acceleration, radiative cooling, transverse force balance, and other processes offers a promising avenue for future, more sophisticated modeling efforts.

\section{Model images for a poloidally dominated magnetic field}
\label{appendix:poloidal_model}

\begin{figure}[t!]
\centering
\includegraphics[width=0.49\linewidth]{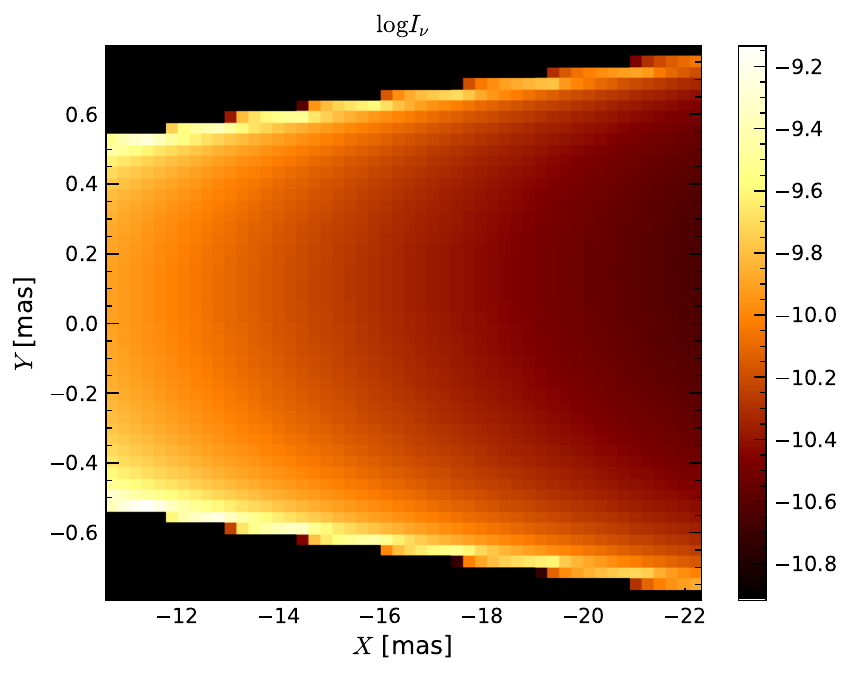}
\includegraphics[width=0.49\linewidth]{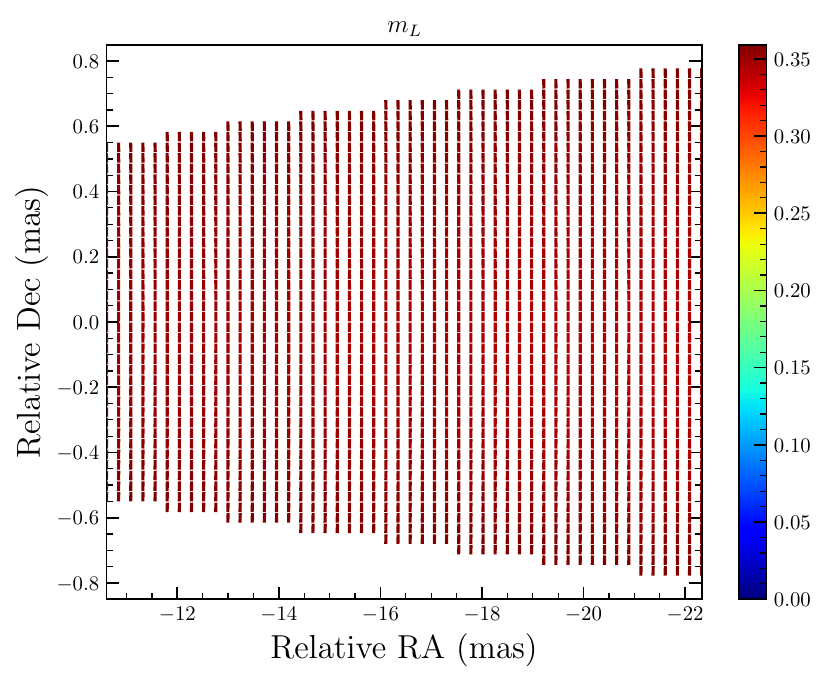}
\caption{Model images for a poloidally dominated magnetic field. Same as Figure~\ref{fig:m87jet}b,c, but for the model with $\Omega_F/|\Omega_{\rm BH}| = -0.01$, resulting in a magnetic field structure dominated by the poloidal component. Both the total intensity and the linear polarization structures are symmetric, with the EVPAs predominantly perpendicular to the jet axis. This configuration is in good agreement with the observed features in the far downstream regions ($\gtrsim100$\,mas; see Figures~\ref{fig:m87jet_large} and \ref{fig:frac_evpa_histo_larger}).}
\label{fig:model_images_poloidal}
\end{figure}

In this appendix, we present model images for a poloidally dominated magnetic field in the jet (Figure~\ref{fig:model_images_poloidal}), which corresponds to the model with $\Omega_{\rm F}/|\Omega_{\rm BH}|=-0.01$. This model naturally explains the downstream increase in symmetry between the northern and southern parts of the jet in both the fractional polarization and the EVPAs, with the EVPAs becoming predominantly perpendicular to the jet axis (Figure~\ref{fig:m87jet_large}).

\bibliography{reference}{}

\begin{thebibliography}{}
\expandafter\ifx\csname natexlab\endcsname\relax\def\natexlab#1{#1}\fi
\providecommand{\url}[1]{\href{#1}{#1}}
\providecommand{\dodoi}[1]{doi:~\href{http://doi.org/#1}{\nolinkurl{#1}}}
\providecommand{\doeprint}[1]{\href{http://ascl.net/#1}{\nolinkurl{http://ascl.net/#1}}}
\providecommand{\doarXiv}[1]{\href{https://arxiv.org/abs/#1}{\nolinkurl{https://arxiv.org/abs/#1}}}

% type= article
\bibitem[{K. {Asada} {et~al.}(2002){Asada}, {Inoue}, {Uchida}, {Kameno},
  {Fujisawa}, {Iguchi}, \& {Mutoh}}]{Asada2002}
{Asada}, K., {Inoue}, M., {Uchida}, Y., {et~al.} 2002, \bibinfo{title}{{A
  Helical Magnetic Field in the Jet of 3C 273},} \pasj, 54, L39,
  \dodoi{10.1093/pasj/54.3.L39}

% type= article
\bibitem[{K. {Asada} \& M. {Nakamura}(2012){Asada} \& {Nakamura}}]{AN2012}
{Asada}, K., \& {Nakamura}, M. 2012, \bibinfo{title}{{The Structure of the M87
  Jet: A Transition from Parabolic to Conical Streamlines},} \apjl, 745, L28,
  \dodoi{10.1088/2041-8205/745/2/L28}

% type= article
\bibitem[{A.-K. {Baczko} {et~al.}(2016){Baczko}, {Schulz}, {Kadler}, {Ros},
  {Perucho}, {Krichbaum}, {B{\"o}ck}, {Bremer}, {Grossberger}, {Lindqvist},
  {Lobanov}, {Mannheim}, {Mart{\'\i}-Vidal}, {M{\"u}ller}, {Wilms}, \&
  {Zensus}}]{Baczko2016}
{Baczko}, A.-K., {Schulz}, R., {Kadler}, M., {et~al.} 2016, \bibinfo{title}{{A
  highly magnetized twin-jet base pinpoints a supermassive black hole},} \aap,
  593, A47, \dodoi{10.1051/0004-6361/201527951}

% type= article
\bibitem[{M.~C. {Begelman} \& Z.-Y. {Li}(1994){Begelman} \& {Li}}]{BL1994}
{Begelman}, M.~C., \& {Li}, Z.-Y. 1994, \bibinfo{title}{{Asymptotic Domination
  of Cold Relativistic MHD Winds by Kinetic Energy Flux},} \apj, 426, 269,
  \dodoi{10.1086/174061}

% type= article
\bibitem[{S. {Bird} {et~al.}(2010){Bird}, {Harris}, {Blakeslee}, \&
  {Flynn}}]{Bird2010}
{Bird}, S., {Harris}, W.~E., {Blakeslee}, J.~P., \& {Flynn}, C. 2010,
  \bibinfo{title}{{The inner halo of M 87: a first direct view of the red-giant
  population},} \aap, 524, A71, \dodoi{10.1051/0004-6361/201014876}

% type= article
\bibitem[{G.~S. {Bisnovatyi-Kogan} \& A.~A. {Ruzmaikin}(1974){Bisnovatyi-Kogan}
  \& {Ruzmaikin}}]{BR1974}
{Bisnovatyi-Kogan}, G.~S., \& {Ruzmaikin}, A.~A. 1974, \bibinfo{title}{{The
  Accretion of Matter by a Collapsing Star in the Presence of a Magnetic
  Field},} \apss, 28, 45, \dodoi{10.1007/BF00642237}

% type= article
\bibitem[{J.~P. {Blakeslee} {et~al.}(2009){Blakeslee}, {Jord{\'a}n}, {Mei},
  {C{\^o}t{\'e}}, {Ferrarese}, {Infante}, {Peng}, {Tonry}, \&
  {West}}]{Blakeslee2009}
{Blakeslee}, J.~P., {Jord{\'a}n}, A., {Mei}, S., {et~al.} 2009,
  \bibinfo{title}{{The ACS Fornax Cluster Survey. V. Measurement and
  Recalibration of Surface Brightness Fluctuations and a Precise Value of the
  Fornax-Virgo Relative Distance},} \apj, 694, 556,
  \dodoi{10.1088/0004-637X/694/1/556}

% type= article
\bibitem[{R. {Blandford} {et~al.}(2019){Blandford}, {Meier}, \&
  {Readhead}}]{Blandford2019}
{Blandford}, R., {Meier}, D., \& {Readhead}, A. 2019,
  \bibinfo{title}{{Relativistic Jets from Active Galactic Nuclei},} \araa, 57,
  467, \dodoi{10.1146/annurev-astro-081817-051948}

% type= article
\bibitem[{R.~D. {Blandford} \& A. {K{\"o}nigl}(1979){Blandford} \&
  {K{\"o}nigl}}]{BK1979}
{Blandford}, R.~D., \& {K{\"o}nigl}, A. 1979, \bibinfo{title}{{Relativistic
  jets as compact radio sources.},} \apj, 232, 34, \dodoi{10.1086/157262}

% type= article
\bibitem[{R.~D. {Blandford} \& R.~L. {Znajek}(1977){Blandford} \&
  {Znajek}}]{BZ1977}
{Blandford}, R.~D., \& {Znajek}, R.~L. 1977, \bibinfo{title}{{Electromagnetic
  extraction of energy from Kerr black holes.},} \mnras, 179, 433,
  \dodoi{10.1093/mnras/179.3.433}

% type= article
\bibitem[{B. {Boccardi} {et~al.}(2016{\natexlab{a}}){Boccardi}, {Krichbaum},
  {Bach}, {Bremer}, \& {Zensus}}]{Boccardi2016a}
{Boccardi}, B., {Krichbaum}, T.~P., {Bach}, U., {Bremer}, M., \& {Zensus},
  J.~A. 2016{\natexlab{a}}, \bibinfo{title}{{First 3 mm-VLBI imaging of the
  two-sided jet in Cygnus A. Zooming into the launching region},} \aap, 588,
  L9, \dodoi{10.1051/0004-6361/201628412}

% type= article
\bibitem[{B. {Boccardi} {et~al.}(2016{\natexlab{b}}){Boccardi}, {Krichbaum},
  {Bach}, {Mertens}, {Ros}, {Alef}, \& {Zensus}}]{Boccardi2016b}
{Boccardi}, B., {Krichbaum}, T.~P., {Bach}, U., {et~al.} 2016{\natexlab{b}},
  \bibinfo{title}{{The stratified two-sided jet of <ASTROBJ>Cygnus A</ASTROBJ>.
  Acceleration and collimation},} \aap, 585, A33,
  \dodoi{10.1051/0004-6361/201526985}

% type= article
\bibitem[{B. {Boccardi} {et~al.}(2021){Boccardi}, {Perucho}, {Casadio},
  {Grandi}, {Macconi}, {Torresi}, {Pellegrini}, {Krichbaum}, {Kadler},
  {Giovannini}, {Karamanavis}, {Ricci}, {Madika}, {Bach}, {Ros}, {Giroletti},
  \& {Zensus}}]{Boccardi2021}
{Boccardi}, B., {Perucho}, M., {Casadio}, C., {et~al.} 2021,
  \bibinfo{title}{{Jet collimation in NGC 315 and other nearby AGN},} \aap,
  647, A67, \dodoi{10.1051/0004-6361/202039612}

% type= article
\bibitem[{G. {Bodo} {et~al.}(2021){Bodo}, {Tavecchio}, \& {Sironi}}]{Bodo2021}
{Bodo}, G., {Tavecchio}, F., \& {Sironi}, L. 2021, \bibinfo{title}{{Kink-driven
  magnetic reconnection in relativistic jets: consequences for X-ray
  polarimetry of BL Lacs},} \mnras, 501, 2836, \dodoi{10.1093/mnras/staa3620}

% type= article
\bibitem[{A.~E. {Broderick} \& A. {Loeb}(2009){Broderick} \&
  {Loeb}}]{BroderickLoeb2009}
{Broderick}, A.~E., \& {Loeb}, A. 2009, \bibinfo{title}{{Imaging the Black Hole
  Silhouette of M87: Implications for Jet Formation and Black Hole Spin},}
  \apj, 697, 1164, \dodoi{10.1088/0004-637X/697/2/1164}

% type= article
\bibitem[{B.~J. {Burn}(1966){Burn}}]{Burn1966}
{Burn}, B.~J. 1966, \bibinfo{title}{{On the depolarization of discrete radio
  sources by Faraday dispersion},} \mnras, 133, 67,
  \dodoi{10.1093/mnras/133.1.67}

% type= article
\bibitem[{M. {Camenzind}(1986){Camenzind}}]{Camenzind1986}
{Camenzind}, M. 1986, \bibinfo{title}{{Centrifugally driven MHD-winds in active
  galactic nuclei},} \aap, 156, 137

% type= article
\bibitem[{M. {Cantiello} {et~al.}(2018){Cantiello}, {Blakeslee}, {Ferrarese},
  {C{\^o}t{\'e}}, {Roediger}, {Raimondo}, {Peng}, {Gwyn}, {Durrell}, \&
  {Cuillandre}}]{Cantiello2018}
{Cantiello}, M., {Blakeslee}, J.~P., {Ferrarese}, L., {et~al.} 2018,
  \bibinfo{title}{{The Next Generation Virgo Cluster Survey (NGVS). XVIII.
  Measurement and Calibration of Surface Brightness Fluctuation Distances for
  Bright Galaxies in Virgo (and Beyond)},} \apj, 856, 126,
  \dodoi{10.3847/1538-4357/aab043}

% type= article
\bibitem[{X. {Cao} \& H.~C. {Spruit}(2002){Cao} \& {Spruit}}]{CS2002}
{Cao}, X., \& {Spruit}, H.~C. 2002, \bibinfo{title}{{Instability of an
  accretion disk with a magnetically driven wind},} \aap, 385, 289,
  \dodoi{10.1051/0004-6361:20011818}

% type= article
\bibitem[{A. {Chael} {et~al.}(2023){Chael}, {Issaoun}, {Pesce}, {Johnson},
  {Ricarte}, {Fromm}, \& {Mizuno}}]{Chael2023}
{Chael}, A., {Issaoun}, S., {Pesce}, D.~W., {et~al.} 2023,
  \bibinfo{title}{{Multifrequency Black Hole Imaging for the Next-generation
  Event Horizon Telescope},} \apj, 945, 40, \dodoi{10.3847/1538-4357/acb7e4}

% type= article
\bibitem[{A.~A. {Chael} {et~al.}(2018){Chael}, {Johnson}, {Bouman},
  {Blackburn}, {Akiyama}, \& {Narayan}}]{Chael2018}
{Chael}, A.~A., {Johnson}, M.~D., {Bouman}, K.~L., {et~al.} 2018,
  \bibinfo{title}{{Interferometric Imaging Directly with Closure Phases and
  Closure Amplitudes},} \apj, 857, 23, \dodoi{10.3847/1538-4357/aab6a8}

% type= article
\bibitem[{A.~A. {Chael} {et~al.}(2016){Chael}, {Johnson}, {Narayan},
  {Doeleman}, {Wardle}, \& {Bouman}}]{Chael2016}
{Chael}, A.~A., {Johnson}, M.~D., {Narayan}, R., {et~al.} 2016,
  \bibinfo{title}{{High-resolution Linear Polarimetric Imaging for the Event
  Horizon Telescope},} \apj, 829, 11, \dodoi{10.3847/0004-637X/829/1/11}

% type= article
\bibitem[{K. {Chatterjee} {et~al.}(2019){Chatterjee}, {Liska}, {Tchekhovskoy},
  \& {Markoff}}]{Chatterjee2019}
{Chatterjee}, K., {Liska}, M., {Tchekhovskoy}, A., \& {Markoff}, S.~B. 2019,
  \bibinfo{title}{{Accelerating AGN jets to parsec scales using general
  relativistic MHD simulations},} \mnras, 490, 2200,
  \dodoi{10.1093/mnras/stz2626}

% type= article
\bibitem[{M.-T. {Chen} {et~al.}(2023){Chen}, {Asada}, {Matsushita}, {Raffin},
  {Inoue}, {Ho}, {Han}, {Kubo}, {Norton}, {Patel}, {Nystrom}, {Huang},
  {Martin-Cocher}, {Yi Koay}, {Romero-Ca{\~n}izales}, {Liu}, {Huang}, {Liu},
  {Wei}, {Chang}, {Chilson}, {Oshiro}, {Jiang}, {Li}, {Bower}, {Shaw},
  {Nishioka}, {Koch}, {Chen}, {Srinivasan}, {Rao}, {Snow}, {Jinchi}, {Han},
  {Chang}, {Lu}, {Ogawa}, {Kimura}, {Hasegawa}, {Pu}, {Koyama}, {Nakamura},
  {Bintley}, {Walther}, {Friberg}, {Dempsey}, {Sriharan}, {Srikanth},
  {Doeleman}, {Brissenden}, {Algaba Marcos}, {Jeter}, {Kuo}, \&
  {Park}}]{Chen2023}
{Chen}, M.-T., {Asada}, K., {Matsushita}, S., {et~al.} 2023,
  \bibinfo{title}{{The Greenland Telescope-Construction, Commissioning, and
  Operations in Pituffik},} \pasp, 135, 095001,
  \dodoi{10.1088/1538-3873/acf072}

% type= article
\bibitem[{E. {Clausen-Brown} {et~al.}(2011){Clausen-Brown}, {Lyutikov}, \&
  {Kharb}}]{Clausen-Brown2011}
{Clausen-Brown}, E., {Lyutikov}, M., \& {Kharb}, P. 2011,
  \bibinfo{title}{{Signatures of large-scale magnetic fields in active galactic
  nuclei jets: transverse asymmetries},} \mnras, 415, 2081,
  \dodoi{10.1111/j.1365-2966.2011.18757.x}

% type= article
\bibitem[{S.~M. {Croke} \& D.~C. {Gabuzda}(2008){Croke} \& {Gabuzda}}]{CG2008}
{Croke}, S.~M., \& {Gabuzda}, D.~C. 2008, \bibinfo{title}{{Aligning VLBI images
  of active galactic nuclei at different frequencies},} \mnras, 386, 619,
  \dodoi{10.1111/j.1365-2966.2008.13087.x}

% type= article
\bibitem[{A. {Cruz-Osorio} {et~al.}(2022){Cruz-Osorio}, {Fromm}, {Mizuno},
  {Nathanail}, {Younsi}, {Porth}, {Davelaar}, {Falcke}, {Kramer}, \&
  {Rezzolla}}]{Cruz-Osorio2022}
{Cruz-Osorio}, A., {Fromm}, C.~M., {Mizuno}, Y., {et~al.} 2022,
  \bibinfo{title}{{State-of-the-art energetic and morphological modelling of
  the launching site of the M87 jet},} Nature Astronomy, 6, 103,
  \dodoi{10.1038/s41550-021-01506-w}

% type= article
\bibitem[{A.~T. {Deller} {et~al.}(2011){Deller}, {Brisken}, {Phillips},
  {Morgan}, {Alef}, {Cappallo}, {Middelberg}, {Romney}, {Rottmann}, {Tingay},
  \& {Wayth}}]{Deller2011}
{Deller}, A.~T., {Brisken}, W.~F., {Phillips}, C.~J., {et~al.} 2011,
  \bibinfo{title}{{DiFX-2: A More Flexible, Efficient, Robust, and Powerful
  Software Correlator},} \pasp, 123, 275, \dodoi{10.1086/658907}

% type= article
\bibitem[{A. {Doi} {et~al.}(2018){Doi}, {Hada}, {Kino}, {Wajima}, \&
  {Nakahara}}]{Doi2018}
{Doi}, A., {Hada}, K., {Kino}, M., {Wajima}, K., \& {Nakahara}, S. 2018,
  \bibinfo{title}{{A Recollimation Shock in a Stationary Jet Feature with
  Limb-brightening in the Gamma-Ray-emitting Narrow-line Seyfert 1 Galaxy 1H
  0323+342},} \apjl, 857, L6, \dodoi{10.3847/2041-8213/aabae2}

% type= article
\bibitem[{ {Event Horizon Telescope Collaboration}
  {et~al.}(2019{\natexlab{a}}){Event Horizon Telescope Collaboration},
  {Akiyama}, {Alberdi}, {Alef}, {Asada}, {Azulay}, {Baczko}, {Ball},
  {Balokovi{\'c}}, {Barrett}, {Bintley}, {Blackburn}, {Boland}, {Bouman},
  {Bower}, {Bremer}, {Brinkerink}, {Brissenden}, {Britzen}, {Broderick},
  {Broguiere}, {Bronzwaer}, {Byun}, {Carlstrom}, {Chael}, {Chan}, {Chatterjee},
  {Chatterjee}, {Chen}, {Chen}, {Cho}, {Christian}, {Conway}, {Cordes}, {Crew},
  {Cui}, {Davelaar}, {De Laurentis}, {Deane}, {Dempsey}, {Desvignes}, {Dexter},
  {Doeleman}, {Eatough}, {Falcke}, {Fish}, {Fomalont}, {Fraga-Encinas},
  {Freeman}, {Friberg}, {Fromm}, {G{\'o}mez}, {Galison}, {Gammie},
  {Garc{\'\i}a}, {Gentaz}, {Georgiev}, {Goddi}, {Gold}, {Gu}, {Gurwell},
  {Hada}, {Hecht}, {Hesper}, {Ho}, {Ho}, {Honma}, {Huang}, {Huang}, {Hughes},
  {Ikeda}, {Inoue}, {Issaoun}, {James}, {Jannuzi}, {Janssen}, {Jeter}, {Jiang},
  {Johnson}, {Jorstad}, {Jung}, {Karami}, {Karuppusamy}, {Kawashima},
  {Keating}, {Kettenis}, {Kim}, {Kim}, {Kim}, {Kino}, {Koay}, {Koch}, {Koyama},
  {Kramer}, {Kramer}, {Krichbaum}, {Kuo}, {Lauer}, {Lee}, {Li}, {Li},
  {Lindqvist}, {Liu}, {Liuzzo}, {Lo}, {Lobanov}, {Loinard}, {Lonsdale}, {Lu},
  {MacDonald}, {Mao}, {Markoff}, {Marrone}, {Marscher}, {Mart{\'\i}-Vidal},
  {Matsushita}, {Matthews}, {Medeiros}, {Menten}, {Mizuno}, {Mizuno}, {Moran},
  {Moriyama}, {Moscibrodzka}, {M{\"u}ller}, {Nagai}, {Nagar}, {Nakamura},
  {Narayan}, {Narayanan}, {Natarajan}, {Neri}, {Ni}, {Noutsos}, {Okino},
  {Olivares}, {Ortiz-Le{\'o}n}, {Oyama}, {{\"O}zel}, {Palumbo}, {Patel}, {Pen},
  {Pesce}, {Pi{\'e}tu}, {Plambeck}, {PopStefanija}, {Porth}, {Prather},
  {Preciado-L{\'o}pez}, {Psaltis}, {Pu}, {Ramakrishnan}, {Rao}, {Rawlings},
  {Raymond}, {Rezzolla}, {Ripperda}, {Roelofs}, {Rogers}, {Ros}, {Rose},
  {Roshanineshat}, {Rottmann}, {Roy}, {Ruszczyk}, {Ryan}, {Rygl},
  {S{\'a}nchez}, {S{\'a}nchez-Arguelles}, {Sasada}, {Savolainen}, {Schloerb},
  {Schuster}, {Shao}, {Shen}, {Small}, {Sohn}, {SooHoo}, {Tazaki}, {Tiede},
  {Tilanus}, {Titus}, {Toma}, {Torne}, {Trent}, {Trippe}, {Tsuda}, {van
  Bemmel}, {van Langevelde}, {van Rossum}, {Wagner}, {Wardle}, {Weintroub},
  {Wex}, {Wharton}, {Wielgus}, {Wong}, {Wu}, {Young}, {Young}, {Younsi},
  {Yuan}, {Yuan}, {Zensus}, {Zhao}, {Zhao}, {Zhu}, {Algaba}, {Allardi},
  {Amestica}, {Anczarski}, {Bach}, {Baganoff}, {Beaudoin}, {Benson},
  {Berthold}, {Blanchard}, {Blundell}, {Bustamente}, {Cappallo},
  {Castillo-Dom{\'\i}nguez}, {Chang}, {Chang}, {Chang}, {Chen}, {Chilson},
  {Chuter}, {C{\'o}rdova Rosado}, {Coulson}, {Crawford}, {Crowley}, {David},
  {Derome}, {Dexter}, {Dornbusch}, {Dudevoir}, {Dzib}, {Eckart}, {Eckert},
  {Erickson}, {Everett}, {Faber}, {Farah}, {Fath}, {Folkers}, {Forbes},
  {Freund}, {G{\'o}mez-Ruiz}, {Gale}, {Gao}, {Geertsema}, {Graham}, {Greer},
  {Grosslein}, {Gueth}, {Haggard}, {Halverson}, {Han}, {Han}, {Hao},
  {Hasegawa}, {Henning}, {Hern{\'a}ndez-G{\'o}mez}, {Herrero-Illana},
  {Heyminck}, {Hirota}, {Hoge}, {Huang}, {Impellizzeri}, {Jiang}, {Kamble},
  {Keisler}, {Kimura}, {Kono}, {Kubo}, {Kuroda}, {Lacasse}, {Laing}, {Leitch},
  {Li}, {Lin}, {Liu}, {Liu}, {Lu}, {Marson}, {Martin-Cocher}, {Massingill},
  {Matulonis}, {McColl}, {McWhirter}, {Messias}, {Meyer-Zhao}, {Michalik},
  {Monta{\~n}a}, {Montgomerie}, {Mora-Klein}, {Muders}, {Nadolski}, {Navarro},
  {Neilsen}, {Nguyen}, {Nishioka}, {Norton}, {Nowak}, {Nystrom}, {Ogawa},
  {Oshiro}, {Oyama}, {Parsons}, {Paine}, {Pe{\~n}alver}, {Phillips}, {Poirier},
  {Pradel}, {Primiani}, {Raffin}, {Rahlin}, {Reiland}, {Risacher}, {Ruiz},
  {S{\'a}ez-Mada{\'\i}n}, {Sassella}, {Schellart}, {Shaw}, {Silva}, {Shiokawa},
  {Smith}, {Snow}, {Souccar}, {Sousa}, {Sridharan}, {Srinivasan}, {Stahm},
  {Stark}, {Story}, {Timmer}, {Vertatschitsch}, {Walther}, {Wei}, {Whitehorn},
  {Whitney}, {Woody}, {Wouterloot}, {Wright}, {Yamaguchi}, {Yu}, {Zeballos},
  {Zhang}, \& {Ziurys}}]{EHT2019a}
{Event Horizon Telescope Collaboration}, {Akiyama}, K., {Alberdi}, A., {et~al.}
  2019{\natexlab{a}}, \bibinfo{title}{{First M87 Event Horizon Telescope
  Results. I. The Shadow of the Supermassive Black Hole},} \apjl, 875, L1,
  \dodoi{10.3847/2041-8213/ab0ec7}

% type= article
\bibitem[{ {Event Horizon Telescope Collaboration}
  {et~al.}(2019{\natexlab{b}}){Event Horizon Telescope Collaboration},
  {Akiyama}, {Alberdi}, {Alef}, {Asada}, {Azulay}, {Baczko}, {Ball},
  {Balokovi{\'c}}, {Barrett}, {Bintley}, {Blackburn}, {Boland}, {Bouman},
  {Bower}, {Bremer}, {Brinkerink}, {Brissenden}, {Britzen}, {Broderick},
  {Broguiere}, {Bronzwaer}, {Byun}, {Carlstrom}, {Chael}, {Chan}, {Chatterjee},
  {Chatterjee}, {Chen}, {Chen}, {Cho}, {Christian}, {Conway}, {Cordes}, {Crew},
  {Cui}, {Davelaar}, {De Laurentis}, {Deane}, {Dempsey}, {Desvignes}, {Dexter},
  {Doeleman}, {Eatough}, {Falcke}, {Fish}, {Fomalont}, {Fraga-Encinas},
  {Friberg}, {Fromm}, {G{\'o}mez}, {Galison}, {Gammie}, {Garc{\'\i}a},
  {Gentaz}, {Georgiev}, {Goddi}, {Gold}, {Gu}, {Gurwell}, {Hada}, {Hecht},
  {Hesper}, {Ho}, {Ho}, {Honma}, {Huang}, {Huang}, {Hughes}, {Ikeda}, {Inoue},
  {Issaoun}, {James}, {Jannuzi}, {Janssen}, {Jeter}, {Jiang}, {Johnson},
  {Jorstad}, {Jung}, {Karami}, {Karuppusamy}, {Kawashima}, {Keating},
  {Kettenis}, {Kim}, {Kim}, {Kim}, {Kino}, {Koay}, {Koch}, {Koyama}, {Kramer},
  {Kramer}, {Krichbaum}, {Kuo}, {Lauer}, {Lee}, {Li}, {Li}, {Lindqvist}, {Liu},
  {Liuzzo}, {Lo}, {Lobanov}, {Loinard}, {Lonsdale}, {Lu}, {MacDonald}, {Mao},
  {Markoff}, {Marrone}, {Marscher}, {Mart{\'\i}-Vidal}, {Matsushita},
  {Matthews}, {Medeiros}, {Menten}, {Mizuno}, {Mizuno}, {Moran}, {Moriyama},
  {Moscibrodzka}, {M{\"u}ller}, {Nagai}, {Nagar}, {Nakamura}, {Narayan},
  {Narayanan}, {Natarajan}, {Neri}, {Ni}, {Noutsos}, {Okino}, {Olivares},
  {Ortiz-Le{\'o}n}, {Oyama}, {{\"O}zel}, {Palumbo}, {Patel}, {Pen}, {Pesce},
  {Pi{\'e}tu}, {Plambeck}, {PopStefanija}, {Porth}, {Prather},
  {Preciado-L{\'o}pez}, {Psaltis}, {Pu}, {Ramakrishnan}, {Rao}, {Rawlings},
  {Raymond}, {Rezzolla}, {Ripperda}, {Roelofs}, {Rogers}, {Ros}, {Rose},
  {Roshanineshat}, {Rottmann}, {Roy}, {Ruszczyk}, {Ryan}, {Rygl},
  {S{\'a}nchez}, {S{\'a}nchez-Arguelles}, {Sasada}, {Savolainen}, {Schloerb},
  {Schuster}, {Shao}, {Shen}, {Small}, {Sohn}, {SooHoo}, {Tazaki}, {Tiede},
  {Tilanus}, {Titus}, {Toma}, {Torne}, {Trent}, {Trippe}, {Tsuda}, {van
  Bemmel}, {van Langevelde}, {van Rossum}, {Wagner}, {Wardle}, {Weintroub},
  {Wex}, {Wharton}, {Wielgus}, {Wong}, {Wu}, {Young}, {Young}, {Younsi},
  {Yuan}, {Yuan}, {Zensus}, {Zhao}, {Zhao}, {Zhu}, {Algaba}, {Allardi},
  {Amestica}, {Bach}, {Beaudoin}, {Benson}, {Berthold}, {Blanchard},
  {Blundell}, {Bustamente}, {Cappallo}, {Castillo-Dom{\'\i}nguez}, {Chang},
  {Chang}, {Chang}, {Chen}, {Chilson}, {Chuter}, {C{\'o}rdova Rosado},
  {Coulson}, {Crawford}, {Crowley}, {David}, {Derome}, {Dexter}, {Dornbusch},
  {Dudevoir}, {Dzib}, {Eckert}, {Erickson}, {Everett}, {Faber}, {Farah},
  {Fath}, {Folkers}, {Forbes}, {Freund}, {G{\'o}mez-Ruiz}, {Gale}, {Gao},
  {Geertsema}, {Graham}, {Greer}, {Grosslein}, {Gueth}, {Halverson}, {Han},
  {Han}, {Hao}, {Hasegawa}, {Henning}, {Hern{\'a}ndez-G{\'o}mez},
  {Herrero-Illana}, {Heyminck}, {Hirota}, {Hoge}, {Huang}, {Impellizzeri},
  {Jiang}, {Kamble}, {Keisler}, {Kimura}, {Kono}, {Kubo}, {Kuroda}, {Lacasse},
  {Laing}, {Leitch}, {Li}, {Lin}, {Liu}, {Liu}, {Lu}, {Marson},
  {Martin-Cocher}, {Massingill}, {Matulonis}, {McColl}, {McWhirter}, {Messias},
  {Meyer-Zhao}, {Michalik}, {Monta{\~n}a}, {Montgomerie}, {Mora-Klein},
  {Muders}, {Nadolski}, {Navarro}, {Nguyen}, {Nishioka}, {Norton}, {Nystrom},
  {Ogawa}, {Oshiro}, {Oyama}, {Padin}, {Parsons}, {Paine}, {Pe{\~n}alver},
  {Phillips}, {Poirier}, {Pradel}, {Primiani}, {Raffin}, {Rahlin}, {Reiland},
  {Risacher}, {Ruiz}, {S{\'a}ez-Mada{\'\i}n}, {Sassella}, {Schellart}, {Shaw},
  {Silva}, {Shiokawa}, {Smith}, {Snow}, {Souccar}, {Sousa}, {Sridharan},
  {Srinivasan}, {Stahm}, {Stark}, {Story}, {Timmer}, {Vertatschitsch},
  {Walther}, {Wei}, {Whitehorn}, {Whitney}, {Woody}, {Wouterloot}, {Wright},
  {Yamaguchi}, {Yu}, {Zeballos}, \& {Ziurys}}]{EHT2019b}
{Event Horizon Telescope Collaboration}, {Akiyama}, K., {Alberdi}, A., {et~al.}
  2019{\natexlab{b}}, \bibinfo{title}{{First M87 Event Horizon Telescope
  Results. II. Array and Instrumentation},} \apjl, 875, L2,
  \dodoi{10.3847/2041-8213/ab0c96}

% type= article
\bibitem[{ {Event Horizon Telescope Collaboration}
  {et~al.}(2019{\natexlab{c}}){Event Horizon Telescope Collaboration},
  {Akiyama}, {Alberdi}, {Alef}, {Asada}, {Azulay}, {Baczko}, {Ball},
  {Balokovi{\'c}}, {Barrett}, {Bintley}, {Blackburn}, {Boland}, {Bouman},
  {Bower}, {Bremer}, {Brinkerink}, {Brissenden}, {Britzen}, {Broderick},
  {Broguiere}, {Bronzwaer}, {Byun}, {Carlstrom}, {Chael}, {Chan}, {Chatterjee},
  {Chatterjee}, {Chen}, {Chen}, {Cho}, {Christian}, {Conway}, {Cordes}, {Crew},
  {Cui}, {Davelaar}, {De Laurentis}, {Deane}, {Dempsey}, {Desvignes}, {Dexter},
  {Doeleman}, {Eatough}, {Falcke}, {Fish}, {Fomalont}, {Fraga-Encinas},
  {Friberg}, {Fromm}, {G{\'o}mez}, {Galison}, {Gammie}, {Garc{\'\i}a},
  {Gentaz}, {Georgiev}, {Goddi}, {Gold}, {Gu}, {Gurwell}, {Hada}, {Hecht},
  {Hesper}, {Ho}, {Ho}, {Honma}, {Huang}, {Huang}, {Hughes}, {Ikeda}, {Inoue},
  {Issaoun}, {James}, {Jannuzi}, {Janssen}, {Jeter}, {Jiang}, {Johnson},
  {Jorstad}, {Jung}, {Karami}, {Karuppusamy}, {Kawashima}, {Keating},
  {Kettenis}, {Kim}, {Kim}, {Kim}, {Kino}, {Koay}, {Koch}, {Koyama}, {Kramer},
  {Kramer}, {Krichbaum}, {Kuo}, {Lauer}, {Lee}, {Li}, {Li}, {Lindqvist}, {Liu},
  {Liuzzo}, {Lo}, {Lobanov}, {Loinard}, {Lonsdale}, {Lu}, {MacDonald}, {Mao},
  {Markoff}, {Marrone}, {Marscher}, {Mart{\'\i}-Vidal}, {Matsushita},
  {Matthews}, {Medeiros}, {Menten}, {Mizuno}, {Mizuno}, {Moran}, {Moriyama},
  {Moscibrodzka}, {M{\"u}ller}, {Nagai}, {Nagar}, {Nakamura}, {Narayan},
  {Narayanan}, {Natarajan}, {Neri}, {Ni}, {Noutsos}, {Okino}, {Olivares},
  {Ortiz-Le{\'o}n}, {Oyama}, {{\"O}zel}, {Palumbo}, {Patel}, {Pen}, {Pesce},
  {Pi{\'e}tu}, {Plambeck}, {PopStefanija}, {Porth}, {Prather},
  {Preciado-L{\'o}pez}, {Psaltis}, {Pu}, {Ramakrishnan}, {Rao}, {Rawlings},
  {Raymond}, {Rezzolla}, {Ripperda}, {Roelofs}, {Rogers}, {Ros}, {Rose},
  {Roshanineshat}, {Rottmann}, {Roy}, {Ruszczyk}, {Ryan}, {Rygl},
  {S{\'a}nchez}, {S{\'a}nchez-Arguelles}, {Sasada}, {Savolainen}, {Schloerb},
  {Schuster}, {Shao}, {Shen}, {Small}, {Sohn}, {SooHoo}, {Tazaki}, {Tiede},
  {Tilanus}, {Titus}, {Toma}, {Torne}, {Trent}, {Trippe}, {Tsuda}, {van
  Bemmel}, {van Langevelde}, {van Rossum}, {Wagner}, {Wardle}, {Weintroub},
  {Wex}, {Wharton}, {Wielgus}, {Wong}, {Wu}, {Young}, {Young}, {Younsi},
  {Yuan}, {Yuan}, {Zensus}, {Zhao}, {Zhao}, {Zhu}, {Cappallo}, {Farah},
  {Folkers}, {Meyer-Zhao}, {Michalik}, {Nadolski}, {Nishioka}, {Pradel},
  {Primiani}, {Souccar}, {Vertatschitsch}, \& {Yamaguchi}}]{EHT2019c}
{Event Horizon Telescope Collaboration}, {Akiyama}, K., {Alberdi}, A., {et~al.}
  2019{\natexlab{c}}, \bibinfo{title}{{First M87 Event Horizon Telescope
  Results. III. Data Processing and Calibration},} \apjl, 875, L3,
  \dodoi{10.3847/2041-8213/ab0c57}

% type= article
\bibitem[{ {Event Horizon Telescope Collaboration}
  {et~al.}(2019{\natexlab{d}}){Event Horizon Telescope Collaboration},
  {Akiyama}, {Alberdi}, {Alef}, {Asada}, {Azulay}, {Baczko}, {Ball},
  {Balokovi{\'c}}, {Barrett}, {Bintley}, {Blackburn}, {Boland}, {Bouman},
  {Bower}, {Bremer}, {Brinkerink}, {Brissenden}, {Britzen}, {Broderick},
  {Broguiere}, {Bronzwaer}, {Byun}, {Carlstrom}, {Chael}, {Chan}, {Chatterjee},
  {Chatterjee}, {Chen}, {Chen}, {Cho}, {Christian}, {Conway}, {Cordes}, {Crew},
  {Cui}, {Davelaar}, {De Laurentis}, {Deane}, {Dempsey}, {Desvignes}, {Dexter},
  {Doeleman}, {Eatough}, {Falcke}, {Fish}, {Fomalont}, {Fraga-Encinas},
  {Freeman}, {Friberg}, {Fromm}, {G{\'o}mez}, {Galison}, {Gammie},
  {Garc{\'\i}a}, {Gentaz}, {Georgiev}, {Goddi}, {Gold}, {Gu}, {Gurwell},
  {Hada}, {Hecht}, {Hesper}, {Ho}, {Ho}, {Honma}, {Huang}, {Huang}, {Hughes},
  {Ikeda}, {Inoue}, {Issaoun}, {James}, {Jannuzi}, {Janssen}, {Jeter}, {Jiang},
  {Johnson}, {Jorstad}, {Jung}, {Karami}, {Karuppusamy}, {Kawashima},
  {Keating}, {Kettenis}, {Kim}, {Kim}, {Kim}, {Kino}, {Koay}, {Koch}, {Koyama},
  {Kramer}, {Kramer}, {Krichbaum}, {Kuo}, {Lauer}, {Lee}, {Li}, {Li},
  {Lindqvist}, {Liu}, {Liuzzo}, {Lo}, {Lobanov}, {Loinard}, {Lonsdale}, {Lu},
  {MacDonald}, {Mao}, {Markoff}, {Marrone}, {Marscher}, {Mart{\'\i}-Vidal},
  {Matsushita}, {Matthews}, {Medeiros}, {Menten}, {Mizuno}, {Mizuno}, {Moran},
  {Moriyama}, {Moscibrodzka}, {M{\"u}ller}, {Nagai}, {Nagar}, {Nakamura},
  {Narayan}, {Narayanan}, {Natarajan}, {Neri}, {Ni}, {Noutsos}, {Okino},
  {Olivares}, {Oyama}, {{\"O}zel}, {Palumbo}, {Patel}, {Pen}, {Pesce},
  {Pi{\'e}tu}, {Plambeck}, {PopStefanija}, {Porth}, {Prather},
  {Preciado-L{\'o}pez}, {Psaltis}, {Pu}, {Ramakrishnan}, {Rao}, {Rawlings},
  {Raymond}, {Rezzolla}, {Ripperda}, {Roelofs}, {Rogers}, {Ros}, {Rose},
  {Roshanineshat}, {Rottmann}, {Roy}, {Ruszczyk}, {Ryan}, {Rygl},
  {S{\'a}nchez}, {S{\'a}nchez-Arguelles}, {Sasada}, {Savolainen}, {Schloerb},
  {Schuster}, {Shao}, {Shen}, {Small}, {Sohn}, {SooHoo}, {Tazaki}, {Tiede},
  {Tilanus}, {Titus}, {Toma}, {Torne}, {Trent}, {Trippe}, {Tsuda}, {van
  Bemmel}, {van Langevelde}, {van Rossum}, {Wagner}, {Wardle}, {Weintroub},
  {Wex}, {Wharton}, {Wielgus}, {Wong}, {Wu}, {Young}, {Young}, {Younsi},
  {Yuan}, {Yuan}, {Zensus}, {Zhao}, {Zhao}, {Zhu}, {Farah}, {Meyer-Zhao},
  {Michalik}, {Nadolski}, {Nishioka}, {Pradel}, {Primiani}, {Souccar},
  {Vertatschitsch}, \& {Yamaguchi}}]{EHT2019d}
{Event Horizon Telescope Collaboration}, {Akiyama}, K., {Alberdi}, A., {et~al.}
  2019{\natexlab{d}}, \bibinfo{title}{{First M87 Event Horizon Telescope
  Results. IV. Imaging the Central Supermassive Black Hole},} \apjl, 875, L4,
  \dodoi{10.3847/2041-8213/ab0e85}

% type= article
\bibitem[{ {Event Horizon Telescope Collaboration}
  {et~al.}(2019{\natexlab{e}}){Event Horizon Telescope Collaboration},
  {Akiyama}, {Alberdi}, {Alef}, {Asada}, {Azulay}, {Baczko}, {Ball},
  {Balokovi{\'c}}, {Barrett}, {Bintley}, {Blackburn}, {Boland}, {Bouman},
  {Bower}, {Bremer}, {Brinkerink}, {Brissenden}, {Britzen}, {Broderick},
  {Broguiere}, {Bronzwaer}, {Byun}, {Carlstrom}, {Chael}, {Chan}, {Chatterjee},
  {Chatterjee}, {Chen}, {Chen}, {Cho}, {Christian}, {Conway}, {Cordes}, {Crew},
  {Cui}, {Davelaar}, {De Laurentis}, {Deane}, {Dempsey}, {Desvignes}, {Dexter},
  {Doeleman}, {Eatough}, {Falcke}, {Fish}, {Fomalont}, {Fraga-Encinas},
  {Friberg}, {Fromm}, {G{\'o}mez}, {Galison}, {Gammie}, {Garc{\'\i}a},
  {Gentaz}, {Georgiev}, {Goddi}, {Gold}, {Gu}, {Gurwell}, {Hada}, {Hecht},
  {Hesper}, {Ho}, {Ho}, {Honma}, {Huang}, {Huang}, {Hughes}, {Ikeda}, {Inoue},
  {Issaoun}, {James}, {Jannuzi}, {Janssen}, {Jeter}, {Jiang}, {Johnson},
  {Jorstad}, {Jung}, {Karami}, {Karuppusamy}, {Kawashima}, {Keating},
  {Kettenis}, {Kim}, {Kim}, {Kim}, {Kino}, {Koay}, {Koch}, {Koyama}, {Kramer},
  {Kramer}, {Krichbaum}, {Kuo}, {Lauer}, {Lee}, {Li}, {Li}, {Lindqvist}, {Liu},
  {Liuzzo}, {Lo}, {Lobanov}, {Loinard}, {Lonsdale}, {Lu}, {MacDonald}, {Mao},
  {Markoff}, {Marrone}, {Marscher}, {Mart{\'\i}-Vidal}, {Matsushita},
  {Matthews}, {Medeiros}, {Menten}, {Mizuno}, {Mizuno}, {Moran}, {Moriyama},
  {Moscibrodzka}, {Mul{\ensuremath{\ddot{}}}ler}, {Nagai}, {Nagar}, {Nakamura},
  {Narayan}, {Narayanan}, {Natarajan}, {Neri}, {Ni}, {Noutsos}, {Okino},
  {Olivares}, {Oyama}, {{\"O}zel}, {Palumbo}, {Patel}, {Pen}, {Pesce},
  {Pi{\'e}tu}, {Plambeck}, {PopStefanija}, {Porth}, {Prather},
  {Preciado-L{\'o}pez}, {Psaltis}, {Pu}, {Ramakrishnan}, {Rao}, {Rawlings},
  {Raymond}, {Rezzolla}, {Ripperda}, {Roelofs}, {Rogers}, {Ros}, {Rose},
  {Roshanineshat}, {Rottmann}, {Roy}, {Ruszczyk}, {Ryan}, {Rygl},
  {S{\'a}nchez}, {S{\'a}nchez-Arguelles}, {Sasada}, {Savolainen}, {Schloerb},
  {Schuster}, {Shao}, {Shen}, {Small}, {Sohn}, {SooHoo}, {Tazaki}, {Tiede},
  {Tilanus}, {Titus}, {Toma}, {Torne}, {Trent}, {Trippe}, {Tsuda}, {van
  Bemmel}, {van Langevelde}, {van Rossum}, {Wagner}, {Wardle}, {Weintroub},
  {Wex}, {Wharton}, {Wielgus}, {Wong}, {Wu}, {Young}, {Young}, {Younsi}, \&
  {Yuan}}]{EHT2019e}
{Event Horizon Telescope Collaboration}, {Akiyama}, K., {Alberdi}, A., {et~al.}
  2019{\natexlab{e}}, \bibinfo{title}{{First M87 Event Horizon Telescope
  Results. V. Physical Origin of the Asymmetric Ring},} \apjl, 875, L5,
  \dodoi{10.3847/2041-8213/ab0f43}

% type= article
\bibitem[{ {Event Horizon Telescope Collaboration}
  {et~al.}(2019{\natexlab{f}}){Event Horizon Telescope Collaboration},
  {Akiyama}, {Alberdi}, {Alef}, {Asada}, {Azulay}, {Baczko}, {Ball},
  {Balokovi{\'c}}, {Barrett}, {Bintley}, {Blackburn}, {Boland}, {Bouman},
  {Bower}, {Bremer}, {Brinkerink}, {Brissenden}, {Britzen}, {Broderick},
  {Broguiere}, {Bronzwaer}, {Byun}, {Carlstrom}, {Chael}, {Chan}, {Chatterjee},
  {Chatterjee}, {Chen}, {Chen}, {Cho}, {Christian}, {Conway}, {Cordes}, {Crew},
  {Cui}, {Davelaar}, {De Laurentis}, {Deane}, {Dempsey}, {Desvignes}, {Dexter},
  {Doeleman}, {Eatough}, {Falcke}, {Fish}, {Fomalont}, {Fraga-Encinas},
  {Friberg}, {Fromm}, {G{\'o}mez}, {Galison}, {Gammie}, {Garc{\'\i}a},
  {Gentaz}, {Georgiev}, {Goddi}, {Gold}, {Gu}, {Gurwell}, {Hada}, {Hecht},
  {Hesper}, {Ho}, {Ho}, {Honma}, {Huang}, {Huang}, {Hughes}, {Ikeda}, {Inoue},
  {Issaoun}, {James}, {Jannuzi}, {Janssen}, {Jeter}, {Jiang}, {Johnson},
  {Jorstad}, {Jung}, {Karami}, {Karuppusamy}, {Kawashima}, {Keating},
  {Kettenis}, {Kim}, {Kim}, {Kim}, {Kino}, {Koay}, {Koch}, {Koyama}, {Kramer},
  {Kramer}, {Krichbaum}, {Kuo}, {Lauer}, {Lee}, {Li}, {Li}, {Lindqvist}, {Liu},
  {Liuzzo}, {Lo}, {Lobanov}, {Loinard}, {Lonsdale}, {Lu}, {MacDonald}, {Mao},
  {Markoff}, {Marrone}, {Marscher}, {Mart{\'\i}-Vidal}, {Matsushita},
  {Matthews}, {Medeiros}, {Menten}, {Mizuno}, {Mizuno}, {Moran}, {Moriyama},
  {Moscibrodzka}, {M{\"u}ller}, {Nagai}, {Nagar}, {Nakamura}, {Narayan},
  {Narayanan}, {Natarajan}, {Neri}, {Ni}, {Noutsos}, {Okino}, {Olivares},
  {Oyama}, {{\"O}zel}, {Palumbo}, {Patel}, {Pen}, {Pesce}, {Pi{\'e}tu},
  {Plambeck}, {PopStefanija}, {Porth}, {Prather}, {Preciado-L{\'o}pez},
  {Psaltis}, {Pu}, {Ramakrishnan}, {Rao}, {Rawlings}, {Raymond}, {Rezzolla},
  {Ripperda}, {Roelofs}, {Rogers}, {Ros}, {Rose}, {Roshanineshat}, {Rottmann},
  {Roy}, {Ruszczyk}, {Ryan}, {Rygl}, {S{\'a}nchez}, {S{\'a}nchez-Arguelles},
  {Sasada}, {Savolainen}, {Schloerb}, {Schuster}, {Shao}, {Shen}, {Small},
  {Sohn}, {SooHoo}, {Tazaki}, {Tiede}, {Tilanus}, {Titus}, {Toma}, {Torne},
  {Trent}, {Trippe}, {Tsuda}, {van Bemmel}, {van Langevelde}, {van Rossum},
  {Wagner}, {Wardle}, {Weintroub}, {Wex}, {Wharton}, {Wielgus}, {Wong}, {Wu},
  {Young}, {Young}, {Younsi}, {Yuan}, {Yuan}, {Zensus}, {Zhao}, {Zhao}, {Zhu},
  {Farah}, {Meyer-Zhao}, {Michalik}, {Nadolski}, {Nishioka}, {Pradel},
  {Primiani}, {Souccar}, {Vertatschitsch}, \& {Yamaguchi}}]{EHT2019f}
{Event Horizon Telescope Collaboration}, {Akiyama}, K., {Alberdi}, A., {et~al.}
  2019{\natexlab{f}}, \bibinfo{title}{{First M87 Event Horizon Telescope
  Results. VI. The Shadow and Mass of the Central Black Hole},} \apjl, 875, L6,
  \dodoi{10.3847/2041-8213/ab1141}

% type= article
\bibitem[{ {Event Horizon Telescope Collaboration}
  {et~al.}(2021{\natexlab{a}}){Event Horizon Telescope Collaboration},
  {Akiyama}, {Algaba}, {Alberdi}, {Alef}, {Anantua}, {Asada}, {Azulay},
  {Baczko}, {Ball}, {Balokovi{\'c}}, {Barrett}, {Benson}, {Bintley},
  {Blackburn}, {Blundell}, {Boland}, {Bouman}, {Bower}, {Boyce}, {Bremer},
  {Brinkerink}, {Brissenden}, {Britzen}, {Broderick}, {Broguiere}, {Bronzwaer},
  {Byun}, {Carlstrom}, {Chael}, {Chan}, {Chatterjee}, {Chatterjee}, {Chen},
  {Chen}, {Chesler}, {Cho}, {Christian}, {Conway}, {Cordes}, {Crawford},
  {Crew}, {Cruz-Osorio}, {Cui}, {Davelaar}, {De Laurentis}, {Deane}, {Dempsey},
  {Desvignes}, {Dexter}, {Doeleman}, {Eatough}, {Falcke}, {Farah}, {Fish},
  {Fomalont}, {Ford}, {Fraga-Encinas}, {Freeman}, {Friberg}, {Fromm},
  {Fuentes}, {Galison}, {Gammie}, {Garc{\'\i}a}, {Gentaz}, {Georgiev}, {Goddi},
  {Gold}, {G{\'o}mez}, {G{\'o}mez-Ruiz}, {Gu}, {Gurwell}, {Hada}, {Haggard},
  {Hecht}, {Hesper}, {Ho}, {Ho}, {Honma}, {Huang}, {Huang}, {Hughes}, {Ikeda},
  {Inoue}, {Issaoun}, {James}, {Jannuzi}, {Janssen}, {Jeter}, {Jiang},
  {Jimenez-Rosales}, {Johnson}, {Jorstad}, {Jung}, {Karami}, {Karuppusamy},
  {Kawashima}, {Keating}, {Kettenis}, {Kim}, {Kim}, {Kim}, {Kim}, {Kino},
  {Koay}, {Kofuji}, {Koch}, {Koyama}, {Kramer}, {Kramer}, {Krichbaum}, {Kuo},
  {Lauer}, {Lee}, {Levis}, {Li}, {Li}, {Lindqvist}, {Lico}, {Lindahl}, {Liu},
  {Liu}, {Liuzzo}, {Lo}, {Lobanov}, {Loinard}, {Lonsdale}, {Lu}, {MacDonald},
  {Mao}, {Marchili}, {Markoff}, {Marrone}, {Marscher}, {Mart{\'\i}-Vidal},
  {Matsushita}, {Matthews}, {Medeiros}, {Menten}, {Mizuno}, {Mizuno}, {Moran},
  {Moriyama}, {Moscibrodzka}, {M{\"u}ller}, {Musoke}, {Mej{\'\i}as},
  {Michalik}, {Nadolski}, {Nagai}, {Nagar}, {Nakamura}, {Narayan}, {Narayanan},
  {Natarajan}, {Nathanail}, {Neilsen}, {Neri}, {Ni}, {Noutsos}, {Nowak},
  {Okino}, {Olivares}, {Ortiz-Le{\'o}n}, {Oyama}, {{\"O}zel}, {Palumbo},
  {Park}, {Patel}, {Pen}, {Pesce}, {Pi{\'e}tu}, {Plambeck}, {PopStefanija},
  {Porth}, {P{\"o}tzl}, {Prather}, {Preciado-L{\'o}pez}, {Psaltis}, {Pu},
  {Ramakrishnan}, {Rao}, {Rawlings}, {Raymond}, {Rezzolla}, {Ricarte},
  {Ripperda}, {Roelofs}, {Rogers}, {Ros}, {Rose}, {Roshanineshat}, {Rottmann},
  {Roy}, {Ruszczyk}, {Rygl}, {S{\'a}nchez}, {S{\'a}nchez-Arguelles}, {Sasada},
  {Savolainen}, {Schloerb}, {Schuster}, {Shao}, {Shen}, {Small}, {Sohn},
  {SooHoo}, {Sun}, {Tazaki}, {Tetarenko}, {Tiede}, {Tilanus}, {Titus}, {Toma},
  {Torne}, {Trent}, {Traianou}, {Trippe}, {van Bemmel}, {van Langevelde}, {van
  Rossum}, {Wagner}, {Ward-Thompson}, {Wardle}, {Weintroub}, {Wex}, {Wharton},
  {Wielgus}, {Wong}, {Wu}, {Yoon}, {Young}, {Young}, {Younsi}, {Yuan}, {Yuan},
  {Zensus}, {Zhao}, \& {Zhao}}]{EHT2021a}
{Event Horizon Telescope Collaboration}, {Akiyama}, K., {Algaba}, J.~C.,
  {et~al.} 2021{\natexlab{a}}, \bibinfo{title}{{First M87 Event Horizon
  Telescope Results. VII. Polarization of the Ring},} \apjl, 910, L12,
  \dodoi{10.3847/2041-8213/abe71d}

% type= article
\bibitem[{ {Event Horizon Telescope Collaboration}
  {et~al.}(2021{\natexlab{b}}){Event Horizon Telescope Collaboration},
  {Akiyama}, {Algaba}, {Alberdi}, {Alef}, {Anantua}, {Asada}, {Azulay},
  {Baczko}, {Ball}, {Balokovi{\'c}}, {Barrett}, {Benson}, {Bintley},
  {Blackburn}, {Blundell}, {Boland}, {Bouman}, {Bower}, {Boyce}, {Bremer},
  {Brinkerink}, {Brissenden}, {Britzen}, {Broderick}, {Broguiere}, {Bronzwaer},
  {Byun}, {Carlstrom}, {Chael}, {Chan}, {Chatterjee}, {Chatterjee}, {Chen},
  {Chen}, {Chesler}, {Cho}, {Christian}, {Conway}, {Cordes}, {Crawford},
  {Crew}, {Cruz-Osorio}, {Cui}, {Davelaar}, {De Laurentis}, {Deane}, {Dempsey},
  {Desvignes}, {Dexter}, {Doeleman}, {Eatough}, {Falcke}, {Farah}, {Fish},
  {Fomalont}, {Ford}, {Fraga-Encinas}, {Friberg}, {Fromm}, {Fuentes},
  {Galison}, {Gammie}, {Garc{\'\i}a}, {Gelles}, {Gentaz}, {Georgiev}, {Goddi},
  {Gold}, {G{\'o}mez}, {G{\'o}mez-Ruiz}, {Gu}, {Gurwell}, {Hada}, {Haggard},
  {Hecht}, {Hesper}, {Himwich}, {Ho}, {Ho}, {Honma}, {Huang}, {Huang},
  {Hughes}, {Ikeda}, {Inoue}, {Issaoun}, {James}, {Jannuzi}, {Janssen},
  {Jeter}, {Jiang}, {Jimenez-Rosales}, {Johnson}, {Jorstad}, {Jung}, {Karami},
  {Karuppusamy}, {Kawashima}, {Keating}, {Kettenis}, {Kim}, {Kim}, {Kim},
  {Kim}, {Kino}, {Koay}, {Kofuji}, {Koch}, {Koyama}, {Kramer}, {Kramer},
  {Krichbaum}, {Kuo}, {Lauer}, {Lee}, {Levis}, {Li}, {Li}, {Lindqvist}, {Lico},
  {Lindahl}, {Liu}, {Liu}, {Liuzzo}, {Lo}, {Lobanov}, {Loinard}, {Lonsdale},
  {Lu}, {MacDonald}, {Mao}, {Marchili}, {Markoff}, {Marrone}, {Marscher},
  {Mart{\'\i}-Vidal}, {Matsushita}, {Matthews}, {Medeiros}, {Menten}, {Mizuno},
  {Mizuno}, {Moran}, {Moriyama}, {Moscibrodzka}, {M{\"u}ller}, {Musoke}, {Mus
  Mej{\'\i}as}, {Michalik}, {Nadolski}, {Nagai}, {Nagar}, {Nakamura},
  {Narayan}, {Narayanan}, {Natarajan}, {Nathanail}, {Neilsen}, {Neri}, {Ni},
  {Noutsos}, {Nowak}, {Okino}, {Olivares}, {Ortiz-Le{\'o}n}, {Oyama},
  {{\"O}zel}, {Palumbo}, {Park}, {Patel}, {Pen}, {Pesce}, {Pi{\'e}tu},
  {Plambeck}, {PopStefanija}, {Porth}, {P{\"o}tzl}, {Prather},
  {Preciado-L{\'o}pez}, {Psaltis}, {Pu}, {Ramakrishnan}, {Rao}, {Rawlings},
  {Raymond}, {Rezzolla}, {Ricarte}, {Ripperda}, {Roelofs}, {Rogers}, {Ros},
  {Rose}, {Roshanineshat}, {Rottmann}, {Roy}, {Ruszczyk}, {Rygl},
  {S{\'a}nchez}, {S{\'a}nchez-Arguelles}, {Sasada}, {Savolainen}, {Schloerb},
  {Schuster}, {Shao}, {Shen}, {Small}, {Sohn}, {SooHoo}, {Sun}, {Tazaki},
  {Tetarenko}, {Tiede}, {Tilanus}, {Titus}, {Toma}, {Torne}, {Trent},
  {Traianou}, {Trippe}, {van Bemmel}, {van Langevelde}, {van Rossum}, {Wagner},
  {Ward-Thompson}, {Wardle}, {Weintroub}, {Wex}, {Wharton}, {Wielgus}, {Wong},
  {Wu}, {Yoon}, {Young}, {Young}, {Younsi}, {Yuan}, {Yuan}, {Zensus}, {Zhao},
  \& {Zhao}}]{EHT2021b}
{Event Horizon Telescope Collaboration}, {Akiyama}, K., {Algaba}, J.~C.,
  {et~al.} 2021{\natexlab{b}}, \bibinfo{title}{{First M87 Event Horizon
  Telescope Results. VIII. Magnetic Field Structure near The Event Horizon},}
  \apjl, 910, L13, \dodoi{10.3847/2041-8213/abe4de}

% type= article
\bibitem[{ {Event Horizon Telescope Collaboration} {et~al.}(2023){Event Horizon
  Telescope Collaboration}, {Akiyama}, {Alberdi}, {Alef}, {Algaba}, {Anantua},
  {Asada}, {Azulay}, {Bach}, {Baczko}, {Ball}, {Balokovi{\'c}}, {Barrett},
  {Baub{\"o}ck}, {Benson}, {Bintley}, {Blackburn}, {Blundell}, {Bouman},
  {Bower}, {Boyce}, {Bremer}, {Brinkerink}, {Brissenden}, {Britzen},
  {Broderick}, {Broguiere}, {Bronzwaer}, {Bustamante}, {Byun}, {Carlstrom},
  {Ceccobello}, {Chael}, {Chan}, {Chang}, {Chatterjee}, {Chatterjee}, {Chen},
  {Chen}, {Cheng}, {Cho}, {Christian}, {Conroy}, {Conway}, {Cordes},
  {Crawford}, {Crew}, {Cruz-Osorio}, {Cui}, {Dahale}, {Davelaar}, {De
  Laurentis}, {Deane}, {Dempsey}, {Desvignes}, {Dexter}, {Dhruv}, {Doeleman},
  {Dougal}, {Dzib}, {Eatough}, {Emami}, {Falcke}, {Farah}, {Fish}, {Fomalont},
  {Ford}, {Foschi}, {Fraga-Encinas}, {Freeman}, {Friberg}, {Fromm}, {Fuentes},
  {Galison}, {Gammie}, {Garc{\'\i}a}, {Gentaz}, {Georgiev}, {Goddi}, {Gold},
  {G{\'o}mez-Ruiz}, {G{\'o}mez}, {Gu}, {Gurwell}, {Hada}, {Haggard}, {Haworth},
  {Hecht}, {Hesper}, {Heumann}, {Ho}, {Ho}, {Honma}, {Huang}, {Huang},
  {Hughes}, {Ikeda}, {Impellizzeri}, {Inoue}, {Issaoun}, {James}, {Jannuzi},
  {Janssen}, {Jeter}, {Jiang}, {Jim{\'e}nez-Rosales}, {Johnson}, {Jorstad},
  {Joshi}, {Jung}, {Karami}, {Karuppusamy}, {Kawashima}, {Keating}, {Kettenis},
  {Kim}, {Kim}, {Kim}, {Kim}, {Kino}, {Koay}, {Kocherlakota}, {Kofuji}, {Koch},
  {Koyama}, {Kramer}, {Kramer}, {Kramer}, {Krichbaum}, {Kuo}, {La Bella},
  {Lauer}, {Lee}, {Lee}, {Leung}, {Levis}, {Li}, {Lico}, {Lindahl},
  {Lindqvist}, {Lisakov}, {Liu}, {Liu}, {Liuzzo}, {Lo}, {Lobanov}, {Loinard},
  {Lonsdale}, {Lowitz}, {Lu}, {MacDonald}, {Mao}, {Marchili}, {Markoff},
  {Marrone}, {Marscher}, {Mart{\'\i}-Vidal}, {Matsushita}, {Matthews},
  {Medeiros}, {Menten}, {Michalik}, {Mizuno}, {Mizuno}, {Moran}, {Moriyama},
  {Moscibrodzka}, {Mulaudzi}, {M{\"u}ller}, {M{\"u}ller}, {Mus}, {Musoke},
  {Myserlis}, {Nadolski}, {Nagai}, {Nagar}, {Nakamura}, {Narayan}, {Narayanan},
  {Natarajan}, {Nathanail}, {Fuentes}, {Neilsen}, {Neri}, {Ni}, {Noutsos},
  {Nowak}, {Oh}, {Okino}, {Olivares}, {Ortiz-Le{\'o}n}, {Oyama}, {{\"O}zel},
  {Palumbo}, {Paraschos}, {Park}, {Parsons}, {Patel}, {Pen}, \&
  {Pesce}}]{EHT2023}
{Event Horizon Telescope Collaboration}, {Akiyama}, K., {Alberdi}, A., {et~al.}
  2023, \bibinfo{title}{{First M87 Event Horizon Telescope Results. IX.
  Detection of Near-horizon Circular Polarization},} \apjl, 957, L20,
  \dodoi{10.3847/2041-8213/acff70}

% type= article
\bibitem[{ {Event Horizon Telescope Collaboration} {et~al.}(2024){Event Horizon
  Telescope Collaboration}, {Akiyama}, {Alberdi}, {Alef}, {Algaba}, {Anantua},
  {Asada}, {Azulay}, {Bach}, {Baczko}, {Ball}, {Balokovi{\'c}},
  {Bandyopadhyay}, {Barrett}, {Baub{\"o}ck}, {Benson}, {Bintley}, {Blackburn},
  {Blundell}, {Bouman}, {Bower}, {Boyce}, {Bremer}, {Brissenden}, {Britzen},
  {Broderick}, {Broguiere}, {Bronzwaer}, {Bustamante}, {Carlstrom}, {Chael},
  {Chan}, {Chang}, {Chatterjee}, {Chatterjee}, {Chen}, {Chen}, {Cheng}, {Cho},
  {Christian}, {Conroy}, {Conway}, {Crawford}, {Crew}, {Cruz-Osorio}, {Cui},
  {Dahale}, {Davelaar}, {De Laurentis}, {Deane}, {Dempsey}, {Desvignes},
  {Dexter}, {Dhruv}, {Dihingia}, {Doeleman}, {Dzib}, {Eatough}, {Emami},
  {Falcke}, {Farah}, {Fish}, {Fomalont}, {Ford}, {Foschi}, {Fraga-Encinas},
  {Freeman}, {Friberg}, {Fromm}, {Fuentes}, {Galison}, {Gammie}, {Garc{\'\i}a},
  {Gentaz}, {Georgiev}, {Goddi}, {Gold}, {G{\'o}mez-Ruiz}, {G{\'o}mez}, {Gu},
  {Gurwell}, {Hada}, {Haggard}, {Hesper}, {Heumann}, {Ho}, {Ho}, {Honma},
  {Huang}, {Huang}, {Hughes}, {Ikeda}, {Violette Impellizzeri}, {Inoue},
  {Issaoun}, {James}, {Jannuzi}, {Janssen}, {Jeter}, {Jiang},
  {Jim{\'e}nez-Rosales}, {Johnson}, {Jorstad}, {Jones}, {Joshi}, {Jung},
  {Karuppusamy}, {Kawashima}, {Keating}, {Kettenis}, {Kim}, {Kim}, {Kim},
  {Kim}, {Kino}, {Koay}, {Kocherlakota}, {Kofuji}, {Koch}, {Koyama}, {Kramer},
  {Kramer}, {Kramer}, {Krichbaum}, {Kuo}, {La Bella}, {Lee}, {Levis}, {Li},
  {Lico}, {Lindahl}, {Lindqvist}, {Lisakov}, {Liu}, {Liu}, {Liuzzo}, {Lo},
  {Lobanov}, {Loinard}, {Lonsdale}, {Lowitz}, {Lu}, {MacDonald}, {Mao},
  {Marchili}, {Markoff}, {Marrone}, {Marscher}, {Mart{\'\i}-Vidal},
  {Matsushita}, {Matthews}, {Medeiros}, {Menten}, {Mizuno}, {Mizuno},
  {Montgomery}, {Moran}, {Moriyama}, {Moscibrodzka}, {Mulaudzi}, {M{\"u}ller},
  {M{\"u}ller}, {Mus}, {Musoke}, {Myserlis}, {Nagai}, {Nagar}, {Nakamura},
  {Narayanan}, {Natarajan}, {Nathanail}, {Fuentes}, {Neilsen}, {Ni}, {Nowak},
  {Oh}, {Okino}, {Olivares}, {Oyama}, {{\"O}zel}, {Palumbo}, {Paraschos},
  {Park}, {Parsons}, {Patel}, {Pen}, {Pesce}, {Pi{\'e}tu}, {PopStefanija},
  {Porth}, {Prather}, {Psaltis}, {Pu}, {Ramakrishnan}, {Rao}, {Rawlings},
  {Raymond}, {Rezzolla}, {Ricarte}, \& {Ripperda}}]{EHT2024}
{Event Horizon Telescope Collaboration}, {Akiyama}, K., {Alberdi}, A., {et~al.}
  2024, \bibinfo{title}{{The persistent shadow of the supermassive black hole
  of M 87. I. Observations, calibration, imaging, and analysis},} \aap, 681,
  A79, \dodoi{10.1051/0004-6361/202347932}

% type= article
\bibitem[{ {Event Horizon Telescope Collaboration} {et~al.}(2025){Event Horizon
  Telescope Collaboration}, {Akiyama}, {Albentosa-Ru{\'\i}z}, {Alberdi},
  {Alef}, {Algaba}, {Anantua}, {Asada}, {Azulay}, {Bach}, {Baczko}, {Ball},
  {Balokovi{\'c}}, {Bandyopadhyay}, {Barrett}, {Baub{\"o}ck}, {Benson},
  {Bintley}, {Blackburn}, {Blundell}, {Bouman}, {Bower}, {Bremer},
  {Brissenden}, {Britzen}, {Broderick}, {Broguiere}, {Bronzwaer}, {Bustamante},
  {Carlstrom}, {Chael}, {Chan}, {Chang}, {Chatterjee}, {Chatterjee}, {Chen},
  {Chen}, {Cheng}, {Cho}, {Christian}, {Conroy}, {Conway}, {Crawford}, {Crew},
  {Cruz-Osorio}, {Cui}, {Curd}, {Dahale}, {Davelaar}, {De Laurentis}, {Deane},
  {Dempsey}, {Desvignes}, {Dexter}, {Dhruv}, {Dihingia}, {Doeleman}, {Dzib},
  {Eatough}, {Emami}, {Falcke}, {Farah}, {Fish}, {Fomalont}, {Ford}, {Foschi},
  {Fraga-Encinas}, {Freeman}, {Friberg}, {Fromm}, {Fuentes}, {Galison},
  {Gammie}, {Garc{\'\i}a}, {Gentaz}, {Georgiev}, {Goddi}, {Gold},
  {G{\'o}mez-Ruiz}, {G{\'o}mez}, {Gu}, {Gurwell}, {Hada}, {Haggard}, {Hesper},
  {Heumann}, {Ho}, {Ho}, {Honma}, {Huang}, {Huang}, {Hughes}, {Ikeda},
  {Impellizzeri}, {Inoue}, {Issaoun}, {James}, {Jannuzi}, {Janssen}, {Jeter},
  {Jiang}, {Jim{\'e}nez-Rosales}, {Johnson}, {Jorstad}, {Jones}, {Joshi},
  {Jung}, {Karuppusamy}, {Kawashima}, {Keating}, {Kettenis}, {Kim}, {Kim},
  {Kim}, {Kim}, {Kino}, {Koay}, {Kocherlakota}, {Kofuji}, {Koch}, {Koyama},
  {Kramer}, {Kramer}, {Kramer}, {Krichbaum}, {Kuo}, {La Bella}, {Lee}, {Levis},
  {Li}, {Lico}, {Lindahl}, {Lindqvist}, {Lisakov}, {Liu}, {Liu}, {Liuzzo},
  {Lo}, {Lobanov}, {Loinard}, {Lonsdale}, {Lowitz}, {Lu}, {MacDonald}, {Mao},
  {Marchili}, {Markoff}, {Marrone}, {Marscher}, {Mart{\'\i}-Vidal},
  {Matsushita}, {Matthews}, {Medeiros}, {Menten}, {Mizuno}, {Mizuno},
  {Montgomery}, {Moran}, {Moriyama}, {Moscibrodzka}, {Mulaudzi}, {M{\"u}ller},
  {M{\"u}ller}, {Mus}, {Musoke}, {Myserlis}, {Nagai}, {Nagar}, {Nair},
  {Nakamura}, {Narayanan}, {Natarajan}, {Nathanail}, {Fuentes}, {Neilsen},
  {Ni}, {Nowak}, {Oh}, {Okino}, {Olivares S{\'a}nchez}, {Oyama}, {{\"O}zel},
  {Palumbo}, {Paraschos}, {Park}, {Parsons}, {Patel}, {Pen}, {Pesce},
  {Pi{\'e}tu}, {PopStefanija}, {Porth}, {Prather}, {Principe}, {Psaltis}, {Pu},
  {Ramakrishnan}, {Rao}, {Rawlings}, \& {Rezzolla}}]{EHT2025}
{Event Horizon Telescope Collaboration}, {Akiyama}, K., {Albentosa-Ru{\'\i}z},
  E., {et~al.} 2025, \bibinfo{title}{{The persistent shadow of the supermassive
  black hole of M87: II. Model comparisons and theoretical interpretations},}
  \aap, 693, A265, \dodoi{10.1051/0004-6361/202451296}

% type= article
\bibitem[{C.~M. {Fromm} {et~al.}(2022){Fromm}, {Cruz-Osorio}, {Mizuno},
  {Nathanail}, {Younsi}, {Porth}, {Olivares}, {Davelaar}, {Falcke}, {Kramer},
  \& {Rezzolla}}]{Fromm2022}
{Fromm}, C.~M., {Cruz-Osorio}, A., {Mizuno}, Y., {et~al.} 2022,
  \bibinfo{title}{{Impact of non-thermal particles on the spectral and
  structural properties of M87},} \aap, 660, A107,
  \dodoi{10.1051/0004-6361/202142295}

% type= article
\bibitem[{D. {Gabuzda}(2018){Gabuzda}}]{Gabuzda2018}
{Gabuzda}, D. 2018, \bibinfo{title}{{Evidence for Helical Magnetic Fields
  Associated with AGN Jets and the Action of a Cosmic Battery},} Galaxies, 7,
  5, \dodoi{10.3390/galaxies7010005}

% type= article
\bibitem[{D.~C. {Gabuzda} {et~al.}(2004){Gabuzda}, {Murray}, \&
  {Cronin}}]{Gabuzda2004}
{Gabuzda}, D.~C., {Murray}, {\'E}., \& {Cronin}, P. 2004,
  \bibinfo{title}{{Helical magnetic fields associated with the relativistic
  jets of four BL Lac objects},} \mnras, 351, L89,
  \dodoi{10.1111/j.1365-2966.2004.08037.x}

% type= article
\bibitem[{K. {Gebhardt} {et~al.}(2011){Gebhardt}, {Adams}, {Richstone},
  {Lauer}, {Faber}, {G{\"u}ltekin}, {Murphy}, \& {Tremaine}}]{Gebhardt2011}
{Gebhardt}, K., {Adams}, J., {Richstone}, D., {et~al.} 2011,
  \bibinfo{title}{{The Black Hole Mass in M87 from Gemini/NIFS Adaptive Optics
  Observations},} \apj, 729, 119, \dodoi{10.1088/0004-637X/729/2/119}

% type= article
\bibitem[{Z. {Gelles} {et~al.}(2025){Gelles}, {Chael}, \&
  {Quataert}}]{Gelles2025}
{Gelles}, Z., {Chael}, A., \& {Quataert}, E. 2025, \bibinfo{title}{{Signatures
  of Black Hole Spin and Plasma Acceleration in Jet Polarimetry},} \apj, 981,
  204, \dodoi{10.3847/1538-4357/adb1aa}

% type= article
\bibitem[{C. {Goddi} {et~al.}(2021){Goddi}, {Mart{\'\i}-Vidal}, {Messias},
  {Bower}, {Broderick}, {Dexter}, {Marrone}, {Moscibrodzka}, {Nagai}, {Algaba},
  {Asada}, {Crew}, {G{\'o}mez}, {Impellizzeri}, {Janssen}, {Kadler},
  {Krichbaum}, {Lico}, {Matthews}, {Nathanail}, {Ricarte}, {Ros}, {Younsi},
  {Akiyama}, {Alberdi}, {Alef}, {Anantua}, {Azulay}, {Baczko}, {Ball},
  {Balokovi{\'c}}, {Barrett}, {Benson}, {Bintley}, {Blackburn}, {Blundell},
  {Boland}, {Bouman}, {Boyce}, {Bremer}, {Brinkerink}, {Brissenden}, {Britzen},
  {Broguiere}, {Bronzwaer}, {Byun}, {Carlstrom}, {Chael}, {Chan}, {Chatterjee},
  {Chatterjee}, {Chen}, {Chen}, {Chesler}, {Cho}, {Christian}, {Conway},
  {Cordes}, {Crawford}, {Cruz-Osorio}, {Cui}, {Davelaar}, {De Laurentis},
  {Deane}, {Dempsey}, {Desvignes}, {Doeleman}, {Eatough}, {Falcke}, {Farah},
  {Fish}, {Fomalont}, {Ford}, {Fraga-Encinas}, {Freeman}, {Friberg}, {Fromm},
  {Fuentes}, {Galison}, {Gammie}, {Garc{\'\i}a}, {Gentaz}, {Georgiev}, {Gold},
  {G{\'o}mez-Ruiz}, {Gu}, {Gurwell}, {Hada}, {Haggard}, {Hecht}, {Hesper},
  {Ho}, {Ho}, {Honma}, {Huang}, {Huang}, {Hughes}, {Inoue}, {Issaoun}, {James},
  {Jannuzi}, {Jeter}, {Jiang}, {Jimenez-Rosales}, {Johnson}, {Jorstad}, {Jung},
  {Karami}, {Karuppusamy}, {Kawashima}, {Keating}, {Kettenis}, {Kim}, {Kim},
  {Kim}, {Kim}, {Kino}, {Koay}, {Kofuji}, {Koch}, {Koyama}, {Kramer}, {Kramer},
  {Kuo}, {Lauer}, {Lee}, {Levis}, {Li}, {Li}, {Lindqvist}, {Lindahl}, {Liu},
  {Liu}, {Liuzzo}, {Lo}, {Lobanov}, {Loinard}, {Lonsdale}, {Lu}, {MacDonald},
  {Mao}, {Marchili}, {Markoff}, {Marscher}, {Matsushita}, {Medeiros}, {Menten},
  {Mizuno}, {Mizuno}, {Moran}, {Moriyama}, {M{\"u}ller}, {Musoke},
  {Mej{\'\i}as}, {Nagar}, {Nakamura}, {Narayan}, {Narayanan}, {Natarajan},
  {Neilsen}, {Neri}, {Ni}, {Noutsos}, {Nowak}, {Okino}, {Olivares},
  {Ortiz-Le{\'o}n}, {Oyama}, {{\"O}zel}, {Palumbo}, {Park}, {Patel}, {Pen},
  {Pesce}, {Pi{\'e}tu}, {Plambeck}, {PopStefanija}, {Porth}, {P{\"o}tzl},
  {Prather}, {Preciado-L{\'o}pez}, {Psaltis}, {Pu}, {Ramakrishnan}, {Rao},
  {Rawlings}, {Raymond}, {Rezzolla}, {Ripperda}, {Roelofs}, {Rogers}, {Rose},
  {Roshanineshat}, {Rottmann}, {Roy}, {Ruszczyk}, {Rygl}, {S{\'a}nchez},
  {S{\'a}nchez-Arguelles}, {Sasada}, {Savolainen}, {Schloerb}, {Schuster},
  {Shao}, {Shen}, {Small}, {Sohn}, {SooHoo}, {Sun}, {Tazaki}, {Tetarenko},
  {Tiede}, {Tilanus}, {Titus}, {Toma}, {Torne}, {Trent}, {Traianou}, {Trippe},
  {van Bemmel}, {van Langevelde}, {van Rossum}, {Wagner}, {Ward-Thompson},
  {Wardle}, {Weintroub}, {Wex}, {Wharton}, {Wielgus}, {Wong}, {Wu}, {Yoon},
  {Young}, {Young}, {Yuan}, {Yuan}, {Zensus}, {Zhao}, {Zhao}, {Bruni},
  {Gopakumar}, {Hern{\'a}ndez-G{\'o}mez}, {Herrero-Illana}, {Ingram},
  {Komossa}, {Kovalev}, {Muders}, {Perucho}, {R{\"o}sch}, \&
  {Valtonen}}]{Goddi2021}
{Goddi}, C., {Mart{\'\i}-Vidal}, I., {Messias}, H., {et~al.} 2021,
  \bibinfo{title}{{Polarimetric Properties of Event Horizon Telescope Targets
  from ALMA},} \apjl, 910, L14, \dodoi{10.3847/2041-8213/abee6a}

% type= article
\bibitem[{C. {Goddi} {et~al.}(2025){Goddi}, {Carlos}, {Crew}, {Matthews},
  {Messias}, {Mus}, {Mart{\'\i}-Vidal}, {Albentosa-Ru{\'\i}z}, {De Laurentis},
  {Liuzzo}, {Marchili}, {Rygl}, {Akiyama}, {Alberdi}, {Alef}, {Carlos Algaba},
  {Anantua}, {Asada}, {Azulay}, {Bach}, {Baczko}, {Ball}, {Balokovi{\'c}},
  {Bandyopadhyay}, {Barrett}, {Baub{\"o}ck}, {Benson}, {Bintley}, {Blackburn},
  {Blundell}, {Bouman}, {Bower}, {Bremer}, {Brissenden}, {Britzen},
  {Broderick}, {Broguiere}, {Bronzwaer}, {Bustamante}, {Carlstrom}, {Chael},
  {Chan}, {Chang}, {Chatterjee}, {Chatterjee}, {Chen}, {Chen}, {Cheng}, {Cho},
  {Christian}, {Conroy}, {Conway}, {Crawford}, {Cruz-Osorio}, {Cui}, {Curd},
  {Dahale}, {Davelaar}, {Deane}, {Dempsey}, {Desvignes}, {Dexter}, {Dhruv},
  {Dihingia}, {Doeleman}, {Dzib}, {Eatough}, {Emami}, {Falcke}, {Farah},
  {Fish}, {Fomalont}, {Alyson Ford}, {Foschi}, {Fraga-Encinas}, {Freeman},
  {Friberg}, {Fromm}, {Fuentes}, {Galison}, {Gammie}, {Garc{\'\i}a}, {Gentaz},
  {Georgiev}, {Gold}, {G{\'o}mez-Ruiz}, {G{\'o}mez}, {Gu}, {Gurwell}, {Hada},
  {Haggard}, {Hesper}, {Heumann}, {Ho}, {Ho}, {Honma}, {Huang}, {Huang},
  {Hughes}, {Ikeda}, {Violette Impellizzeri}, {Inoue}, {Issaoun}, {James},
  {Jannuzi}, {Janssen}, {Jeter}, {Jiang}, {Jim{\'e}nez-Rosales}, {Johnson},
  {Jorstad}, {Jones}, {Joshi}, {Jung}, {Karuppusamy}, {Kawashima}, {Keating},
  {Kettenis}, {Kim}, {Kim}, {Kim}, {Kim}, {Kino}, {Koay}, {Kocherlakota},
  {Kofuji}, {Koyama}, {Kramer}, {Kramer}, {Kramer}, {Krichbaum}, {Kuo}, {La
  Bella}, {Lee}, {Levis}, {Li}, {Lico}, {Lindahl}, {Lindqvist}, {Lisakov},
  {Liu}, {Liu}, {Lo}, {Lobanov}, {Loinard}, {Lonsdale}, {Lowitz}, {Lu},
  {MacDonald}, {Mao}, {Markoff}, {Marrone}, {Marscher}, {Matsushita},
  {Medeiros}, {Menten}, {Mizuno}, {Mizuno}, {Montgomery}, {Moran}, {Moriyama},
  {Moscibrodzka}, {Mulaudzi}, {M{\"u}ller}, {M{\"u}ller}, {Musoke}, {Myserlis},
  {Nagai}, {Nagar}, {Nair}, {Nakamura}, {Narayanan}, {Natarajan}, {Nathanail},
  {Fuentes}, {Neilsen}, {Ni}, {Nowak}, {Oh}, {Okino}, {S{\'a}nchez}, {Oyama},
  {{\"O}zel}, {Palumbo}, {Paraschos}, {Park}, {Parsons}, {Patel}, {Pen},
  {Pesce}, {Pi{\'e}tu}, {PopStefanija}, {Porth}, {Prather}, {Principe},
  {Psaltis}, {Pu}, {Ramakrishnan}, {Rao}, \& {Rawlings}}]{Goddi2025}
{Goddi}, C., {Carlos}, D.~F., {Crew}, G.~B., {et~al.} 2025,
  \bibinfo{title}{{First polarization study of the M87 jet and active galactic
  nuclei at submillimeter wavelengths with ALMA},} \aap, 699, A265,
  \dodoi{10.1051/0004-6361/202554140}

% type= article
\bibitem[{J.~L. {G{\'o}mez} {et~al.}(2016){G{\'o}mez}, {Lobanov}, {Bruni},
  {Kovalev}, {Marscher}, {Jorstad}, {Mizuno}, {Bach}, {Sokolovsky}, {Anderson},
  {Galindo}, {Kardashev}, \& {Lisakov}}]{Gomez2016}
{G{\'o}mez}, J.~L., {Lobanov}, A.~P., {Bruni}, G., {et~al.} 2016,
  \bibinfo{title}{{Probing the Innermost Regions of AGN Jets and Their Magnetic
  Fields with RadioAstron. I. Imaging BL Lacertae at 21 Microarcsecond
  Resolution},} \apj, 817, 96, \dodoi{10.3847/0004-637X/817/2/96}

% type= inbook
\bibitem[{E.~W. {Greisen}(2003){Greisen}}]{Greisen2003}
{Greisen}, E.~W. 2003, Astrophysics and Space Science Library, Vol. 285, {AIPS,
  the VLA, and the VLBA}, ed. A.~{Heck}, 109, \dodoi{10.1007/0-306-48080-8_7}

% type= article
\bibitem[{K. {Hada} {et~al.}(2024){Hada}, {Asada}, {Nakamura}, \&
  {Kino}}]{Hada2024}
{Hada}, K., {Asada}, K., {Nakamura}, M., \& {Kino}, M. 2024, \bibinfo{title}{{M
  87: a cosmic laboratory for deciphering black hole accretion and jet
  formation},} \aapr, 32, 5, \dodoi{10.1007/s00159-024-00155-y}

% type= article
\bibitem[{K. {Hada} {et~al.}(2011){Hada}, {Doi}, {Kino}, {Nagai}, {Hagiwara},
  \& {Kawaguchi}}]{Hada2011}
{Hada}, K., {Doi}, A., {Kino}, M., {et~al.} 2011, \bibinfo{title}{{An origin of
  the radio jet in M87 at the location of the central black hole},} \nat, 477,
  185, \dodoi{10.1038/nature10387}

% type= article
\bibitem[{K. {Hada} {et~al.}(2013){Hada}, {Doi}, {Nagai}, {Inoue}, {Honma},
  {Giroletti}, \& {Giovannini}}]{Hada2013}
{Hada}, K., {Doi}, A., {Nagai}, H., {et~al.} 2013, \bibinfo{title}{{Evidence
  for a Nuclear Radio Jet and its Structure down to lsim100 Schwarzschild Radii
  in the Center of the Sombrero Galaxy (M 104, NGC 4594)},} \apj, 779, 6,
  \dodoi{10.1088/0004-637X/779/1/6}

% type= article
\bibitem[{K. {Hada} {et~al.}(2016){Hada}, {Kino}, {Doi}, {Nagai}, {Honma},
  {Akiyama}, {Tazaki}, {Lico}, {Giroletti}, {Giovannini}, {Orienti}, \&
  {Hagiwara}}]{Hada2016}
{Hada}, K., {Kino}, M., {Doi}, A., {et~al.} 2016,
  \bibinfo{title}{{High-sensitivity 86 GHz (3.5 mm) VLBI Observations of M87:
  Deep Imaging of the Jet Base at a Resolution of 10 Schwarzschild Radii},}
  \apj, 817, 131, \dodoi{10.3847/0004-637X/817/2/131}

% type= article
\bibitem[{K. {Hada} {et~al.}(2018){Hada}, {Doi}, {Wajima}, {D'Ammand o},
  {Orienti}, {Giroletti}, {Giovannini}, {Nakamura}, \& {Asada}}]{Hada2018}
{Hada}, K., {Doi}, A., {Wajima}, K., {et~al.} 2018,
  \bibinfo{title}{{Collimation, Acceleration, and Recollimation Shock in the
  Jet of Gamma-Ray Emitting Radio-loud Narrow-line Seyfert 1 Galaxy
  1H0323+342},} \apj, 860, 141, \dodoi{10.3847/1538-4357/aac49f}

% type= article
\bibitem[{M.~J. {Hardcastle} {et~al.}(2016){Hardcastle}, {Lenc}, {Birkinshaw},
  {Croston}, {Goodger}, {Marshall}, {Perlman}, {Siemiginowska}, {Stawarz}, \&
  {Worrall}}]{Hardcastle2016}
{Hardcastle}, M.~J., {Lenc}, E., {Birkinshaw}, M., {et~al.} 2016,
  \bibinfo{title}{{Deep Chandra observations of Pictor A},} \mnras, 455, 3526,
  \dodoi{10.1093/mnras/stv2553}

% type= article
\bibitem[{K. {Hirotani}(2005){Hirotani}}]{Hirotani2005}
{Hirotani}, K. 2005, \bibinfo{title}{{Kinetic Luminosity and Composition of
  Active Galactic Nuclei Jets},} \apj, 619, 73, \dodoi{10.1086/426497}

% type= article
\bibitem[{M.~A. {Hodge} {et~al.}(2018){Hodge}, {Lister}, {Aller}, {Aller},
  {Kovalev}, {Pushkarev}, \& {Savolainen}}]{Hodge2018}
{Hodge}, M.~A., {Lister}, M.~L., {Aller}, M.~F., {et~al.} 2018,
  \bibinfo{title}{{MOJAVE XVI: Multiepoch Linear Polarization Properties of
  Parsec-scale AGN Jet Cores},} \apj, 862, 151,
  \dodoi{10.3847/1538-4357/aacb2f}

% type= article
\bibitem[{T. {Hovatta} {et~al.}(2012){Hovatta}, {Lister}, {Aller}, {Aller},
  {Homan}, {Kovalev}, {Pushkarev}, \& {Savolainen}}]{Hovatta2012}
{Hovatta}, T., {Lister}, M.~L., {Aller}, M.~F., {et~al.} 2012,
  \bibinfo{title}{{MOJAVE: Monitoring of Jets in Active Galactic Nuclei with
  VLBA Experiments. VIII. Faraday Rotation in Parsec-scale AGN Jets},} \aj,
  144, 105, \dodoi{10.1088/0004-6256/144/4/105}

% type= article
\bibitem[{I.~V. {Igumenshchev} {et~al.}(2003){Igumenshchev}, {Narayan}, \&
  {Abramowicz}}]{Igumenshchev2003}
{Igumenshchev}, I.~V., {Narayan}, R., \& {Abramowicz}, M.~A. 2003,
  \bibinfo{title}{{Three-dimensional Magnetohydrodynamic Simulations of
  Radiatively Inefficient Accretion Flows},} \apj, 592, 1042,
  \dodoi{10.1086/375769}

% type= article
\bibitem[{S.~G. {Jorstad} {et~al.}(2013){Jorstad}, {Marscher}, {Smith},
  {Larionov}, {Agudo}, {Gurwell}, {Wehrle}, {L{\"a}hteenm{\"a}ki},
  {Nikolashvili}, {Schmidt}, {Arkharov}, {Blinov}, {Blumenthal}, {Casadio},
  {Chigladze}, {Efimova}, {Eggen}, {G{\'o}mez}, {Grupe}, {Hagen-Thorn},
  {Joshi}, {Kimeridze}, {Konstantinova}, {Kopatskaya}, {Kurtanidze},
  {Kurtanidze}, {Larionova}, {Larionova}, {Sigua}, {MacDonald}, {Maune},
  {McHardy}, {Miller}, {Molina}, {Morozova}, {Scott}, {Taylor}, {Tornikoski},
  {Troitsky}, {Thum}, {Walker}, {Williamson}, {Sallum}, {Consiglio}, \&
  {Strelnitski}}]{Jorstad2013}
{Jorstad}, S.~G., {Marscher}, A.~P., {Smith}, P.~S., {et~al.} 2013,
  \bibinfo{title}{{A Tight Connection between Gamma-Ray Outbursts and
  Parsec-scale Jet Activity in the Quasar 3C 454.3},} \apj, 773, 147,
  \dodoi{10.1088/0004-637X/773/2/147}

% type= inproceedings
\bibitem[{W. {Junor} {et~al.}(2001){Junor}, {Biretta}, \& {Wardle}}]{junor2001}
{Junor}, W., {Biretta}, J.~A., \& {Wardle}, J.~F.~C. 2001,
  \bibinfo{title}{{VLBA lambda lambda 6, 4 cm polarimetry of Vir A},} in IAU
  Symposium, Vol. 205, Galaxies and their Constituents at the Highest Angular
  Resolutions, ed. R.~T. {Schilizzi}, 136

% type= article
\bibitem[{J.-S. {Kim} {et~al.}(2025){Kim}, {M{\"u}ller}, {Nikonov}, {Lu},
  {Knollm{\"u}ller}, {En{\ss}lin}, {Wielgus}, \& {Lobanov}}]{Kim2025}
{Kim}, J.-S., {M{\"u}ller}, H., {Nikonov}, A.~S., {et~al.} 2025,
  \bibinfo{title}{{Imaging a ring-like structure and the extended jet of M87 at
  86 GHz},} \aap, 696, A169, \dodoi{10.1051/0004-6361/202452038}

% type= article
\bibitem[{M. {Kino} {et~al.}(2022){Kino}, {Takahashi}, {Kawashima}, {Park},
  {Hada}, {Ro}, \& {Cui}}]{Kino2022}
{Kino}, M., {Takahashi}, M., {Kawashima}, T., {et~al.} 2022,
  \bibinfo{title}{{Implications from the Velocity Profile of the M87 Jet: A
  Possibility of a Slowly Rotating Black Hole Magnetosphere},} \apj, 939, 83,
  \dodoi{10.3847/1538-4357/ac8c2f}

% type= article
\bibitem[{S.~S. {Komissarov} {et~al.}(2007){Komissarov}, {Barkov}, {Vlahakis},
  \& {K{\"o}nigl}}]{Komissarov2007}
{Komissarov}, S.~S., {Barkov}, M.~V., {Vlahakis}, N., \& {K{\"o}nigl}, A. 2007,
  \bibinfo{title}{{Magnetic acceleration of relativistic active galactic
  nucleus jets},} \mnras, 380, 51, \dodoi{10.1111/j.1365-2966.2007.12050.x}

% type= article
\bibitem[{S.~S. {Komissarov} {et~al.}(2009){Komissarov}, {Vlahakis},
  {K{\"o}nigl}, \& {Barkov}}]{Komissarov2009}
{Komissarov}, S.~S., {Vlahakis}, N., {K{\"o}nigl}, A., \& {Barkov}, M.~V. 2009,
  \bibinfo{title}{{Magnetic acceleration of ultrarelativistic jets in gamma-ray
  burst sources},} \mnras, 394, 1182, \dodoi{10.1111/j.1365-2966.2009.14410.x}

% type= article
\bibitem[{Y.~Y. {Kovalev} {et~al.}(2020){Kovalev}, {Pushkarev}, {Nokhrina},
  {Plavin}, {Beskin}, {Chernoglazov}, {Lister}, \& {Savolainen}}]{Kovalev2020}
{Kovalev}, Y.~Y., {Pushkarev}, A.~B., {Nokhrina}, E.~E., {et~al.} 2020,
  \bibinfo{title}{{A transition from parabolic to conical shape as a common
  effect in nearby AGN jets},} \mnras, 495, 3576,
  \dodoi{10.1093/mnras/staa1121}

% type= article
\bibitem[{E. {Kravchenko} {et~al.}(2020){Kravchenko}, {Giroletti}, {Hada},
  {Meier}, {Nakamura}, {Park}, \& {Walker}}]{Kravchenko2020}
{Kravchenko}, E., {Giroletti}, M., {Hada}, K., {et~al.} 2020,
  \bibinfo{title}{{Linear polarization in the nucleus of M87 at 7 mm and 1.3
  cm},} \aap, 637, L6, \dodoi{10.1051/0004-6361/201937315}

% type= article
\bibitem[{R. {Kuze} {et~al.}(2024){Kuze}, {Kimura}, \& {Toma}}]{Kuze2024}
{Kuze}, R., {Kimura}, S.~S., \& {Toma}, K. 2024,
  \bibinfo{title}{{Multiwavelength Emission from Jets and Magnetically Arrested
  Disks in Nearby Radio Galaxies: Application to M87},} \apj, 977, 22,
  \dodoi{10.3847/1538-4357/ad88f4}

% type= article
\bibitem[{R.~A. {Laing}(1980){Laing}}]{Laing1980}
{Laing}, R.~A. 1980, \bibinfo{title}{{A model for the magnetic-field structure
  in extended radio sources.},} \mnras, 193, 439,
  \dodoi{10.1093/mnras/193.3.439}

% type= article
\bibitem[{R.~A. {Laing}(1981){Laing}}]{Laing1981}
{Laing}, R.~A. 1981, \bibinfo{title}{{Magnetic fields in extragalactic radio
  sources.},} \apj, 248, 87, \dodoi{10.1086/159132}

% type= article
\bibitem[{M.~L. {Lister} {et~al.}(2018){Lister}, {Aller}, {Aller}, {Hodge},
  {Homan}, {Kovalev}, {Pushkarev}, \& {Savolainen}}]{Lister2018}
{Lister}, M.~L., {Aller}, M.~F., {Aller}, H.~D., {et~al.} 2018,
  \bibinfo{title}{{MOJAVE. XV. VLBA 15 GHz Total Intensity and Polarization
  Maps of 437 Parsec-scale AGN Jets from 1996 to 2017},} \apjs, 234, 12,
  \dodoi{10.3847/1538-4365/aa9c44}

% type= article
\bibitem[{J.~D. {Livingston} {et~al.}(2025){Livingston}, {Nikonov}, {Dzib},
  {Debbrecht}, {Kovalev}, {Lisakov}, {MacDonald}, {Paraschos}, {R{\"o}der}, \&
  {Wielgus}}]{Livingston2025}
{Livingston}, J.~D., {Nikonov}, A.~S., {Dzib}, S.~A., {et~al.} 2025,
  \bibinfo{title}{{A helical magnetic field in quasar NRAO 150 revealed by
  Faraday rotation},} \aap, 695, A260, \dodoi{10.1051/0004-6361/202453056}

% type= article
\bibitem[{A.~P. {Lobanov}(1998){Lobanov}}]{Lobanov1998}
{Lobanov}, A.~P. 1998, \bibinfo{title}{{Ultracompact jets in active galactic
  nuclei},} \aap, 330, 79.
\newblock \doarXiv{astro-ph/9712132}

% type= article
\bibitem[{R.-S. {Lu} {et~al.}(2023){Lu}, {Asada}, {Krichbaum}, {Park},
  {Tazaki}, {Pu}, {Nakamura}, {Lobanov}, {Hada}, {Akiyama}, {Kim},
  {Marti-Vidal}, {G{\'o}mez}, {Kawashima}, {Yuan}, {Ros}, {Alef}, {Britzen},
  {Bremer}, {Broderick}, {Doi}, {Giovannini}, {Giroletti}, {Ho}, {Honma},
  {Hughes}, {Inoue}, {Jiang}, {Kino}, {Koyama}, {Lindqvist}, {Liu}, {Marscher},
  {Matsushita}, {Nagai}, {Rottmann}, {Savolainen}, {Schuster}, {Shen}, {de
  Vicente}, {Walker}, {Yang}, {Zensus}, {Algaba}, {Allardi}, {Bach},
  {Berthold}, {Bintley}, {Byun}, {Casadio}, {Chang}, {Chang}, {Chang}, {Chen},
  {Chen}, {Chilson}, {Chuter}, {Conway}, {Crew}, {Dempsey}, {Dornbusch},
  {Faber}, {Friberg}, {Garc{\'\i}a}, {Garrido}, {Han}, {Han}, {Hasegawa},
  {Herrero-Illana}, {Huang}, {Huang}, {Impellizzeri}, {Jiang}, {Jinchi},
  {Jung}, {Kallunki}, {Kirves}, {Kimura}, {Koay}, {Koch}, {Kramer}, {Kraus},
  {Kubo}, {Kuo}, {Li}, {Lin}, {Liu}, {Liu}, {Lo}, {Lu}, {MacDonald},
  {Martin-Cocher}, {Messias}, {Meyer-Zhao}, {Minter}, {Nair}, {Nishioka},
  {Norton}, {Nystrom}, {Ogawa}, {Oshiro}, {Patel}, {Pen}, {Pidopryhora},
  {Pradel}, {Raffin}, {Rao}, {Ruiz}, {Sanchez}, {Shaw}, {Snow}, {Sridharan},
  {Srinivasan}, {Tercero}, {Torne}, {Traianou}, {Wagner}, {Walther}, {Wei},
  {Yang}, \& {Yu}}]{Lu2023}
{Lu}, R.-S., {Asada}, K., {Krichbaum}, T.~P., {et~al.} 2023, \bibinfo{title}{{A
  ring-like accretion structure in M87 connecting its black hole and jet},}
  \nat, 616, 686, \dodoi{10.1038/s41586-023-05843-w}

% type= article
\bibitem[{Y. {Lyubarsky}(2009){Lyubarsky}}]{Lyubarsky2009}
{Lyubarsky}, Y. 2009, \bibinfo{title}{{Asymptotic Structure of
  Poynting-Dominated Jets},} \apj, 698, 1570,
  \dodoi{10.1088/0004-637X/698/2/1570}

% type= article
\bibitem[{M. {Lyutikov} {et~al.}(2005){Lyutikov}, {Pariev}, \&
  {Gabuzda}}]{Lyutikov2005}
{Lyutikov}, M., {Pariev}, V.~I., \& {Gabuzda}, D.~C. 2005,
  \bibinfo{title}{{Polarization and structure of relativistic parsec-scale AGN
  jets},} \mnras, 360, 869, \dodoi{10.1111/j.1365-2966.2005.08954.x}

% type= article
\bibitem[{A.~P. {Marscher} {et~al.}(2008){Marscher}, {Jorstad}, {D'Arcangelo},
  {Smith}, {Williams}, {Larionov}, {Oh}, {Olmstead}, {Aller}, {Aller},
  {McHardy}, {L{\"a}hteenm{\"a}ki}, {Tornikoski}, {Valtaoja}, {Hagen-Thorn},
  {Kopatskaya}, {Gear}, {Tosti}, {Kurtanidze}, {Nikolashvili}, {Sigua},
  {Miller}, \& {Ryle}}]{Marscher2008}
{Marscher}, A.~P., {Jorstad}, S.~G., {D'Arcangelo}, F.~D., {et~al.} 2008,
  \bibinfo{title}{{The inner jet of an active galactic nucleus as revealed by a
  radio-to-{\ensuremath{\gamma}}-ray outburst},} \nat, 452, 966,
  \dodoi{10.1038/nature06895}

% type= article
\bibitem[{J.~C. {McKinney}(2006){McKinney}}]{McKinney2006}
{McKinney}, J.~C. 2006, \bibinfo{title}{{General relativistic
  magnetohydrodynamic simulations of the jet formation and large-scale
  propagation from black hole accretion systems},} \mnras, 368, 1561,
  \dodoi{10.1111/j.1365-2966.2006.10256.x}

% type= article
\bibitem[{F. {Mertens} {et~al.}(2016){Mertens}, {Lobanov}, {Walker}, \&
  {Hardee}}]{Mertens2016}
{Mertens}, F., {Lobanov}, A.~P., {Walker}, R.~C., \& {Hardee}, P.~E. 2016,
  \bibinfo{title}{{Kinematics of the jet in M 87 on scales of 100-1000
  Schwarzschild radii},} \aap, 595, A54, \dodoi{10.1051/0004-6361/201628829}

% type= article
\bibitem[{Y. {Mizuno} {et~al.}(2012){Mizuno}, {Lyubarsky}, {Nishikawa}, \&
  {Hardee}}]{Mizuno2012}
{Mizuno}, Y., {Lyubarsky}, Y., {Nishikawa}, K.-I., \& {Hardee}, P.~E. 2012,
  \bibinfo{title}{{Three-dimensional Relativistic Magnetohydrodynamic
  Simulations of Current-driven Instability. III. Rotating Relativistic Jets},}
  \apj, 757, 16, \dodoi{10.1088/0004-637X/757/1/16}

% type= article
\bibitem[{E. {Murphy} {et~al.}(2013){Murphy}, {Cawthorne}, \&
  {Gabuzda}}]{Murphy2013}
{Murphy}, E., {Cawthorne}, T.~V., \& {Gabuzda}, D.~C. 2013,
  \bibinfo{title}{{Analysing the transverse structure of the relativistic jets
  of active galactic nuclei},} \mnras, 430, 1504, \dodoi{10.1093/mnras/sts561}

% type= article
\bibitem[{M. {Nakamura} \& K. {Asada}(2013){Nakamura} \& {Asada}}]{NA2013}
{Nakamura}, M., \& {Asada}, K. 2013, \bibinfo{title}{{The Parabolic Jet
  Structure in M87 as a Magnetohydrodynamic Nozzle},} \apj, 775, 118,
  \dodoi{10.1088/0004-637X/775/2/118}

% type= article
\bibitem[{M. {Nakamura} {et~al.}(2007){Nakamura}, {Li}, \& {Li}}]{Nakamura2007}
{Nakamura}, M., {Li}, H., \& {Li}, S. 2007, \bibinfo{title}{{Stability
  Properties of Magnetic Tower Jets},} \apj, 656, 721, \dodoi{10.1086/510361}

% type= article
\bibitem[{M. {Nakamura} {et~al.}(2018){Nakamura}, {Asada}, {Hada}, {Pu},
  {Noble}, {Tseng}, {Toma}, {Kino}, {Nagai}, {Takahashi}, {Algaba}, {Orienti},
  {Akiyama}, {Doi}, {Giovannini}, {Giroletti}, {Honma}, {Koyama}, {Lico},
  {Niinuma}, \& {Tazaki}}]{Nakamura2018}
{Nakamura}, M., {Asada}, K., {Hada}, K., {et~al.} 2018,
  \bibinfo{title}{{Parabolic Jets from the Spinning Black Hole in M87},} \apj,
  868, 146, \dodoi{10.3847/1538-4357/aaeb2d}

% type= article
\bibitem[{R. {Narayan} {et~al.}(2003){Narayan}, {Igumenshchev}, \&
  {Abramowicz}}]{Narayan2003}
{Narayan}, R., {Igumenshchev}, I.~V., \& {Abramowicz}, M.~A. 2003,
  \bibinfo{title}{{Magnetically Arrested Disk: an Energetically Efficient
  Accretion Flow},} \pasj, 55, L69, \dodoi{10.1093/pasj/55.6.L69}

% type= article
\bibitem[{A.~S. {Nikonov} {et~al.}(2023){Nikonov}, {Kovalev}, {Kravchenko},
  {Pashchenko}, \& {Lobanov}}]{Nikonov2023}
{Nikonov}, A.~S., {Kovalev}, Y.~Y., {Kravchenko}, E.~V., {Pashchenko}, I.~N.,
  \& {Lobanov}, A.~P. 2023, \bibinfo{title}{{Properties of the jet in M87
  revealed by its helical structure imaged with the VLBA at 8 and 15 GHz},}
  \mnras, 526, 5949, \dodoi{10.1093/mnras/stad3061}

% type= article
\bibitem[{E.~E. {Nokhrina} {et~al.}(2019){Nokhrina}, {Gurvits}, {Beskin},
  {Nakamura}, {Asada}, \& {Hada}}]{Nokhrina2019}
{Nokhrina}, E.~E., {Gurvits}, L.~I., {Beskin}, V.~S., {et~al.} 2019,
  \bibinfo{title}{{M87 black hole mass and spin estimate through the position
  of the jet boundary shape break},} \mnras, 489, 1197,
  \dodoi{10.1093/mnras/stz2116}

% type= article
\bibitem[{E.~E. {Nokhrina} {et~al.}(2020){Nokhrina}, {Kovalev}, \&
  {Pushkarev}}]{Nokhrina2020}
{Nokhrina}, E.~E., {Kovalev}, Y.~Y., \& {Pushkarev}, A.~B. 2020,
  \bibinfo{title}{{Physical parameters of active galactic nuclei derived from
  properties of the jet geometry transition region},} \mnras, 498, 2532,
  \dodoi{10.1093/mnras/staa2458}

% type= article
\bibitem[{E.~E. {Nokhrina} \& A.~B. {Pushkarev}(2024){Nokhrina} \&
  {Pushkarev}}]{NP2024}
{Nokhrina}, E.~E., \& {Pushkarev}, A.~B. 2024, \bibinfo{title}{{Core shift in
  parabolic accelerating jets},} \mnras, 528, 2523,
  \dodoi{10.1093/mnras/stae179}

% type= article
\bibitem[{F.~N. {Owen} {et~al.}(1990){Owen}, {Eilek}, \& {Keel}}]{Owen1990}
{Owen}, F.~N., {Eilek}, J.~A., \& {Keel}, W.~C. 1990,
  \bibinfo{title}{{Detection of Large Faraday Rotation in the Inner 2
  Kiloparsecs of M87},} \apj, 362, 449, \dodoi{10.1086/169282}

% type= article
\bibitem[{V.~I. {Pariev} {et~al.}(2003){Pariev}, {Istomin}, \&
  {Beresnyak}}]{Pariev2003}
{Pariev}, V.~I., {Istomin}, Y.~N., \& {Beresnyak}, A.~R. 2003,
  \bibinfo{title}{{Relativistic parsec-scale jets: II. Synchrotron emission},}
  \aap, 403, 805, \dodoi{10.1051/0004-6361:20030350}

% type= article
\bibitem[{J. Park \& J.~C. Algaba(2022)Park \& Algaba}]{PA2022}
Park, J., \& Algaba, J.~C. 2022, \bibinfo{title}{Polarization Observations of
  AGN Jets: Past and Future,} Galaxies, 10, \dodoi{10.3390/galaxies10050102}

% type= article
\bibitem[{J. {Park} {et~al.}(2023){Park}, {Asada}, \& {Byun}}]{Park2023}
{Park}, J., {Asada}, K., \& {Byun}, D.-Y. 2023, \bibinfo{title}{{Calibrating
  VLBI Polarization Data Using GPCAL. II. Time-dependent Calibration},} \apj,
  958, 28, \dodoi{10.3847/1538-4357/acfd30}

% type= article
\bibitem[{J. {Park} {et~al.}(2021{\natexlab{a}}){Park}, {Asada}, {Nakamura},
  {Kino}, {Pu}, {Hada}, {Kravchenko}, \& {Giroletti}}]{Park2021c}
{Park}, J., {Asada}, K., {Nakamura}, M., {et~al.} 2021{\natexlab{a}},
  \bibinfo{title}{{A Revised View of the Linear Polarization in the Subparsec
  Core of M87 at 7 mm},} \apj, 922, 180, \dodoi{10.3847/1538-4357/ac26bf}

% type= article
\bibitem[{J. {Park} {et~al.}(2021{\natexlab{b}}){Park}, {Byun}, {Asada}, \&
  {Yun}}]{Park2021a}
{Park}, J., {Byun}, D.-Y., {Asada}, K., \& {Yun}, Y. 2021{\natexlab{b}},
  \bibinfo{title}{{GPCAL: A Generalized Calibration Pipeline for Instrumental
  Polarization in VLBI Data},} \apj, 906, 85, \dodoi{10.3847/1538-4357/abcc6e}

% type= article
\bibitem[{J. {Park} {et~al.}(2019{\natexlab{a}}){Park}, {Hada}, {Kino},
  {Nakamura}, {Ro}, \& {Trippe}}]{Park2019a}
{Park}, J., {Hada}, K., {Kino}, M., {et~al.} 2019{\natexlab{a}},
  \bibinfo{title}{{Faraday Rotation in the Jet of M87 inside the Bondi Radius:
  Indication of Winds from Hot Accretion Flows Confining the Relativistic
  Jet},} \apj, 871, 257, \dodoi{10.3847/1538-4357/aaf9a9}

% type= article
\bibitem[{J. {Park} {et~al.}(2021{\natexlab{c}}){Park}, {Hada}, {Nakamura},
  {Asada}, {Zhao}, \& {Kino}}]{Park2021b}
{Park}, J., {Hada}, K., {Nakamura}, M., {et~al.} 2021{\natexlab{c}},
  \bibinfo{title}{{Jet Collimation and Acceleration in the Giant Radio Galaxy
  NGC 315},} \apj, 909, 76, \dodoi{10.3847/1538-4357/abd6ee}

% type= article
\bibitem[{J. {Park} {et~al.}(2018){Park}, {Kam}, {Trippe}, {Kang}, {Byun},
  {Kim}, {Algaba}, {Lee}, {Zhao}, {Kino}, {Shin}, {Hada}, {Lee}, {Oh},
  {Hodgson}, \& {Sohn}}]{Park2018}
{Park}, J., {Kam}, M., {Trippe}, S., {et~al.} 2018, \bibinfo{title}{{Revealing
  the Nature of Blazar Radio Cores through Multifrequency Polarization
  Observations with the Korean VLBI Network},} \apj, 860, 112,
  \dodoi{10.3847/1538-4357/aac490}

% type= article
\bibitem[{J. {Park} {et~al.}(2019{\natexlab{b}}){Park}, {Hada}, {Kino},
  {Nakamura}, {Hodgson}, {Ro}, {Cui}, {Asada}, {Algaba}, {Sawada-Satoh}, {Lee},
  {Cho}, {Shen}, {Jiang}, {Trippe}, {Niinuma}, {Sohn}, {Jung}, {Zhao},
  {Wajima}, {Tazaki}, {Honma}, {An}, {Akiyama}, {Byun}, {Kim}, {Zhang},
  {Cheng}, {Kobayashi}, {Shibata}, {Lee}, {Roh}, {Oh}, {Yeom}, {Jung}, {Oh},
  {Kim}, {Hwang}, \& {Hagiwara}}]{Park2019b}
{Park}, J., {Hada}, K., {Kino}, M., {et~al.} 2019{\natexlab{b}},
  \bibinfo{title}{{Kinematics of the M87 Jet in the Collimation Zone: Gradual
  Acceleration and Velocity Stratification},} \apj, 887, 147,
  \dodoi{10.3847/1538-4357/ab5584}

% type= article
\bibitem[{J. {Park} {et~al.}(2024){Park}, {Zhao}, {Nakamura}, {Mizuno}, {Pu},
  {Asada}, {Takahashi}, {Toma}, {Kino}, {Cho}, {Hada}, {Edwards}, {Ro}, {Kam},
  {Yi}, {Lee}, {Koyama}, {Byun}, {Phillips}, {Reynolds}, {Hodgson}, \&
  {Lee}}]{Park2024}
{Park}, J., {Zhao}, G.-Y., {Nakamura}, M., {et~al.} 2024,
  \bibinfo{title}{{Discovery of Limb Brightening in the Parsec-scale Jet of NGC
  315 through Global Very Long Baseline Interferometry Observations and Its
  Implications for Jet Models},} \apjl, 973, L45,
  \dodoi{10.3847/2041-8213/ad7137}

% type= article
\bibitem[{A. {Pasetto} {et~al.}(2021){Pasetto}, {Carrasco-Gonz{\'a}lez},
  {G{\'o}mez}, {Mart{\'\i}}, {Perucho}, {O'Sullivan}, {Anderson},
  {D{\'\i}az-Gonz{\'a}lez}, {Fuentes}, \& {Wardle}}]{Pasetto2021}
{Pasetto}, A., {Carrasco-Gonz{\'a}lez}, C., {G{\'o}mez}, J.~L., {et~al.} 2021,
  \bibinfo{title}{{Reading M87's DNA: A Double Helix Revealing a Large-scale
  Helical Magnetic Field},} \apjl, 923, L5, \dodoi{10.3847/2041-8213/ac3a88}

% type= article
\bibitem[{E.~S. {Perlman} {et~al.}(1999){Perlman}, {Biretta}, {Zhou}, {Sparks},
  \& {Macchetto}}]{Perlman1999}
{Perlman}, E.~S., {Biretta}, J.~A., {Zhou}, F., {Sparks}, W.~B., \&
  {Macchetto}, F.~D. 1999, \bibinfo{title}{{Optical and Radio Polarimetry of
  the M87 Jet at 0.2'' Resolution},} \aj, 117, 2185, \dodoi{10.1086/300844}

% type= article
\bibitem[{E.~S. {Perlman} {et~al.}(2011){Perlman}, {Adams}, {Cara}, {Bourque},
  {Harris}, {Madrid}, {Simons}, {Clausen-Brown}, {Cheung}, {Stawarz},
  {Georganopoulos}, {Sparks}, \& {Biretta}}]{Perlman2011}
{Perlman}, E.~S., {Adams}, S.~C., {Cara}, M., {et~al.} 2011,
  \bibinfo{title}{{Optical Polarization and Spectral Variability in the M87
  Jet},} \apj, 743, 119, \dodoi{10.1088/0004-637X/743/2/119}

% type= article
\bibitem[{H.-Y. {Pu} \& M. {Takahashi}(2020){Pu} \& {Takahashi}}]{PT2020}
{Pu}, H.-Y., \& {Takahashi}, M. 2020, \bibinfo{title}{{Properties of Trans-fast
  Magnetosonic Jets in Black Hole Magnetospheres},} \apj, 892, 37,
  \dodoi{10.3847/1538-4357/ab77ab}

% type= article
\bibitem[{A.~B. {Pushkarev} {et~al.}(2012){Pushkarev}, {Hovatta}, {Kovalev},
  {Lister}, {Lobanov}, {Savolainen}, \& {Zensus}}]{Pushkarev2012}
{Pushkarev}, A.~B., {Hovatta}, T., {Kovalev}, Y.~Y., {et~al.} 2012,
  \bibinfo{title}{{MOJAVE: Monitoring of Jets in Active galactic nuclei with
  VLBA Experiments. IX. Nuclear opacity},} \aap, 545, A113,
  \dodoi{10.1051/0004-6361/201219173}

% type= article
\bibitem[{L. {Ricci} {et~al.}(2024){Ricci}, {Perucho}, {L{\'o}pez-Miralles},
  {Mart{\'\i}}, \& {Boccardi}}]{Ricci2024}
{Ricci}, L., {Perucho}, M., {L{\'o}pez-Miralles}, J., {Mart{\'\i}}, J.~M., \&
  {Boccardi}, B. 2024, \bibinfo{title}{{Magnetic and thermal acceleration in
  extragalactic jets. An application to NGC 315},} \aap, 683, A235,
  \dodoi{10.1051/0004-6361/202346870}

% type= article
\bibitem[{L. {Ricci} {et~al.}(2022){Ricci}, {Boccardi}, {Nokhrina}, {Perucho},
  {MacDonald}, {Mattia}, {Grandi}, {Madika}, {Krichbaum}, \&
  {Zensus}}]{Ricci2022}
{Ricci}, L., {Boccardi}, B., {Nokhrina}, E., {et~al.} 2022,
  \bibinfo{title}{{Exploring the disk-jet connection in NGC 315},} \aap, 664,
  A166, \dodoi{10.1051/0004-6361/202243958}

% type= article
\bibitem[{F.~M. {Rieger}(2019){Rieger}}]{Rieger2019}
{Rieger}, F.~M. 2019, \bibinfo{title}{{An Introduction to Particle Acceleration
  in Shearing Flows},} Galaxies, 7, 78, \dodoi{10.3390/galaxies7030078}

% type= article
\bibitem[{H. {Ro} {et~al.}(2023){Ro}, {Kino}, {Sohn}, {Hada}, {Park},
  {Nakamura}, {Cui}, {Yi}, {Chung}, {Hodgson}, {Kawashima}, {An}, {Trippe},
  {Algaba}, {Kim}, {Sawada-Satoh}, {Wajima}, {Shen}, {Cheng}, {Cho}, {Jiang},
  {Jung}, {Lee}, {Niinuma}, {Oh}, {Tazaki}, {Zhao}, {Akiyama}, {Honma}, {Lee},
  {Lu}, {Zhang}, {Asada}, {Cui}, {Hagiwara}, {Hirota}, {Kawaguchi}, {Koyama},
  {Lee}, {Oh}, {Sugiyama}, {Takamura}, {Wang}, {Hwang}, {Jung}, {Kim}, {Kim},
  {Kobayashi}, {Oh}, {Oyama}, {Roh}, \& {Yeom}}]{Ro2023}
{Ro}, H., {Kino}, M., {Sohn}, B.~W., {et~al.} 2023, \bibinfo{title}{{Spectral
  analysis of a parsec-scale jet in M 87: Observational constraint on the
  magnetic field strengths in the jet},} \aap, 673, A159,
  \dodoi{10.1051/0004-6361/202142988}

% type= inproceedings
\bibitem[{M.~C. {Shepherd}(1997){Shepherd}}]{Shepherd1997}
{Shepherd}, M.~C. 1997, \bibinfo{title}{{Difmap: an Interactive Program for
  Synthesis Imaging},} in Astronomical Society of the Pacific Conference
  Series, Vol. 125, Astronomical Data Analysis Software and Systems VI, ed.
  G.~{Hunt} \& H.~{Payne}, 77

% type= article
\bibitem[{L. {Sironi} {et~al.}(2021){Sironi}, {Rowan}, \&
  {Narayan}}]{Sironi2021}
{Sironi}, L., {Rowan}, M.~E., \& {Narayan}, R. 2021,
  \bibinfo{title}{{Reconnection-driven Particle Acceleration in Relativistic
  Shear Flows},} \apjl, 907, L44, \dodoi{10.3847/2041-8213/abd9bc}

% type= article
\bibitem[{D.~D. {Sokoloff} {et~al.}(1998){Sokoloff}, {Bykov}, {Shukurov},
  {Berkhuijsen}, {Beck}, \& {Poezd}}]{Sokoloff1998}
{Sokoloff}, D.~D., {Bykov}, A.~A., {Shukurov}, A., {et~al.} 1998,
  \bibinfo{title}{{Depolarization and Faraday effects in galaxies},} \mnras,
  299, 189, \dodoi{10.1046/j.1365-8711.1998.01782.x}

% type= article
\bibitem[{H.~C. {Spruit}(1996){Spruit}}]{Spruit1996}
{Spruit}, H.~C. 1996, \bibinfo{title}{{Magnetohydrodynamic winds and jets from
  accretion disks},} arXiv e-prints, astro,
  \dodoi{10.48550/arXiv.astro-ph/9602022}

% type= article
\bibitem[{K. {Takahashi} {et~al.}(2018){Takahashi}, {Toma}, {Kino}, {Nakamura},
  \& {Hada}}]{Takahashi2018}
{Takahashi}, K., {Toma}, K., {Kino}, M., {Nakamura}, M., \& {Hada}, K. 2018,
  \bibinfo{title}{{Fast-spinning Black Holes Inferred from Symmetrically
  Limb-brightened Radio Jets},} \apj, 868, 82, \dodoi{10.3847/1538-4357/aae832}

% type= article
\bibitem[{A. {Tchekhovskoy} {et~al.}(2010){Tchekhovskoy}, {Narayan}, \&
  {McKinney}}]{Tchekhovskoy2010}
{Tchekhovskoy}, A., {Narayan}, R., \& {McKinney}, J.~C. 2010,
  \bibinfo{title}{{Black Hole Spin and The Radio Loud/Quiet Dichotomy of Active
  Galactic Nuclei},} \apj, 711, 50, \dodoi{10.1088/0004-637X/711/1/50}

% type= article
\bibitem[{A. {Tchekhovskoy} {et~al.}(2011){Tchekhovskoy}, {Narayan}, \&
  {McKinney}}]{Tchekhovskoy2011}
{Tchekhovskoy}, A., {Narayan}, R., \& {McKinney}, J.~C. 2011,
  \bibinfo{title}{{Efficient generation of jets from magnetically arrested
  accretion on a rapidly spinning black hole},} \mnras, 418, L79,
  \dodoi{10.1111/j.1745-3933.2011.01147.x}

% type= article
\bibitem[{ {The Event Horizon Telescope Collaboration}(2025){The Event Horizon
  Telescope Collaboration}}]{EHT2025b}
{The Event Horizon Telescope Collaboration}. 2025,
  \bibinfo{title}{{Horizon-scale variability of M87* from 2017--2021 EHT
  observations},} arXiv e-prints, arXiv:2509.24593,
  \dodoi{10.48550/arXiv.2509.24593}

% type= article
\bibitem[{K. {Toma} \& F. {Takahara}(2013){Toma} \& {Takahara}}]{TT2013}
{Toma}, K., \& {Takahara}, F. 2013, \bibinfo{title}{{Efficient acceleration of
  relativistic magnetohydrodynamic jets},} Progress of Theoretical and
  Experimental Physics, 2013, 083E02, \dodoi{10.1093/ptep/ptt058}

% type= article
\bibitem[{A. {Tomimatsu} \& M. {Takahashi}(2003){Tomimatsu} \&
  {Takahashi}}]{TT2003}
{Tomimatsu}, A., \& {Takahashi}, M. 2003, \bibinfo{title}{{Relativistic
  Acceleration of Magnetically Driven Jets},} \apj, 592, 321,
  \dodoi{10.1086/375579}

% type= article
\bibitem[{Y. {Tsunetoe} {et~al.}(2025){Tsunetoe}, {Pesce}, {Narayan}, {Chael},
  {Gelles}, {Gammie}, {Quataert}, \& {Palumbo}}]{Tsunetoe2025}
{Tsunetoe}, Y., {Pesce}, D.~W., {Narayan}, R., {et~al.} 2025,
  \bibinfo{title}{{Limb-brightened Jet in M87 from Anisotropic Nonthermal
  Electrons},} \apj, 984, 35, \dodoi{10.3847/1538-4357/adc37a}

% type= inbook
\bibitem[{N. {Vlahakis}(2015){Vlahakis}}]{Vlahakis2015}
{Vlahakis}, N. 2015, Astrophysics and Space Science Library, Vol. 414, {Theory
  of Relativistic Jets}, ed. I.~{Contopoulos}, D.~{Gabuzda}, \& N.~{Kylafis},
  177, \dodoi{10.1007/978-3-319-10356-3_7}

% type= article
\bibitem[{N. {Vlahakis} \& A. {K{\"o}nigl}(2003){Vlahakis} \&
  {K{\"o}nigl}}]{VK2003}
{Vlahakis}, N., \& {K{\"o}nigl}, A. 2003, \bibinfo{title}{{Relativistic
  Magnetohydrodynamics with Application to Gamma-Ray Burst Outflows. I. Theory
  and Semianalytic Trans-Alfv{\'e}nic Solutions},} \apj, 596, 1080,
  \dodoi{10.1086/378226}

% type= article
\bibitem[{N. {Vlahakis} \& A. {K{\"o}nigl}(2004){Vlahakis} \&
  {K{\"o}nigl}}]{VK2004}
{Vlahakis}, N., \& {K{\"o}nigl}, A. 2004, \bibinfo{title}{{Magnetic Driving of
  Relativistic Outflows in Active Galactic Nuclei. I. Interpretation of
  Parsec-Scale Accelerations},} \apj, 605, 656, \dodoi{10.1086/382670}

% type= article
\bibitem[{R.~C. {Walker} {et~al.}(2018){Walker}, {Hardee}, {Davies}, {Ly}, \&
  {Junor}}]{Walker2018}
{Walker}, R.~C., {Hardee}, P.~E., {Davies}, F.~B., {Ly}, C., \& {Junor}, W.
  2018, \bibinfo{title}{{The Structure and Dynamics of the Subparsec Jet in M87
  Based on 50 VLBA Observations over 17 Years at 43 GHz},} \apj, 855, 128,
  \dodoi{10.3847/1538-4357/aaafcc}

% type= article
\bibitem[{K. {Yi} {et~al.}(2024){Yi}, {Park}, {Nakamura}, {Hada}, \&
  {Trippe}}]{Yi2024}
{Yi}, K., {Park}, J., {Nakamura}, M., {Hada}, K., \& {Trippe}, S. 2024,
  \bibinfo{title}{{Jet collimation and acceleration in the flat spectrum radio
  quasar 1928+738},} \aap, 688, A94, \dodoi{10.1051/0004-6361/202449790}

% type= article
\bibitem[{F. {Yuan} {et~al.}(2022){Yuan}, {Wang}, \& {Yang}}]{Yuan2022}
{Yuan}, F., {Wang}, H., \& {Yang}, H. 2022, \bibinfo{title}{{The Accretion Flow
  in M87 is Really MAD},} \apj, 924, 124, \dodoi{10.3847/1538-4357/ac4714}

% type= article
\bibitem[{M. {Zamaninasab} {et~al.}(2013){Zamaninasab}, {Savolainen},
  {Clausen-Brown}, {Hovatta}, {Lister}, {Krichbaum}, {Kovalev}, \&
  {Pushkarev}}]{Zamaninasab2013}
{Zamaninasab}, M., {Savolainen}, T., {Clausen-Brown}, E., {et~al.} 2013,
  \bibinfo{title}{{Evidence for a large-scale helical magnetic field in the
  quasar 3C 454.3},} \mnras, 436, 3341, \dodoi{10.1093/mnras/stt1816}

% type= article
\bibitem[{R.~T. {Zavala} \& G.~B. {Taylor}(2002){Zavala} \& {Taylor}}]{ZT2002}
{Zavala}, R.~T., \& {Taylor}, G.~B. 2002, \bibinfo{title}{{Faraday Rotation
  Measures in the Parsec-Scale Jets of the Radio Galaxies M87, 3C 111, and 3C
  120},} \apjl, 566, L9, \dodoi{10.1086/339441}

% type= article
\bibitem[{R.~T. {Zavala} \& G.~B. {Taylor}(2004){Zavala} \& {Taylor}}]{ZT2004}
{Zavala}, R.~T., \& {Taylor}, G.~B. 2004, \bibinfo{title}{{A View through
  Faraday's Fog. II. Parsec-Scale Rotation Measures in 40 Active Galactic
  Nuclei},} \apj, 612, 749, \dodoi{10.1086/422741}

\end{thebibliography}
\bibliographystyle{aasjournalv7}

\end{document}